\documentclass[11pt]{article}
\usepackage[T1]{fontenc}
\usepackage[utf8]{inputenc}
\usepackage{authblk}
\usepackage{fullpage}
\usepackage{amssymb}
\usepackage{amsmath}
\usepackage{pifont}
\usepackage{upgreek}
\usepackage{color}
\usepackage{graphicx}
\usepackage{mathtools}
\usepackage{bm}
\usepackage{subfigure}
\usepackage{url}
\usepackage{multirow}
\usepackage{stackrel}
\usepackage{paralist}
\usepackage{xspace}
\usepackage{slashed}
\usepackage{cancel}
\usepackage{footmisc}
\usepackage{cite}
\usepackage[colorlinks = true,
            linkcolor = blue,
            urlcolor  = blue,
            citecolor = blue,
            anchorcolor = blue]{hyperref}

\newcommand{\ndbd}{$0\nu\beta\beta$ }
\newcommand{\twonudbd}{$2\nu\beta\beta$ }

\title{Report of the Topical Group on Neutrino Properties for Snowmass 2021}

\author[1]{Carlo Giunti\thanks{carlo.giunti@to.infn.it (NF05 convener and editor)}}
\affil[1]{Istituto Nazionale di Fisica Nucleare (INFN), Via P. Giuria 1, I-10125 Torino, Italy}

\author[2]{Julieta Gruszko\thanks{jgruszko@unc.edu (NF05 convener and editor)}}
\affil[2]{Physics and Astronomy Department, University of North Carolina at Chapel Hill, 120 E. Cameron Ave., Chapel Hill, NC 27599}

\author[3]{Benjamin Jones\thanks{ben.jones@uta.edu (NF05 convener and editor)}}
\affil[3]{Department of Physics, University of Texas at Arlington,
Arlington, Texas 76019, USA}

\author[4]{Lisa Kaufman\thanks{ljkauf@slac.stanford.edu (NF05 convener and editor)}}  
\affil[4]{SLAC National Accelerator Laboratory, Menlo Park, California 94025, USA}

\author[4]{Diana Parno\thanks{dparno@cmu.edu (NF05 convener and editor)}}
\affil[4]{Department of Physics, Carnegie Mellon University, Pittsburgh PA 15213, USA}

\author[5]{Andrea Pocar\thanks{pocar@umass.edu, (co-editor and RF04 convener)}}
\affil[5]{Department of Physics, University of Massachusetts at Amherst, Amherst, MA 01003, USA}

\date{\today}

\begin{document}

\maketitle

\begin{abstract}
Neutrinos are the most elusive among the known elementary particles,
because of their feeble interactions with ordinary matter.
They are also the most mysterious, because of their tiny masses that suggest a novel mass generating mechanism, 
their unknown Dirac or Majorana nature,
and their big quantum mixing leading to large-amplitude flavor oscillations. 
This Topical Group focuses on neutrino properties that are not directly investigated in other Topical Groups of the Neutrino Frontier:
in particular, the absolute value of the neutrino masses, the Dirac or Majorana nature of neutrinos,
their electromagnetic properties,
their lifetime, and hypothetical exotic properties.

\end{abstract}

\setcounter{section}{-1}
\newpage
\section{Executive Summary}
\label{sec:execsummary}

The existence of non-zero neutrino mass is, to date, the only laboratory-based observation of beyond-Standard-Model physics. The disparity of mass scales between neutrinos and other fundamental particles suggests a high-energy scale for neutrino-mass generation. In models where this mechanism lies at or below the TeV scale, the physics of neutrino mass may be accompanied by complementary signatures at colliders.
In models where the scale is higher, experiments probing the nature of neutrino mass are the only feasible way of exploring this new physics. Meanwhile, the observation of neutrinoless double beta decay would provide direct evidence that lepton number is violated, opening a path to baryogenesis via leptogenesis.
As such, direct tests of the scale or nature of neutrino mass target some of the most central open questions in fundamental physics today.

The absolute mass scale of the neutrino is accessible through several complementary measurements though notably does not impact neutrino oscillations; laboratory probes measure the kinematics of beta decay, a field that has recently seen substantial technical and scientific advances. These measurements are complementary to astrophysical and cosmological approaches. Searches for neutrinoless double beta decay investigate the Majorana or Dirac nature of the neutrino. The next generation of these experiments at the ton-scale is prepared to begin construction early in the coming P5 period. Completion of these experiments is a continuing focus of the neutrino physics community. Pursuing the physics associated with neutrino mass was a key Science Driver in the 2014 P5 report, and the timely development and deployment of a US-led ton-scale neutrinoless double beta decay experiment was a top priority item in the 2015 Nuclear-Physics Long-Range Plan, a commitment that continues today under the stewardship of the DOE Office of Nuclear Physics. A rich research and development program toward beyond-ton-scale sensitivities is underway. The envisioned experiments would be sensitive to wide range of neutrino-physics phenomena, and the technologies under development have broad applications in particle physics. 

Other neutrino properties may be connected to extensions of the Standard Model, yet are not observable via oscillations.
Neutrino electromagnetic properties are of fundamental interest, and  the nascent program measuring coherent elastic neutrino-nucleus scattering (CEvNS) offers intriguing sensitivity.  Lorentz and quantum mechanical properties of neutrinos may also illuminate physics beyond the standard model.  

\subsubsection{Neutrino-mass scale}
\textit{What is the absolute mass scale of the neutrino? Complementary experimental approaches currently set limits on this mass scale; when a non-zero mass is measured, will it appear in other probes in a way consistent with our current understanding?}

Direct, kinematic measurements of the neutrino-mass scale are essential to disentangle this property from the model dependence of cosmological concordances, supernova dynamics, or neutrinoless double-beta decay. The improving sensitivity of all neutrino-mass-measurement techniques raises the possibility of a fruitful disagreement between methods. A measured neutrino mass $m_{\beta}$ within the projected KATRIN sensitivity would constrain the available model space for neutrinoless double-beta decay, and require the introduction of new physics to be reconciled with current cosmological data. Any positive measurement should be followed up using a different experimental technique and/or a different decaying isotope. There is a strong consensus to pursue realization both of cyclotron-radiation emission spectroscopy for a next-generation tritium experiment such as Project~8, and of microcalorimetry with embedded \textsuperscript{163}Ho as developed by ECHo and HOLMES, to ensure this flexibility. Continued effort to identify additional isotopes for kinematic $m_{\beta}$ measurements could open up new experimental possibilities.

\paragraph{Synergies} Improved, direct neutrino-mass measurements can be related to cosmological observations (CF7) and neutrinoless double-beta decay results (below) via neutrino oscillation parameters (NF01). Experimental efforts require developments in instrumentation (IF) and computation (CompF), and offer sensitivity to possible sterile neutrinos (NF02) and other beyond-SM physics (NF03). The neutrino-mass scale has strong implications for theory (NF08/TF11).

\subsubsection{The nature of neutrino mass}
\textit{What is the mechanism that generates neutrino mass? Is the neutrino a Majorana fermion or a Dirac fermion? How can the sensitivity of neutrinoless double-beta decay searches best be improved beyond the inverted-ordering region targeted by the next generation of experiments?} 

Detection of neutrinoless double beta decay is the only known method with plausible sensitivity to the Majorana nature of the neutrino, one of the most important open questions in particle physics. Techniques aimed at its discovery have been developed in several isotopes with tens-to-hundreds~kg scale demonstrators paving the way for ton-scale and larger discovery-class experiments over the last Snowmass period. The proposed experiments use a range of detection techniques: loaded liquid scintillator, gas and liquid time projection chamber, bolometer, and solid-state detectors are all competitive approaches. 
Certain experimental challenges, notably isotope procurement and background measurement and control, are broadly shared among experimental schemes. 
The coming ton-scale generation of experiments will probe effective Majorana neutrino masses, $m_{\beta\beta}$, as small as 18\,meV in discovery mode, fully exploring the parameter space favored by the inverted mass ordering and covering a large fraction of the parameter space associated with normal neutrino mass ordering under the light neutrino exchange mechanism.

A portfolio of international experiments, representing multiple isotopes and detection technologies, will both probe this parameter space and, importantly, permit confirmation of any discovery. There is strong consensus for the goal of building at least two ton-scale experiments with US leadership, multi-agency, and international support, as well as continuing participation in other programs worldwide.  

The neutrino community is pursuing a thriving R\&D program for beyond-ton-scale neutrinoless double-beta decay searches. This program offers exciting opportunities for upgrades that would improve the sensitivity of ton-scale experiments, and for future sensitive, ultra-low-background detectors that can access a broad physics program. Support for R\&D in multiple isotopes and with multiple technologies is recognized as vital to ensure readiness to fully explore the range of half-lives associated with normal neutrino mass ordering, or to confirm and pursue precision measurements in the case that neutrinoless double-beta decay is discovered at the ton scale.

\paragraph{Synergies} Advanced clean materials, radio-assay facilities, and underground laboratories (UF) are needed in order to support all experiments, and their continued development is a crucial part of the program.  Advances in cross-cutting areas including advanced light detectors, cryogenic detectors, and time projection chamber techniques (IF), as well as machine-learning techniques for data analysis (CompF) will have obvious impact. The central physics question is highly relevant to theory (NF08/TF11), cosmology (CF7), and collider searches for heavy neutral leptons (EF09). Although presently stewarded by DOE-Nuclear Physics (NP), many proposed beyond-ton-scale detectors would have significant HEP capabilities and synergies with HEP instrumentation and analysis approaches.  There is strong consensus that maximizing synergies between HEP and NP for subsequent neutrinoless double beta decay phases would be highly scientifically beneficial.

\subsubsection{Neutrino electromagnetic properties}

\textit{Are the electromagnetic properties of the neutrino consistent with standard-model predictions arising from radiative effects?}

In the Standard Model neutrinos have small charge radii induced by radiative corrections.
The predicted values of the electron and muon neutrino charge radii are less than an order of magnitude smaller than the
current experimental upper limits and can be tested in the next generation of accelerator and reactor experiments
through the observation of neutrino-electron elastic scattering and CEvNS.
Precision measurements of the neutrino charge radii would either be an important confirmation of the Standard Model, or would discover new physics.
The same types of experimental measurements
are also sensitive to more exotic neutrino electromagnetic properties:
magnetic moments and millicharges,
which would be certainly due to new physics beyond the Standard Model.
The discovery of millicharges or anomalously large neutrino magnetic moments 
would have also important implications for astrophysics and cosmology.

\paragraph{Synergies}
The existence of non-standard electromagnetic and other properties of neutrinos
would have many phenomenological implications for astrophysics and cosmology (CF7),
and
fundamental implications for theories beyond the Standard Model (NF03, NF08/TF11).

\subsubsection{Other neutrino properties}
\textit{Can probes of other neutrino properties, such as the neutrino lifetime or Lorentz-invariance violation in the neutrino sector, shed light on physics beyond the standard model?}

Since neutrinos are very elusive,
they can have very exotic properties that have not been discovered so far,
e.g., properties that violate the Lorentz and CPT symmetries
and gravitational interactions that violate the equivalence principle.
There is no specific experimental plan aimed at investigating these properties,
but experimentalists and phenomenologists should be alert to use new experimental data
for the exploration of all non-standard neutrino properties.

\newpage

\tableofcontents


\section{Direct neutrino-mass measurements}
\label{sec:direct-mass}

When the Standard Model was developed, neutrinos were introduced as massless leptons in well-defined flavor states $\{ \nu_e, \nu_\mu, \nu_\tau\}$ corresponding to the massive leptons. The subsequent discovery of neutrino oscillations~\cite{Super-Kamiokande:1998kpq,SNO:2002tuh} established that these three flavor states are actually quantum superpositions of three well-defined mass states $\{ \nu_1, \nu_2, \nu_3 \}$. The flavor-oscillation frequency depends on the splittings between two squared mass values, $\Delta m_{ij}^2 \equiv m^2_i - m^2_j$, so the existence of flavor oscillation ensures that there are three distinct mass values in the active neutrino sector. However, oscillation experiments are sensitive only to these squared mass differences, and not to the absolute neutrino-mass scale -- that is, the offset of the lightest neutrino-mass value from zero. In this section we discuss the use of direct kinematic measurements to probe this mass scale. Determination of the ordering of the mass values is covered in the NF01 report. Theoretical efforts are covered in the TF11 report. Here, we briefly highlight a few relevant theoretical white papers. Neutrino-mass sum rules arise from a wide range of theoretical explorations of neutrino-mass generation, and will affect the observables that probe the neutrino-mass scale\footnote{LOI: \href{https://www.snowmass21.org/docs/files/summaries/NF/SNOWMASS21-NF1_NF5-TF11_TF0_Julia_Gehrlein-025.pdf}{Gehrlein et al., Leptonic Sum Rules}}. Several possible energy-frontier testing frameworks for neutrino-mass-generating theories are explored\footnote{LOI: \href{https://www.snowmass21.org/docs/files/summaries/EF/SNOWMASS21-EF9_EF0-NF3_NF5-TF11_TF0_Manimala_Mitra-099.pdf}{Ruiz et al., Scrutinising Left Right Symmetric Extensions at LHC and Beyond}} \footnote{LOI: \href{https://www.snowmass21.org/docs/files/summaries/NF/SNOWMASS21-NF8_NF3-TF11_TF8_Julia_Gehrlein-114.pdf}{Abada et al., Testable neutrino mass models}}
\footnote{LOI: \href{https://www.snowmass21.org/docs/files/summaries/RF/SNOWMASS21-RF5_RF0-EF8_EF9-NF2_NF3-TF7_TF0-CompF2_CompF0_Ruiz_Richard-118.pdf}{Ruiz, Neutrino mass models at colliders in a post-ESU era}}. Possible time modulation of neutrino masses, arising from interaction with ultralight bosonic fields, is explored\footnote{LOI: \href{https://www.snowmass21.org/docs/files/summaries/NF/SNOWMASS21-NF3_NF1-CF2_CF0-TF11_TF0_Pedro_Machado-203.pdf}{Dev et al., Ultralight dark matter and neutrinos}}.

\subsection{General approach}
\label{sub:GeneralMass}

Recently surveyed in Ref.~\cite{Formaggio:2021nfz}, ``direct'' probes infer the neutrino-mass scale from the kinematics of $\upbeta$ decay. Since this approach relies primarily on conservation of energy, it offers the minimum model-dependence of any such probe (Sec.~\ref{sub:OtherMass}). A $\upbeta$ decay creates a $\upbeta^-$($\upbeta^+$) along with a $\bar{\nu}_e$ ($\nu_e$), which -- together with the recoil of the daughter nucleus -- share the energy corresponding to the $Q$-value of the decay. This energy goes into the kinetic energy of all three particles, as well as the mass energy of the neutrino. Since that mass energy cannot be carried away by the other particles, the presence of a non-zero neutrino mass imprints a signature on the shape of the high-energy tail of the $\upbeta$ spectrum. From Fermi's golden rule, the differential decay rate of a $\upbeta$-decaying nucleus is given by
\begin{equation}
        \frac{d\Gamma}{dE} = \frac{G_{\text{F}}^{2}\cos^{2}\left(\theta_{\text{C}}\right)}{2\pi^{3}}|M_{\text{nuc}}|^{2}F\left(Z,E\right)p\left(E+m_{e}\right)
        \cdot\sum_{f}P_{f}\epsilon_{f}\sqrt{\epsilon_{f}^{2}-m_{\nu}^{2}}\Theta\left(\epsilon_{f}-m_{\nu}\right),
    \label{Eq:DifferentialSpectrum}
\end{equation}
where $G_{\text{F}}$ is the Fermi constant, $\cos^{2}\left(\theta_{\text{C}}\right)$ is the Cabibbo angle, $|M_{\text{nuc}}|^{2}$ is the nuclear matrix element, and $F\left(Z,E\right)$ is the Fermi function with the atomic charge $Z$ of the daughter nucleus. $\epsilon_{f} = E_{0} - V_{f} - E$ is the corrected neutrino energy, where $E_{0}$ is the effective decay endpoint and $V_{f}$ are atomic or molecular excitation energy levels populated with probabilities $P_{f}$, which may affect the total lepton energy. In the special case of an electron-capture decay, there is no $\upbeta$ in the final state; rather, the decay energy is shared between the neutrino and the daughter atom, so that the high-energy tail of the de-excitation spectrum carries that same signature. The differential spectrum is similar to Eq.~\ref{Eq:DifferentialSpectrum}, but includes resonances for the capture of specific bound atomic electrons.

The experimental observable, an effective neutrino-mass squared, is an incoherent sum of the squares of the mass values $m_i^2$, each weighted by its contribution to the electron-flavor state produced in $\upbeta$ decay:
\begin{equation}
\label{eq:mbeta}
    m_\upbeta^2 = \sum_i|U_{ei}|^2 m_i^2.
\end{equation}
The weights $|U_{ei}|^2$ are given by the elements of the Pontecorvo-Maki-Nakagawa-Sakata (PMNS) matrix. Typically, we assume CPT symmetry and treat neutrinos and antineutrinos interchangeably in determining the mass scale; this assumption can in principle be tested by comparing $\bar{\nu}_e$ emitters, such as ${}^3$H, with $\nu_e$ emitters such as ${}^{163}$Ho. In this discussion, we treat only measurements in the quasi-degenerate regime, where the neutrino mass splittings are small compared to the absolute mass scale; there are presently no direct plans for extending to the hierarchical regime. 

Since the neutrino is so light, only a very small fraction of decays carry information about the neutrino mass. Competitive experimental sensitivity thus requires a high specific activity of the decaying isotope, as well as a low $Q$-value to ensure that a comparatively larger fraction of the spectrum is useful. Table~\ref{tab:numass-isotopes} lists candidate isotopes that have been considered for neutrino-mass measurements in this century. ${}^{187}$Re was used for the MANU~\cite{Galeazzi:2001ih, Gatti:2001ty} and MIBETA~\cite{sisti2004new} experiments, but was ultimately abandoned due to significant challenges with the necessary detector technology~\cite{Nucciotti:2015rsl}. The decay of the ground state of ${}^{115}$In into the first excited state of ${}^{115}$Sn has the lowest known $Q$-value of any $\upbeta$ decay, but any neutrino-mass measurement would have to contend with the higher-$Q$ decay branch to the ${}^{115}$Sn ground state, which is six orders of magnitude more likely~\cite{Andreotti:2011zza, Urban:2016duq}; a similar problem afflicts ${}^{135}$Cs~\cite{deRoubin:2020eol}. This year, three decays to excited states of ${}^{75}$As were confirmed to be kinematically allowed~\cite{Ramalho:2022fdf}. The electron-capture decays of ${}^{75}$Se are sensitive to nuclear-structure details, but the $\upbeta^-$ decay of ${}^{75}$Ge to the 1172.00(60)~keV excited state of ${}^{75}$As could be an allowed transition, depending on its spin~\cite{Ramalho:2022fdf}. This would make it more appealing as a candidate for neutrino-mass measurements, although it suffers from a short half-life as well as the higher-$Q$ ground-state-to-ground-state decay branch.

\begin{table}[tb]
    \centering
    \caption{$\upbeta$ candidate isotopes for direct kinematic measurement of the absolute neutrino-mass scale, where the transition is confirmed to be kinematically allowed, along with current experimental projects aimed at measuring $m_\upbeta$. Electron-capture decays are denoted EC. }
\begin{tabular}{|c|c|c|c|c|}
\hline
    Isotope & Decay type & $Q$-value (keV) & Half-life (yr) & Current $m_\upbeta$ efforts \\
    \hline
     ${}^3$H & $\upbeta^-$ & 18.57 & 12.3 & KATRIN, Project 8 \\
     ${}^{75}$Se & EC & 0.64(51) & 0.3279(1) & N/A\\
     ${}^{75}$Se & EC & 6.04(41) & 0.3279(1) & N/A\\
     ${}^{75}$Ge & $\upbeta^-$ &  6.56(60) & 
     0.00015750(8) & N/A\\     
     ${}^{115}$In & $\upbeta^-$ & 0.173(12) & $4.3(5) \times 10^{20}$ & N/A \\
     ${}^{135}$Cs & $\upbeta^-$ & 0.440 & $1.33\times 10^{6}$ & N/A \\
     ${}^{163}$Ho & EC & 2.833(34) & 4570 & ECHo, HOLMES \\
     ${}^{187}$Re & $\upbeta^-$ & 2470(4) & $4.12(23) \times 10^{10}$~yr & N/A \\
     \hline
\end{tabular}
    \label{tab:numass-isotopes}
\end{table}

Additional candidates for ultra-low $Q$-value decays have been proposed based on literature searches~\cite{Kopp:2009yp,Gamage:2019xvx,Keblbeck:2022twm}, but a program of precision measurements of the parent and daughter atomic masses, and of specific excitation energy levels, is required in order to establish whether these candidate transitions are energetically possible. 
Such measurements have recently ruled out the isotopes ${}^{89}\mathrm{Sr}$~\cite{Sandler:2019iws}, ${}^{76}\mathrm{As}$~\cite{Ge:2022}, ${}^{155}\mathrm{Tb}$~\cite{Ge:2022}, ${}^{112}\mathrm{Ag}$~\cite{Gamage:2022thf}, and ${}^{115}\mathrm{Cd}$~\cite{Gamage:2022thf}  as candidates for neutrino-mass measurements; more precise spectroscopic data are needed for daughter excited states of ${}^{139}\mathrm{Ba}$~\cite{Sandler:2019iws} and ${}^{113}\mathrm{Ag}$~\cite{Gamage:2022thf}. When a transition is possible, measurement of the level spins is needed to understand the importance of nuclear structure to the spectrum.

The two current isotopes of choice are ${}^3$H and ${}^{163}$Ho:
\begin{align}
    {}^3\mathrm{H} &\rightarrow {^3\mathrm{He}}^+ + \upbeta^{-} + \overline{\nu}_{e} \\
    {}^{163}\mathrm{Ho} &\rightarrow {}^{163}\mathrm{Dy}^* + \nu_{e}
\end{align}
Each of these approaches has a distinct set of challenges. For example, ${}^{163}$Ho enjoys a lower $Q$-value than ${}^3$H, but has a more complex spectral shape, and it requires a detection technique that will capture the energy from multiple de-excitation channels. ${}^3$H experiments need only be concerned with determining the energy of a single electron, but this simplicity also leaves them vulnerable to systematic effects resulting in energy loss. 

The fundamental analysis strategy for such a search involves fitting a predicted spectral shape, with $m_\beta^2$ as a free parameter, to the acquired data. Because the rate falls off steeply with energy in this region of the spectrum, improperly handled systematic effects tend to mis-assign low-energy events to high-energy bins, altering the spectral shape in such a way as to drive $m_\beta^2$ more negative~\cite{Robertson:1988xca}. Indeed, several negative $m_\beta^2$ results from the early 1990s were traced to previously underappreciated systematics~\cite{Formaggio:2021nfz}. The achievable sensitivity and discovery potential depends, of course, both on systematics and statistics; a recent discussion of these issues, in the context of a Bayesian framework, is informative~\cite{AshtariEsfahani:2020bfp}.

\subsection{Tritium-based experiments: status and prospects}
\label{sub:Tritium}

Historically, the world-leading limits on $m_\beta$ have always been set by tritium-based experiments. As a $\upbeta$-decaying isotope, ${}^3$H (henceforth abbreviated T) decay always produces two leptons in the final state, and information about $m_\beta$ can be obtained through a precise measurement of the $\upbeta$ spectrum. The first challenge of such a measurement arises from the chemical reactivity of T: the most technically straightforward way to ensure a stable source is to bind the T into molecules, but in that case a T decay may excite rotational, vibrational or electronic modes that will distort the measured energy spectrum; in fact, two past negative $m_\beta^2$ results, from the Los Alamos~\cite{Robertson:1991vn} and Livermore~\cite{Stoeffl:1995wm} experiments, were later explained by updated calculations of these excitations~\cite{Bodine:2015sma}. Both the energy levels and the corresponding probabilities of the molecular final-state distribution (FSD) must be understood in detail for the spectral calculation (cf. Eq.~\ref{Eq:DifferentialSpectrum}). Precise, detailed quantum-chemical calculations, using a geminal basis with explicit correlations between the two electrons, are now available for all isotopic molecules of T$_2$~\cite{Saenz:2000dul, Doss:2006zv, KATRIN:2021fgc}. The FSD cannot be measured directly, but the calculation agrees well with spectroscopic data as well as a recent measurement of the dissociation probabilities following $\upbeta$ decay in T$_2$ and HT~\cite{TRIMS:2020nsv}. To avoid residual uncertainties and broadening from the FSD, the Project~8 collaboration is working to realize an atomic tritium source (Sec.~\ref{ssb:Project8}).

\subsubsection{KATRIN}
\label{ssb:KATRIN}

\begin{figure}[tb]
    \centering
    \includegraphics[width=0.6\textwidth]{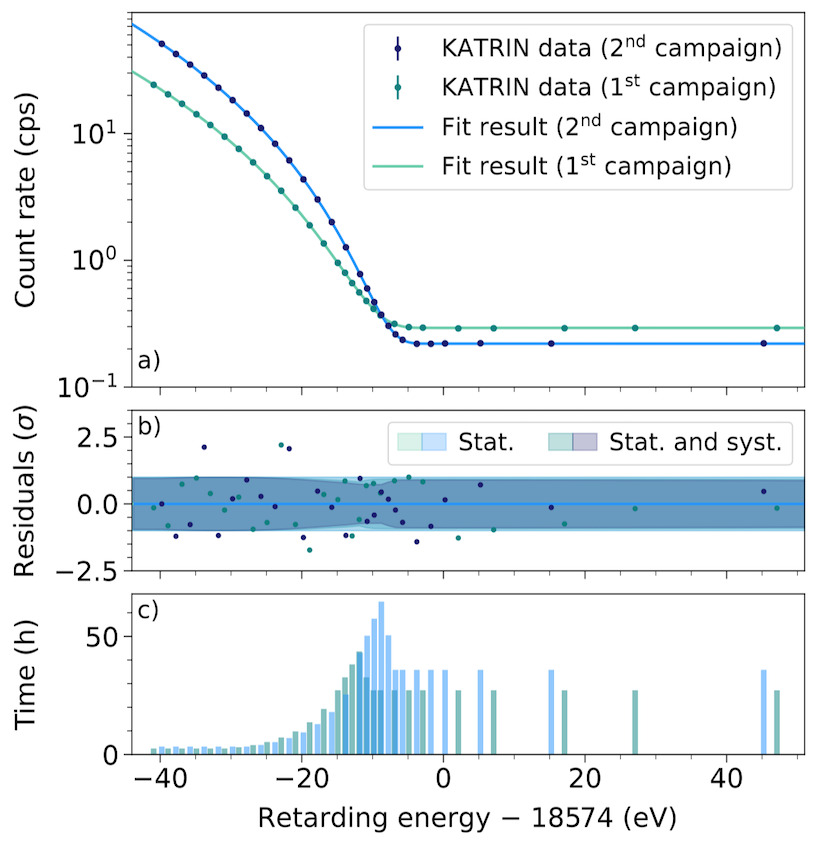}
    \caption{(a) Measured KATRIN spectra and joint-fit results from the first (green) and second (blue) neutrino-mass measurement campaigns. The reduced activity of the first campaign is clearly visible. (b) Residuals from the joint fit, in which $m_\beta^2$ was a shared parameter between the two campaigns (along with some systematics), but the amplitude, endpoint, background and other systematics were permitted to vary between campaigns. (c) Measurement-time distribution for the two campaigns, differing largely because of the reduced background for the second campaign. The retarding energy is $qU$. Reproduced from Ref.~\cite{KATRIN:2021uub}.}
    \label{fig:katrinspect}
\end{figure}

The Karlsruhe Tritium Neutrino (KATRIN) experiment, currently running in Karlsruhe, Germany, has set the world's best direct limit on the neutrino-mass scale -- $m_\beta < 0.8$~eV (90\% C.L.) -- based on its first calendar year of operations, in 2019~\cite{KATRIN:2021uub}. Within KATRIN's 70~m beamline~\cite{KATRIN:2021dfa}, $\upbeta$s are adiabatically guided along magnetic field lines from an intense, gaseous T$_2$ source ($\mathcal{O}(10^{11}$~Bq)) to a spectrometer that analyzes their energy via magnetic adiabatic collimation with an electrostatic filter (MAC-E filter). This filter essentially sets a threshold for the $\upbeta$ kinetic energy via its retarding potential $U$: if the $\upbeta$ has energy $E > qU$, where $q$ is the electron charge, it will be transmitted downstream to the detector; otherwise, it will be reflected upstream. The filter width $\Delta E$, a measure of how well the $\upbeta$s are collimated, varies as the magnetic-field ratio $B_{\mathrm{min}}/B_{\mathrm{max}}$.
The resulting integral spectrum is recorded in a segmented, Si p-i-n diode while scanning $U$ according to a pre-determined measurement-time distribution. 

KATRIN has been acquiring data since 2019; Fig.~\ref{fig:katrinspect} shows a joint fit of the two published measurement campaigns, both acquired in 2019. In the first, the source activity was limited by a burn-in phase of the tritium system~\cite{Aker2019knm1, Aker2021Knm1}; in the second, a more effective regeneration of the liquid-nitrogen-cooled baffles reduced a pernicious, non-Poissonian background source~\cite{KATRIN:2021uub}. In combination, these data result in a slightly positive best-fit value of $m_\beta^2$, and a limit of $m_\beta < 0.8$~eV (90\% C.L.). This result is strongly statistics-limited.

As detailed in Ref.\footnote{\label{WP:KATRIN}White paper: KATRIN: Status and Prospects for the Neutrino Mass and Beyond~\cite{aker2022katrin}}, the KATRIN collaboration is actively addressing its systematic uncertainties -- including plasma effects in the source, energy-loss through scattering~\cite{KATRIN:2021rqj}, and molecular final states -- through computation, simulation, and dedicated measurement campaigns. The spectrometer backgrounds, elevated substantially above first predictions, are the largest obstacle to the experiment's design sensitivity of 0.2~eV (90\% C.L.). Since signal $\upbeta$s are decelerated almost to rest within the MAC-E filter before being re-accelerated on their way to the detector, low-energy secondary electrons created in the central region of the MAC-E filter are energetically indistinguishable from the signal. The two primary sources of these secondary electrons are the ionization interactions of magnetically trapped shakeoff electrons from ${}^{219}$Rn decay, and the blackbody-induced ionization of highly excited Rydberg atoms sputtered from the spectrometer walls by $\upalpha$ decay of ${}^{210}$Po. The former source can be greatly mitigated by proper regeneration of liquid-nitrogen-cooled baffles between the primary spectrometer volume, and the ${}^{219}$Rn-emanating nonevaporable getter strips that maintain the spectrometer vacuum. The latter source is more difficult, but KATRIN is pursuing several promising lines of R\&D, including a pre-detector filter that would discriminate on the different angular distributions of the signal and background electrons, and a scheme for de-exciting the Rydberg atoms with THz radiation before they can be ionized. KATRIN plans about 3 more calendar years of running, followed by a second phase in which it will search for sterile neutrinos at the keV scale\footref{WP:KATRIN}.

\subsubsection{Project 8}
\label{ssb:Project8}

\begin{figure}[tb]
    \centering
    \includegraphics[width=0.45\textwidth]{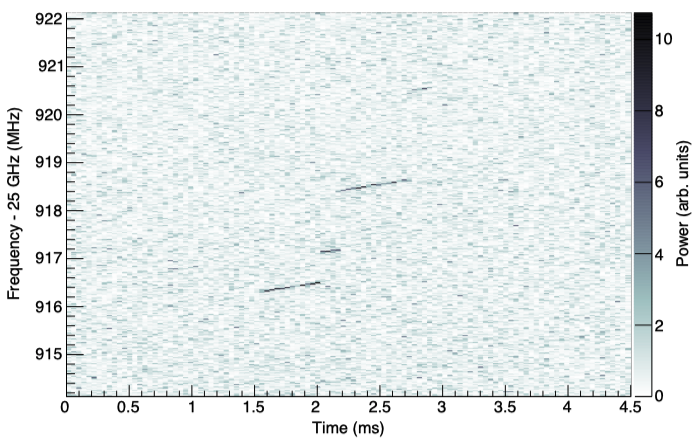}
    \includegraphics[width=0.45\textwidth]{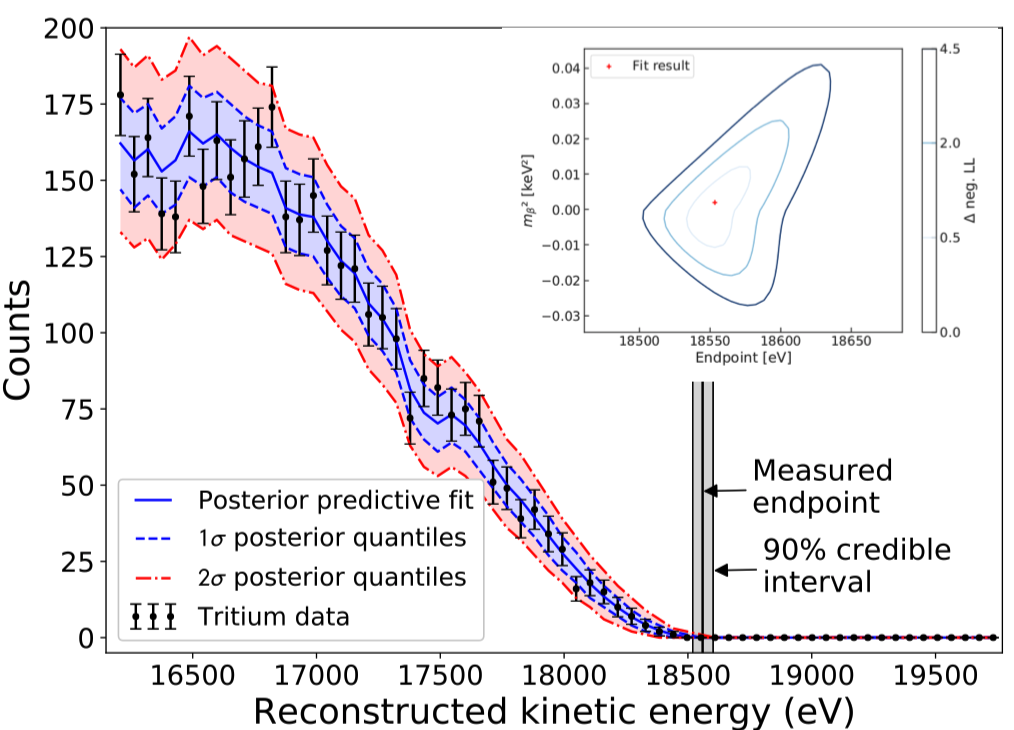}  
    \caption{\textbf{Left:} Example Project 8 spectrogram of a 17.8-keV conversion electron from ${}^{83m}$Kr decay. The frequency axis gives the output after 24.2 GHz down-conversion. The color scale gives the power in dBm (-120~dBm=1~fW). A higher cyclotron frequency corresponds to a lower electron energy. The gradual slope of each line segment shows energy loss due to cyclotron radiation; each discrete jump in energy corresponds to a scattering interaction with the gas in the source. The initial electron kinetic energy, at the lower left extreme of the track, is extracted to build the spectrum.  Reproduced from Ref.~\cite{esfahani2022project}. \textbf{Right:} Measured T$_2$ endpoint spectrum, with overlaid Bayesian fit~\cite{esfahani2022project}. The inset shows the neutrino-mass and endpoint contours from the maximum-likelihood fit to the spectrum.}
    \label{fig:p8-spectrogram}
\end{figure}

The Project~8 experiment, currently in development in Seattle, Washington, aims to measure the $\upbeta$ spectrum without transporting the $\upbeta$s out of the source gas. Instead, the $\upbeta$s are magnetically trapped within the source, so that radiofrequency antennae or waveguides can measure their cyclotron frequencies, which (for relativistic particles) depend on their kinetic energy. This innovative technique~\cite{Monreal:2009za} has been dubbed cyclotron radiation emission spectroscopy (CRES). Near the tritium endpoint, a $\upbeta$ radiates at about 1~fW, making signal pickup a technical challenge: trapping the particle under observation is a necessity for this method. The observed spectrogram is complex (Fig.~\ref{fig:p8-spectrogram}, left), but a sophisticated analysis can uncover rich information about the $\upbeta$ kinematics~\cite{AshtariEsfahani:2019yva}. Early tests with a ${}^{83m}$Kr source, with a small source volume viewed by a waveguide, enabled testing and optimization of different magnetic traps, achieving an energy resolution as good as $\sim 3$~eV~\cite{Project8:2017nal}. With an upgraded, small-volume apparatus, Project 8 has acquired its first tritium spectra (Fig.~\ref{fig:p8-spectrogram}, right) and inferred the first preliminary CRES-based neutrino-mass limit, $m_\beta < 185$~eV (90\% C.L.)~\cite{P8:TAUP21}.

One immediate near-term challenge, as Project~8 aims to scale up to achieve a competitive sensitivity to $m_\beta$, is to scale up the sensitive volume in which electrons can be trapped and their cyclotron radiation observed. This requires moving away from waveguide technology. The collaboration plans to build and test a mode-filtered resonant cavity for this next phase; a free-space cavity viewed by rings of antennae is a possible backup option. Reducing the operating frequency from 26~GHz to 1~GHz may improve the systematic picture, e.g. by reducing dipolar spin-flip losses. In all cases the trapping and detection efficiencies may vary based on frequency, effects which must be carefully studied. Radiation from trapped electrons is also affected by magnetic-field variations along their paths.

A relatively small percentage of created $\upbeta$s have pitch angles appropriate for trapping, which also significantly limits the background. The Project~8 collaboration has identified only one possible background so far, in which cosmic rays passing through the source gas eject $\delta$ electrons. In Phase II, this background is limited by tritium measurements to $< 3 \times 10^{-10}$~eV$^{-1}$s$^{-1}$ (90\% CL)~\cite{esfahani2022project}.

\begin{figure}[tb]
    \centering
    \includegraphics[width=0.4\textwidth]{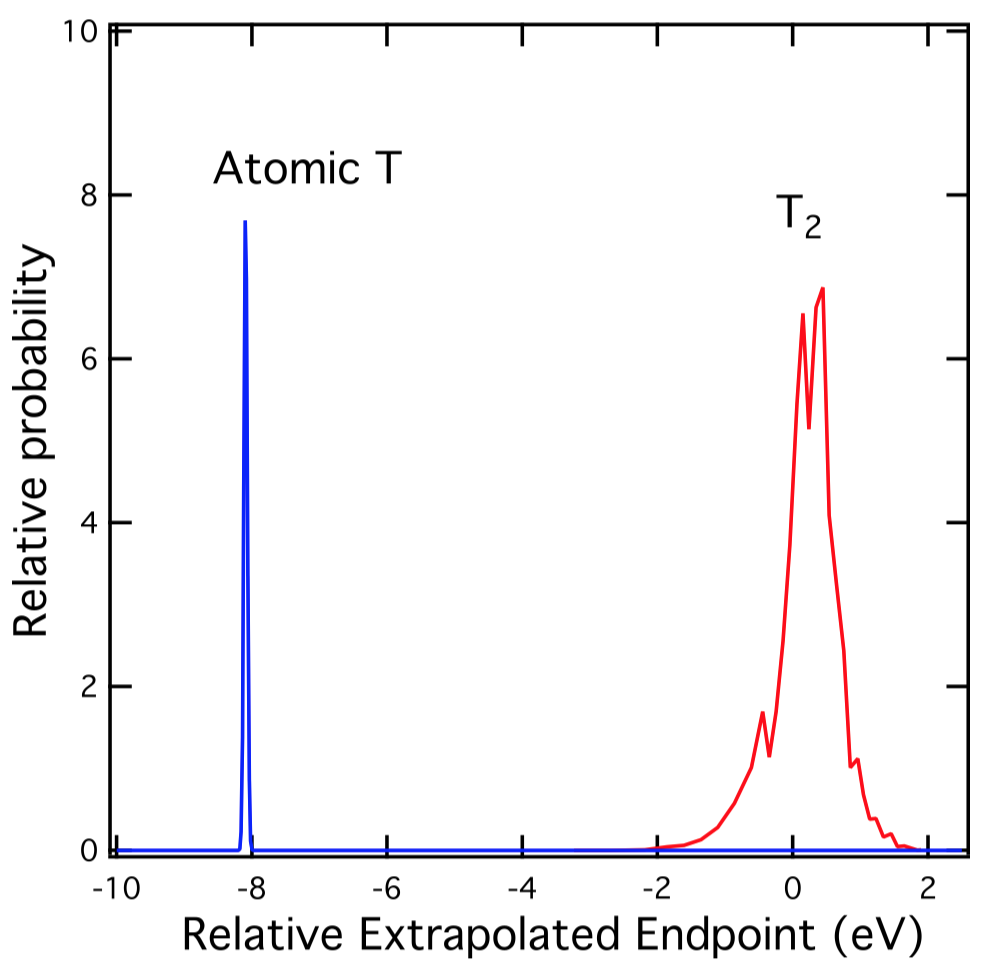}
    \caption{Final-state distributions for atomic T (blue) and molecular T$_2$ (red; electronic ground state only). The atomic line includes Doppler broadening typical of 1~K. Because the atomic endpoint is below the molecular endpoint, residual T$_2$ represents a dangerous background for spectra measured with an atomic source. Reproduced from Ref.~\cite{Formaggio:2021nfz}.}
    \label{fig:atomic-molecular-T2}
\end{figure}

Project~8's use of magnetic trapping offers an intriguing possibility for achieving an atomic tritium source, which would significantly reduce the influence of final-state effects (Fig.~\ref{fig:atomic-molecular-T2}). Ref.~\cite{Formaggio:2021nfz} gives an overview of the challenges inherent in this task. Project~8 is actively working to overcome these challenges in pursuit of an ultimate, cold atomic phase of the experiment. First, T$_2$ gas must be dissociated into its individual atoms, a process which gives them high kinetic energies; an accommodator is thus required in order to cool them to the few-mK level. The cooled T atoms must be selectively transported into a large source volume, where only low-field-seeking states can be trapped using Ioffe or Halbach array magnetic traps. Early design progress has been encouraging, but the final system of course also relies on a demonstration of large-volume CRES\footnote{White paper: The Project 8 Neutrino Mass Experiment~\cite{esfahani2022project}}.


\subsection{\textsuperscript{163}Ho-based experiments: status and prospects}
\label{sub:Capture}

To capture the diffuse atomic de-excitation energy following the decay of ${}^{163}$Ho, the Electron Capture ${}^{163}$Ho (ECHo) and HOLMES experiments  use microcalorimetric techniques originating in X-ray astronomy. The ${}^{163}$Ho source is embedded in a gold absorber with a weak thermal link to a low-temperature heat bath. With the exception of the escaping $\nu_{e}$, the decay energy is absorbed by the gold, and the resulting temperature increase $\Delta T$ is recorded. ECHo and HOLMES use two different strategies for encoding the temperature increase as a change in magnetic field, enabling readout by inherently low-noise SQUIDs (superconducting quantum interference devices). 

The temperature readout of each channel may be different, but ECHo and HOLMES share a variety of experimental challenges. First, ${}^{163}$Ho does not have sufficient natural abundance for the required activity, and must be manufactured. There are two fundamental strategies for this manufacture. Neutron irradiation of ${}^{162}\mathrm{Er}_2\mathrm{O}_3$ has a high cross-section for ${}^{163}$Ho production, but also creates significant radioactive impurities that can produce backgrounds in a measurement -- including ${}^{166m}$Ho, which is especially pernicious because it cannot be chemically separated from ${}^{163}$Ho. Meanwhile, proton irradiation of ${}^{\mathrm{nat}}$Dy creates high-purity samples, at the cost of a much lower cross section~\cite{Croce:2015kwa}.

Once a sufficiently large and purified ${}^{163}$Ho sample has been obtained, its implantation onto the first layer of the absorber, and its subsequent coverage by a second sandwich layer of absorber material, must be carefully controlled in order to optimize the activity of each detector while providing complete encapsulation of the decaying nuclei. Then, as the activity of each detector is increased, pileup becomes an insidious source of background, superimposing low-energy structures on the sensitive portion of the spectrum. Finally, the balance of specific activity with pileup requires a large number of detectors, which must all be read out and combined into a single spectrum. Thanks to its well-defined resonances, the ${}^{163}$Ho spectrum is self-calibrating to a large degree, but sophisticated multiplexing techniques are required to read out hundreds or thousands of channels in an economical way. Software-driven radio provides a promising framework for such multiplexing schemes.

A strong theoretical understanding of the ${}^{163}$Ho spectrum is essential. The recent establishment of the $Q$-value at 2.83~keV~\cite{ECHo:2015qgh}, via Penning-trap mass-difference measurements, means that the endpoint region of the spectrum is well separated from the primary resonances, at some cost in rate. The importance of shakeoff electrons, sometimes resulting in multiple vacancies, has recently been recognized~\cite{Robertson:2014fka, Faessler:2015txa, DeRujula:2016fdu}. Finally, the measured spectrum will be modified by the solid-state environment of the isotope decaying in its absorber. Substantial progress has been made with solid-state methods~\cite{Brass:2017kov, Brab:2020uzx}. Detailed tests of microcalorimeter response to calibration sources~\cite{Koehler:2018hzx} will be an important validation of these calculations.

Both ECHo (Sec.~\ref{ssb:ECHo}) and HOLMES (Sec.~\ref{ssb:HOLMES}) are currently working towards eV-scale neutrino-mass sensitivity measurements with prototype experiments, while conducting R\&D work targeted at larger deployments. As necessary milestones on the way to a sub-eV sensitivity, the community must demonstrate successful analysis of spectra with sizeable ($\mathcal{O}(50)$ detectors) microcalorimeter arrays; high-resolution spectral measurement with multiplexed detectors; and large-scale, high-yield detector fabrication. In the long term, a sub-eV measurement would benefit from a joint effort between the two extant collaborations, and from significantly increased US involvement to help address challenges around scaled detector fabrication and readout\footnote{White Paper: Measuring the electron neutrino mass using the electron capture decay of ${}^{163}$Ho~\cite{Ullom:2022kai}}.

\subsubsection{ECHo}
\label{ssb:ECHo}
\begin{figure}[tb]
    \centering
    \includegraphics[width=0.6\textwidth]{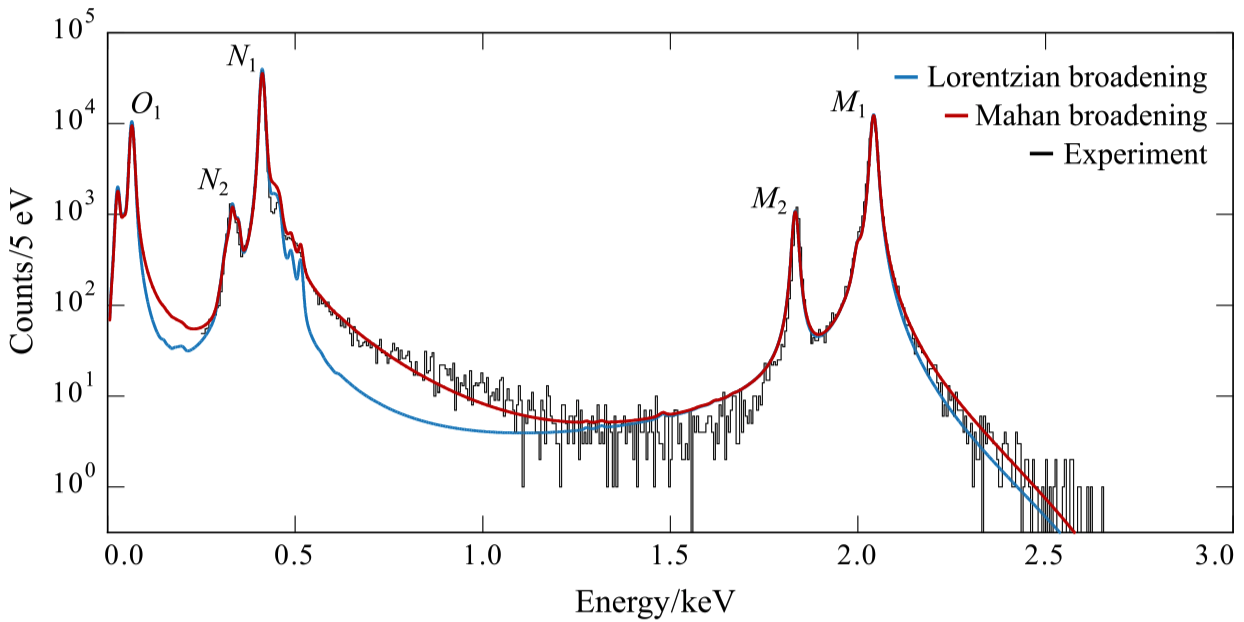}
    \caption{Experimental ${}^{163}$Ho spectrum (black) measured with four microcalorimeters over four days by the ECHo collaboration. The spectrum is overlaid with an ab initio calculation, broadened by Lorentzian (blue) and Mahan (red) lineshapes. In both cases the theoretical spectrum is shifted by +6~eV and stretched by 1\% in energy scale.  Reproduced from Ref.~\cite{Velte:2019jvx}.}
    \label{fig:echospect}
\end{figure}

In the ECHo approach of a metallic magnetic calorimeter (MMC) placed in a uniform magnetic field, $\Delta T$ alters the magnetization of an attached paramagnetic sensor, leading to a change in magnetic flux~\cite{Gastaldo:2017edk}. Fig.~\ref{fig:echospect} shows a recent spectrum acquired with four ECHo MMC detector pixels. ECHo is currently analyzing data from the ECHo-1k demonstrator, which uses an array of 58 pixels with average activity about 0.5~Bq and energy resolution about 5--6~keV FWHM~\cite{Mantegazzini:2021yed} to target a neutrino-mass sensitivity of $m_\beta \lesssim 20$~eV. Each pixel detector is read out with an individual SQUID circuit, with a two-level data-reduction scheme to reduce spurious events~\cite{Hammann:2021nuj}. Background from natural radioactivity in the detector materials is adequately controlled~\cite{Goggelmann:2022nin}.

The collaboration has performed extensive work on ${}^{163}$Ho production by neutron bombardment at a nuclear research reactor~\cite{Dorrer:2018apa}, along with isotopic purification and implantation at an optimized RISIKO facility~\cite{Kieck:2018sap, Kieck:2019fcr}. R\&D efforts presently focus on the ECHo-100k phase, with a projected 2~eV sensitivity achieved by thousands of multiplexed detector pixels, operated at mK temperatures with a per-pixel activity of order 10~Bq.

\subsubsection{HOLMES}
\label{ssb:HOLMES}
HOLMES~\cite{Alpert:2014lfa} attaches a transition-edge sensor (TES) to each microcalorimeter. This sensor is held near its superconducting transition temperature, so that $\Delta T$ induces a large change in resistance, altering the current through a small electromagnet and thereby inducing a change of magnetic flux that can be measured by a SQUID. HOLMES envisions significantly higher activity in each detector than does ECHo, $\mathcal{O}(100)$~Bq, along with higher rise times.  

HOLMES has demonstrated satisfactory production of ${}^{163}$Ho by neutron bombardment at a nuclear research reactor, as well as chemical separation~\cite{Heinitz:2018bav}. The collaboration is now developing an ion beam for performing implantation on the detectors, with an anticipated mass-separation efficiency of $\mathcal{O}(10^5)$ for reducing the quantity of ${}^{166m}$Ho in a ${}^{163}$Ho beam. After tests with ${}^{\mathrm{nat}}$Ho, the collaboration aims to switch to enriched beams later this year, which will allow production of a 64-detector array with per-detector activity of $\mathcal{O}(1)$~Bq. Early multiplexing tests have been successful on a small, two-detector prototype, and the scheme can be extended to the larger array~\cite{HOLMES:2019ykt}. A novel data-reduction scheme, built on machine-learning techniques, could improve the effective time resolution despite relatively long rise times~\cite{Borghesi:2021fda}. In parallel with analysis of the resulting high-statistics spectrum, further improvements on the ion beam will allow increased detector production with an increased per-detector activity.

\subsection{Comment on additional physics reach}
\label{sub:CommentMass}

In each of the efforts described above, the detailed spectral measurement provides sensitivity to physics targets beyond the absolute neutrino-mass scale. For example, if the measurement extends to sufficiently low energies, it will be sensitive to the characteristic spectral distortion from a fourth neutrino-mass state $m_4$. This distortion would be located at $E_0 - m_4$ with an amplitude set by the mixing angle $\theta_{14}$. This appealing, non-oscillation sterile-neutrino search is addressed in more detail by the NF02 topical group. KATRIN has already set limits on eV-scale sterile neutrinos~\cite{KATRIN:2020dpx, KATRIN:2022ith}, and plans an upgrade that will allow high-rate scans with sensitivity at the keV scale.

Additional beyond-Standard-Model physics topics,  accessible in principle in these spectral measurements, include Lorentz-invariance violation; non-standard neutrino interactions; right-handed weak currents; and emission of new particles such as light bosons. These topics are covered in the NF03 topical group.

Neutrino capture on a radioactive nucleus is a threshold-free process, making it an attractive candidate for detecting relic neutrinos from the earliest moments of the universe. In the endpoint region of the $\upbeta$ or electron-capture spectrum, the signal from these captures would be a small peak, one neutrino mass above $E_0$. Based on its early data sets, KATRIN has set limits on a local relic-neutrino overdensity~\cite{KATRIN:2022kkv}. The PTOLEMY collaboration\footnote{LOI: \href{https://www.snowmass21.org/docs/files/summaries/NF/SNOWMASS21-NF4_NF10_PTOLEMY-021.pdf}{PTOLEMY: Towards Direct Detection of the Cosmic Neutrino Background}}~\cite{PTOLEMY:2018jst, PTOLEMY:2019hkd} plans to detect relic neutrinos with an exceptionally large tritium target and extremely precise measurement of the resulting $\upbeta$ spectrum. The NF04 topical group addresses this in more detail.

In addition, each experiment will be able to probe the atomic and molecular dynamics of the targeted decay, and may make contributions to the spectroscopy of its calibration sources. 

\subsection{Other methods of neutrino mass measurement}
\label{sub:OtherMass}

Direct, kinematic spectral measurements, as described above, are one of four well-developed, complementary approaches for probing the absolute neutrino-mass scale. We briefly describe the others below. Note that the relationships between observables depend explicitly on the standard 3-neutrino oscillation framework and on the standard cosmological concordance; if these assumptions are not valid, then comprehensive measurement and comparison of these observables will help to illuminate new physics.

\paragraph{Astrophysics}
Beyond $\upbeta$-decay measurements, an alternate kinematic probe arises from the detection of a population of neutrinos arising from a galactic supernova. The arrival time of a massive neutrino at a distant detector on Earth will depend on its kinetic energy; this dispersion in the time-of-flight data yields information about the effective mass across the three mass states. The timing distribution of SN1987A neutrinos gives an upper limit $m_\nu^{\mathrm{TOF}}<5.7$~eV (95\% C.L.)~\cite{LoredoLamb:1987a:2002}. Current and future neutrino detectors are designed to collect substantially greater statistics from the next galactic supernova, although some limitations arise from imprecise knowledge of the core collapse itself. Ref.~\cite{Hansen:2019giq} explores the achievable sensitivity, including sensitivity from a time-of-flight comparison between neutrinos and gravitational waves. The NF04 topical group is providing a more comprehensive summary of supernova-neutrino science and prospects. We highlight forthcoming white papers on this topic\footnote{LOI: \href{https://www.snowmass21.org/docs/files/summaries/NF/SNOWMASS21-NF4_NF5-TF11_TF0-158.pdf}{P. Dedin and M. Santos, Neutrinos from Supernovae}},\footnote{LOI: \href{https://www.snowmass21.org/docs/files/summaries/NF/SNOWMASS21-NF8_NF4-CF3_CF7-TF9_TF8_Johns-121.pdf}{Johns et al., Supernova neutrinos and particle-physics opportunities}}.

\begin{figure}[tb]
    \centering
    \includegraphics[width=0.45\textwidth]{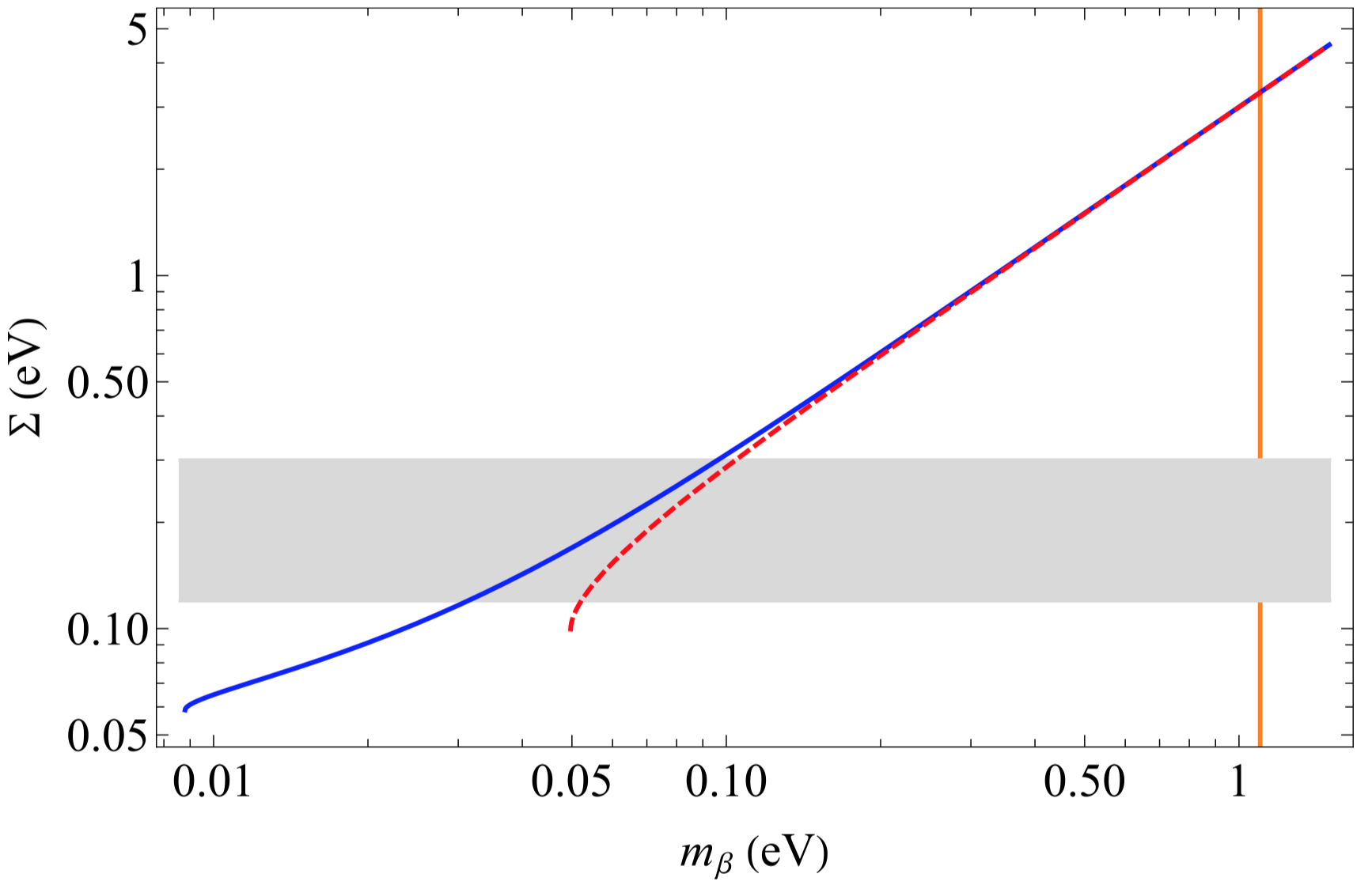}
    \includegraphics[width=0.45\textwidth]{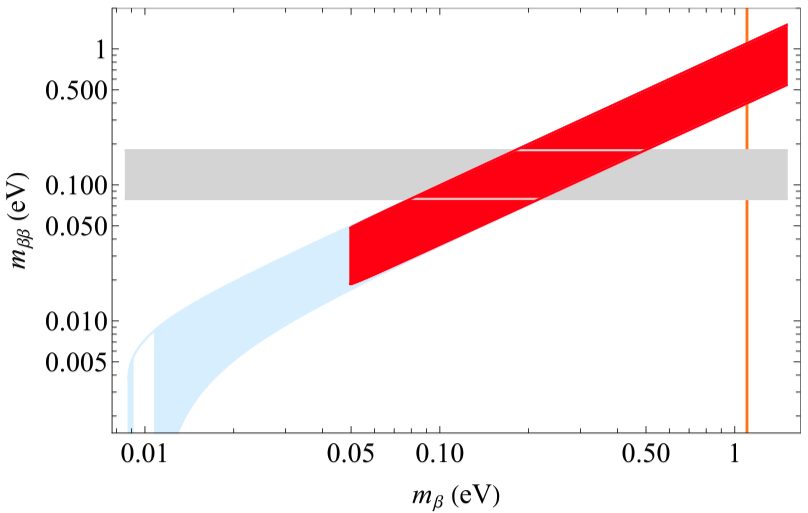}
    \caption{\textbf{Left:} The cosmological neutrino-mass observable $\Sigma m_i$ plotted against $m_\beta$ for normal (solid, blue) and inverted (dashed, red) orderings. The gray band shows the range of 95\% C.L. $\Sigma m_i$ limits discussed in Ref.~\cite{RoyChoudhury:2019hls}.  \textbf{Right:} The neutrinoless-double-beta-decay neutrino-mass observable plotted against $m_\beta$ for the normal (light blue) and inverted (dark red) mass orderings; the widths of these curves arise from the range of possible Majorana phases. The horizontal gray band gives the 95\% upper bound from GERDA, with width from the variation in nuclear-matrix elements~\cite{GERDA:2020xhi}.  Both figures, reproduced from Ref.~\cite{Formaggio:2021nfz}, relate the observables using the best-fit oscillation parameters from Ref.~\cite{Esteban:2020cvm}, and show a vertical orange line at KATRIN's first neutrino-mass limit of $m_\nu < 1.1$~eV (90\% C.L.)~\cite{Aker2019knm1}.}
    \label{fig:massobservable-comp}
\end{figure}

\paragraph{Cosmology}
Neutrinos are the most abundant known massive particle in the universe, and relic neutrinos from the moments following the Big Bang therefore had a significant collective effect on large-scale-structure formation in the early universe. The $\Lambda$ cold-dark-matter ($\Lambda$CDM) model, or alternative concordances, provides a framework for inferring the sum of the neutrino-mass values -- $\Sigma m_i$ -- from cosmological data. Recently, the Dark Energy Survey has inferred a limit of $\Sigma m_i < 0.13$~eV (95\% C.L.) based on a combination of their own gravitational-lensing and galaxy-clustering data with data sets from baryon acoustic oscillations, Type Ia supernovae, redshift-space distortions, and the Planck cosmic-microwave-background measurement~\cite{DES:2021wwk}. Fig.~\ref{fig:massobservable-comp} (left) shows the relationship between the cosmological observable $\Sigma m_i$ and the kinematic observable $m_\beta$. Next-generation observations target a sensitivity allowing $>3\sigma$ detection of $\Sigma m_i$, even at the normal-ordering floor of $\Sigma m_i = 0.06$~eV. Future plans and prospects for improving the cosmological sensitivity are discussed in detail within the CF7 topical group.

\paragraph{Neutrinoless double-beta decay}
As Sec.~\ref{sec:Majorana} will describe in detail, a wide variety of current and planned experiments are searching for the hypothesized neutrinoless double-beta decay process, which can occur only if the neutrino has a Majorana nature. If this decay occurs and is predominantly driven by light-neutrino exchange, its rate will depend on the effective neutrino-mass observable $\langle m_{\beta \beta} \rangle = \left| \sum U^2_{ei} m_i \right|$; Fig.~\ref{fig:massobservable-comp} (right) displays the relationship between $\langle m_{\beta \beta} \rangle$ and $m_\beta$. The most sensitive present limits on $\langle m_{\beta \beta} \rangle$ are set at 90\% C.L. by the GERDA (0.08--0.18~eV~\cite{GERDA:2020xhi}) and KamLAND-Zen (0.04--0.16~eV~\cite{KamLAND-Zen:2022tow}) experiments. The next generation of experiments aims to probe the entire region allowed by the inverted mass ordering. However, if neutrinos are \textit{not} Majorana fermions, than the neutrinoless double-beta decay rate will be zero and these experiments will have no sensitivity to the neutrino-mass scale. Even in the case of Majorana fermions, it must also be noted that, if the lightest neutrino mass value is less than about 0.002~eV, $\langle m_{\beta \beta} \rangle$ is relatively insensitive to the mass scale in both the inverted and normal orderings.

\paragraph{Other} Recently, a novel atomic-physics idea has arisen, relying on radiative emission of a neutrino pair from an atomic de-excitation, stimulated by collective, coherent Raman scattering~\cite{Hara:2019bur,Tashiro:2019ghs}. Neutrino-pair emission has not yet been observed, but in principle it could be possible to probe the neutrino-mass scale by measuring the angular spectrum of the resulting pairs. Separately, the use of levitated, optomechanical quantum sensors has been proposed to search for heavy sterile-neutrino states, and preliminary investigations suggest that such sensors could in principle measure the neutrino-mass scale if a suitable, ultra-low $Q$-value decay can be identified (cf Table~\ref{tab:numass-isotopes})~\cite{Carney:2022pku}. 
In both cases, substantial effort would be required to make a credible attempt at such a measurement. 

\subsubsection{Complementarity of neutrino-mass measurement approaches}
\label{sec:complement}

Each of the three most sensitive means of probing the neutrino mass -- direct kinematic measurements, cosmology, and neutrinoless double beta decay -- accesses a different observable related to the absolute neutrino-mass scale. The complementary information from these probes allows further inference, as explored by Abazajian et al.\footnote{White Paper: Synergy between cosmological and laboratory searches in neutrino physics: a white paper~\cite{Abazajian:2022ofy}}. For example, a measured neutrino mass of $m_\beta \approx 0.5$~eV, within the projected KATRIN sensitivity, would severely limit the available parameter space for neutrinoless double-beta decay, and require the introduction of new physics to be reconciled with current cosmological data. A positive detection of $\Sigma m_i$ at a sufficiently large value, with no observation of neutrinoless double beta decay in the corresponding parameter space, would imply that neutrinos have a Dirac nature or that light-neutrino exchange is not the primary mechanism of neutrinoless double beta decay; conversely, null results for $\Sigma m_i$, coupled with the observation of neutrinoless double beta decay, would suggest either a problem with the $\Lambda$CDM concordance or new BSM physics in the neutrino sector. Even statistically significant observations of the neutrino mass in all three channels could signal a discordance, if the signals are not consistent with the contours defined by neutrino-oscillation experiments and shown in Fig.~\ref{fig:massobservable-comp}. Direct kinematic measurements have special value as a clean, model-independent probe. By pursuing robust, independent measurements in each channel, however, we attain the ability to make precision tests of the entire neutrino-mass landscape, with corresponding sensitivity to unexpected physics.

\newpage

\section{Neutrinoless double beta decay and Majorana neutrinos}
\label{sec:Majorana}

The search for neutrinoless double beta decay (\ndbd) is the most sensitive known way to test for the Majorana nature of the neutrino~\cite{Agostini:2022zub}.

In this process, possible in certain even-even nuclei where single beta decay is forbidden, two neutrons are transformed into two protons and two electrons ($0\nu\beta^-\beta^-$):
\begin{equation}
     ^A_{Z}X \rightarrow ^A_{Z+2}X + 2e^-
\end{equation}
or two protons are transformed into two neutrons and two positrons ($0\nu\beta^+\beta^+$):
\begin{equation}
     ^A_{Z}X \rightarrow ^A_{Z-2}X + 2e^+
\end{equation}

Both of the above processes violate lepton number by two units, and via the Schechter Valle theorem~\cite{schechter1982neutrinoless}, their observation would imply a Majorana nature of the neutrino. A similar pair of Standard Model processes are possible with two antineutrinos (neutrinos) in the final state, called two neutrino double beta decay, or \twonudbd.  These processes do not violate lepton number, and are at least 10$^6$ times faster than the neutrinoless process in practical isotopes:
\begin{eqnarray}
     ^A_{Z}X \rightarrow ^A_{Z+2}X + 2e^- + 2\bar{\nu}\\
     ^A_{Z}X \rightarrow ^A_{Z-2}X + 2e^+ + 2\nu
\end{eqnarray}

While there are tens of known double beta decay isotopes in both $0\nu\beta^-\beta^-$ and $0\nu\beta^+\beta^+$ modes~\cite{haxton1984double}, competitive experimental limits are found for only a smaller subset, constrained by practical considerations.  A high Q value for this decay is a substantial advantage, both to enable detectability above radiogenic backgrounds, and since by phase space arguments higher Q-values tend to favor shorter (and hence more accessible) decay half-lives.  Because of the proton to neutron mass difference, all practically studied \ndbd isotopes are therefore $0\nu\beta^-\beta^-$ emitters.  Since availability of isotope is often a limiting concern,  practical considerations favor elements where the relative isotope has a high natural abundance.  Primarily because of these two considerations, recent and contemporary experiments have focused primarily on the isotopes $^{76}$Ge, $^{100}$Mo, $^{130}$Te, and $^{136}$Xe.

\ndbd is observed via detection of the emergent electrons from the decay. Since \ndbd and \twonudbd are distinguished only by presence or absence of unobservable neutrinos, the only way to experimentally identify a \ndbd event above a far more copious \twonudbd background is via precise calorimetry.  In ton-scale experiments, resolutions of at most 2\% FWHM are required to suppress backgrounds from \twonudbd to negligible levels in most isotopes. Thus energy resolution is a critical feature of a sensitive \ndbd technology.  

Once energy resolution sufficient for suppression of \twonudbd backgrounds is achieved,  other backgrounds become limiting. Most 100~kg-scale experiments to date have been primarily limited by backgrounds of radiogenic origin, usually from $\gamma$ emitting isotopes in the uranium and thorium decay chains, in particular $^{208}$Tl and $^{214}$Bi.  These gamma rays produce photoelectric and Compton events that can introduce energy deposits in the energy region of interest (ROI), usually one or two FWHM around the \ndbd Q-value.  The more precise the energy resolution, the smaller contribution these non-monoenergetic backgrounds will make in the ROI, and improving energy resolution will always be one compelling way to reduce backgrounds.    Beyond precision energy measurement, other observables may also be employed to suppress these backgrounds, including identifying the topological nature of the event (single site vs multi-site, or in some technologies double electron vs-single electron), or identification of the decay daughter nucleus.  In addition to radiogenic backgrounds, large-scale experiments will also be confronted by increasingly problematic backgrounds of cosmogenic origin, ($\alpha,n$) reactions followed by neutron capture, and eventually even solar neutrinos. These must be mitigated by whatever means necessary.

\begin{figure}[tb]
    \centering
    \includegraphics[width=0.99\textwidth]{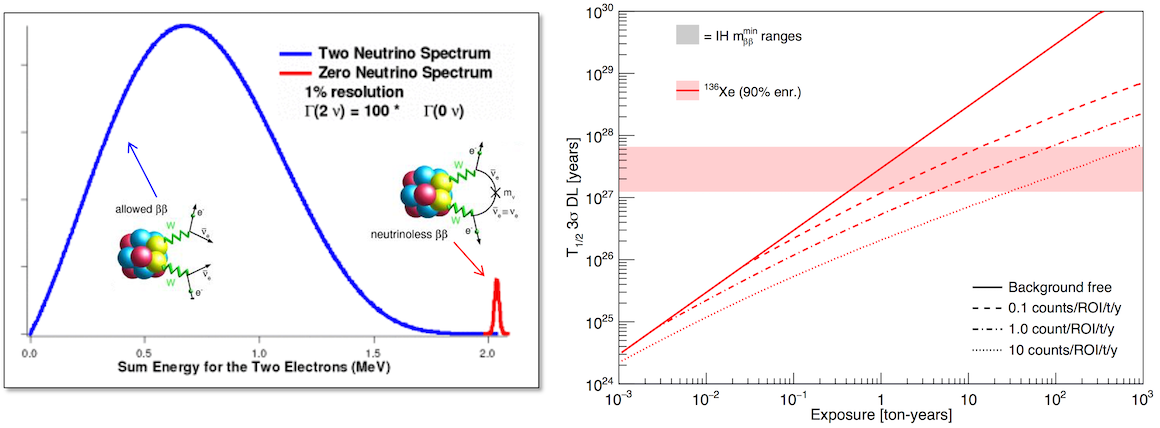}
   
    \caption{Left: Energy spectra of electron energy in two neutrino and neutrinoess double beta decay. Right: Sensitivity vs exposure in a \ndbd experiment conducted as a counting experiment with different background levels (from ~\cite{agostini2017discovery}).}
    \label{fig:dbdintro}
\end{figure}

The motivation for driving backgrounds to extremely low levels is illustrated by Fig.~\ref{fig:dbdintro}, right, adapted from Ref.~\cite{agostini2017discovery}.  This calculation imagines a simplified \ndbd analysis where events in the ROI are counted in a single bin, and a statistical excess is sought to establish a positive signal in the presence of various rates of experimental background. The plot shows the $3\sigma$ discovery potential - the half-life for which 50\% of experimental trials would claim a 3$\sigma$ discovery - as a function of exposure in ton-years. In real experiments this construction is often an oversimplification, as additional power may be gained from making fits in various detector parameters; but this simple thought experiment is  nevertheless informative, as it illustrates well the transition from background-free to background-limited regimes. 

For  low exposures,  all sensitivities, no matter what the level of background, are proportional to
$T_{1/2}$. In this regime, the expected background
count is much less than one; thus observation of a single
event would be a high-significance a discovery, since it must be signal.
Doubling the exposure in this regime effectively doubles the half-life
that corresponds to a 50\% chance of an event in this time window. Since $T_{1/2}$ is inversely proportional
to $m_{\beta\beta}^{2}$, sensitivity initially grows with time as $T_{1/2}\propto\epsilon$, or $m_{\beta\beta}\propto\epsilon^{-1/2}$. 

At larger exposures the experiment has been running
sufficiently long that some background events are expected, and the task at hand
becomes one of establishing the significance of statistical excess of signal events over a finite background. In the high-exposure
limit, the number of expected background events increases proportionally
to $\epsilon$, as does the number of expected signal events, so sensitivity
grows like $T_{1/2}\propto\sqrt{\epsilon}$, or $m_{\beta\beta}\propto\epsilon^{-1/4}$.
When experiments reach this regime, progress in sensitivity is slowed substantially. The determining
factor in how soon a given experiment will reach this turning point
is the level of background.

The required background rate to stave off the background-limited regime depends, naturally, upon detector mass.  Larger experiments are more difficult to realize not only because of the technical difficulties involved in building them, but also because of the need to achieve ever lower levels of background to avoid the $m_{\beta\beta}\propto\epsilon^{-1/4}$ regime.  Scaling laws associated with self shielding and improved surface-to-volume of larger detectors provide a strong advantage.  But these beneficial scaling behaviours must also be accompanied by improvements in precision, material radio-purity, and laboratory depth if experiments are to make continued progress in the low background regime at large scales.

At the time of writing this report, the \ndbd field is undergoing a transition between 100~kg-scale (``current generation'') experiments with backgrounds in the range of few-to-hundreds of counts per ton per year in the ROI, to next-generation ``ton-scale'' experiments that require backgrounds in the regime of 0.1-1 count per ton per year in the ROI.   Realization of at least one such ton-scale experiment will be a major focus in the coming Snowmass period.  In parallel, continuing R\&D must be pursued, in order for technologies to be proven that can enable a class of ``beyond-ton-scale'' detection methodologies when they are needed.  Ongoing theoretical work must also be supported to connect limits or observations from these experiments to statements about the underlying physics that drives the decay.  

\begin{figure}[tb]
    \centering
    \includegraphics[width=0.5\textwidth]{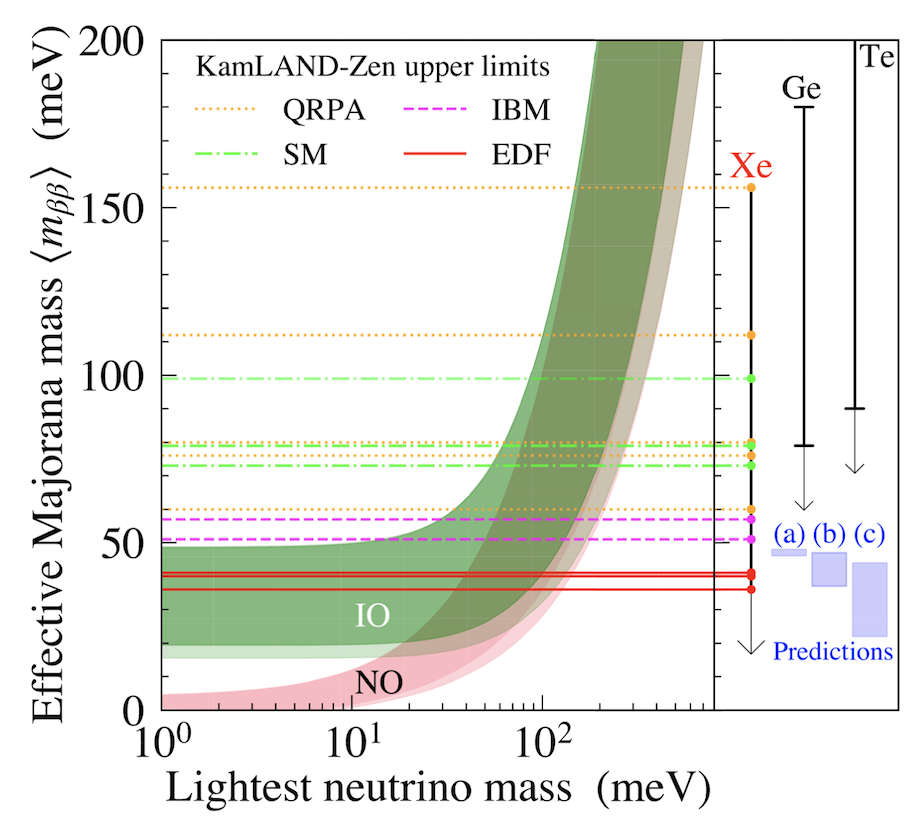}    \caption{Recent result from the KamLAND-Zen experiment, the most sensitive search for \ndbd at the time of writing.\cite{KamLAND-Zen:2022tow}}
    \label{fig:KZ}
\end{figure}

The expected rate of \ndbd depends on the mechanism by which the decay occurs, as well as the properties of the isotope chosen in which to search for it.  Under the ``minimal'' light Majorana neutrino exchange mechanism, the decay rate $\Gamma$ factorizes approximately as:
\begin{equation}
\Gamma=G\left\Vert M\right\Vert {}^{2}m_{\beta\beta}^{2}.\label{eq:RateOfBB}
\end{equation}
Here, $G$ is the phase space factor, which depends on the phase space available to the decay products; $\left\Vert M\right\Vert$ is the nuclear matrix element, encoding the effects of the structure of the decaying nucleus; and $m_{\beta\beta}$ is the effective Majorana mass, which is related to all of the mass splittings $\Delta m^2_{ij}$, all the mixing angles ($c_{ij}=\cos\theta_{ij}$,$s_{ij}=\sin\theta_{ij}$), the lightest neutrino mass $m_1$, and the Majorana CP phases ($\lambda_a$,$\lambda_b$) in the neutrino sector, via:
\begin{equation}
m_{\beta\beta}=c_{12}^{2}c_{13}^{2}m_{1}e^{2i\lambda_{a}}+s_{12}^{2}c_{13}^{2}e^{2i\lambda_{b}}\sqrt{m_{1}^{2}+\Delta m_{12}^{2}}+s_{13}^{2}\sqrt{m_{1}^{2}\pm|\Delta m_{23}^{2}|}\label{eq:mbb}.
\end{equation}
Uncertainties in these parameters, and especially the complete inaccessibility of  $\lambda_a$,$\lambda_b$ due to their irrelevance in neutrino oscillations, imply range of possible decay lifetimes depending strongly on the lightest neutrino mass and the neutrino mass ordering.  

The strongest limit on the half life for \ndbd from any isotope presently comes from the Kamland-Zen experiment~\cite{KamLAND-Zen:2022tow}, which has used an exposure of 970 kg yr of $^{136}$Xe dissolved in liquid scintillator to establish T$_{1/2}\geq2.3\times10^{26}$~yr. This corresponds to an upper limit on the lightest neutrino mass under the light Majorana neutrino exchange mechanism of 36 - 156 meV, depending on the adopted nuclear matrix element calculation.  This result is shown in Fig~\ref{fig:KZ} alongside the allowed bands of values for m$_{\beta\beta}$ in the inverted ordering (IO) or normal ordering (NO) regions of parameter space.  Proposed ton-scale experiments aim to probe half-lives up to around 10$^{28}$~yr to explore to the lower parts of the inverted mass ordering parameter space, while beyond-ton-scale programs must achieve sensitivity to 10$^{29}$~yr and beyond in order to access  the normal ordering regime.

This section of the report is structured as follows.  Sec.~\ref{sub:bbthy} briefly discussed ongoing theoretical work.  Sec.~\ref{sub:bbcurton} outlines the current generation and ton-scale experiments. Sec.~\ref{sub:bblarge} discusses technologies under active exploration for beyond-ton-scale experiments. Sec.~\ref{sub:bbimproved} presents a brief overview of work ongoing to understand and optimize the performance of existing detector technologies.  Sec.~\ref{sub:bbisotope} discusses challenges associated with isotope production in different elements and Sec.~\ref{sub:facilities} outlines the needs for near-future facilities to support the \ndbd program.  Finally, in the interest of completeness, Sec~\ref{sub:bbother} briefly touches upon other methods that have been discussed for establishing the Majorana nature of the neutrino, outside of \ndbd.

\subsection{Open questions in \ndbd theory}
\label{sub:bbthy}

\ndbd decay plays a key role in understanding the role of lepton number in fundamental particle physics and cosmology and size and origin of neutrino masses \cite{Dolinski:2019nrj,Agostini:2022zub}. 
The advent of the ton-scale program for \ndbd decay and emerging experimental techniques require a modern and state-of-the-art theoretical framework to interpret the future results. A comprehensive white paper has been prepared on theoretical aspects of neutrinoless double beta decay~\footnote{White Paper: Neutrinoless Double-Beta Decay: A Roadmap for Matching Theory to Experiment~\cite{cirigliano2022neutrinoless}}.  Some recent developments and open problems in the theoretical aspects of \ndbd decay searches can be framed within the following three general categories:

\begin{enumerate}
    \item Decay rates calculated within the minimal mechanism, where \ndbd decay is induced by the exchange of the three known active light neutrinos, are still rather uncertain. Recent years have seen tremendous progress in the use of {\it ab initio} techniques to calculate \ndbd decay rates \cite{Pastore:2017ofx, Cirigliano:2019vdj, Yao:2019rck, Yao:2020olm,Novario:2020dmr, Weiss:2021rig} from microscopic nuclear forces and electroweak currents derived from chiral effective field theory (EFT)~\cite{Gysbers:2019uyb,Hammer:2019poc}. However, the same EFT methods have identified a new leading contribution to \ndbd decay that involves new QCD and nuclear matrix elements~\cite{Cirigliano:2018hja,Cirigliano:2019vdj}. First calculations of both matrix elements have recently appeared~\cite{Cirigliano:2020dmx, Cirigliano:2021qko,  Wirth:2021pij, Jokiniemi:2021qqv, Weiss:2021rig} indicating an enhancement of the $0\nu\beta\beta$ decay rate. That being said, their consistent implementation into complete calculations of heavy isotopes is still a major source of large uncertainties and requires further study.
    \item \ndbd decay experiments are often interpreted in terms of ``lobster plots'' (e.g. the red and green bands of Fig.~\ref{fig:KZ}) that illustrate the reach of various experiments as a function of the lightest neutrino mass (or, equivalently, as a function of the sum of light neutrino masses). Within the minimal mechanism these plots visually enhance an apparent ``dead-zone'' between the inverted and normal hierarchy where no signal is expected. As a result, it is often asserted that ton-scale experiments cannot make a detection if neutrinos masses are normally ordered. This point of view is misleading. A plethora of \ndbd decay mechanisms exist beyond the minimal mechanism that may populate this space of \ndbd lifetimes. These mechanisms have been classified using EFT methods \cite{Prezeau:2003xn, Cirigliano:2017djv,Cirigliano:2018yza,Graf:2018ozy,Dekens:2020ttz,Deppisch:2020ztt} which can be easily matched to specific particle physics scenarios. These studies indicate, in particular, that ton-scale experiments are sensitive to new lepton-number-violating mechanisms that appear at very high energy scales (up to hundreds of TeV). Next-generation experiments can thus viably obverse a nonzero signal even under the assumption of normal mass ordering.  Sensitive \ndbd decay experiments that extend the half-life sensitivity in any given isotope thus probe uncharted territory and are discovery experiments. Their interpretation solely in terms of the ``lobster-plot'' is only one window into a much wider landscape of lepton number violating physics. 
    \item At the same time, much less is known about the calculations of \ndbd decay rates in non-minimal mechanism. In particular, the required QCD and nuclear atomic matrix elements have been classified at leading order in chiral EFT~\cite{Prezeau:2003xn, Cirigliano:2017djv,Cirigliano:2018yza}, but many are still poorly understood. Obtaining these matrix elements requires a close collaboration between the lattice-QCD, chiral EFT, {\it ab-initio}, and nuclear structure communities. Recent lattice-QCD efforts have greatly reduced the uncertainties on $\pi\pi\rightarrow ee$ matrix elements \cite{Nicholson:2018mwc, Feng:2018pdq, Tuo:2019bue,Detmold:2020jqv}, and methods are being developed to also tackle the more complicated $nn \rightarrow pp + ee$ matrix elements \cite{Davoudi:2020gxs, Davoudi:2021noh}.
\end{enumerate}

\subsection{Current generation and the ton-scale}
\label{sub:bbcurton}
The current generation of ton-scale \ndbd experiments is being stewarded by the Department of Energy's Office of Nuclear Physics.  Here we include a brief summary of the status of the experiments with significant US involvement that are planned for construction and operation within this coming decade.  At a meeting held in the fall of 2021 at INFN\footnote{\href{https://agenda.infn.it/event/27143/timetable/\#20210929.detailed}{https://agenda.infn.it/event/27143/timetable/\#20210929.detailed}} for European and North American stakeholders (in particular the DOE), the near-term programmatic future of \ndbd efforts was discussed.  The conclusions of this meeting are summarized here in three main points.  First, the science regarding searches for neutrinoless double beta decay is recognized as ``very compelling'' and ``capable of reshaping our current understanding of nature.''  Secondly, the international stakeholders in \ndbd research agree in principle that ``the best chance for success is an international campaign with more than one large ton-scale experiment implemented in the next decade with one in Europe and the other in North America.''  Finally, the international community in \ndbd is interested in exploring ``whether a more formal structure for international coordination of this research would be beneficial not only for experiments of the next decade but also for future multi-ton and/or multi-site experiments.''

The major ongoing experimental programs will now be discussed.

\begin{figure}[b!]
\begin{centering}
\includegraphics[width=0.99\columnwidth]{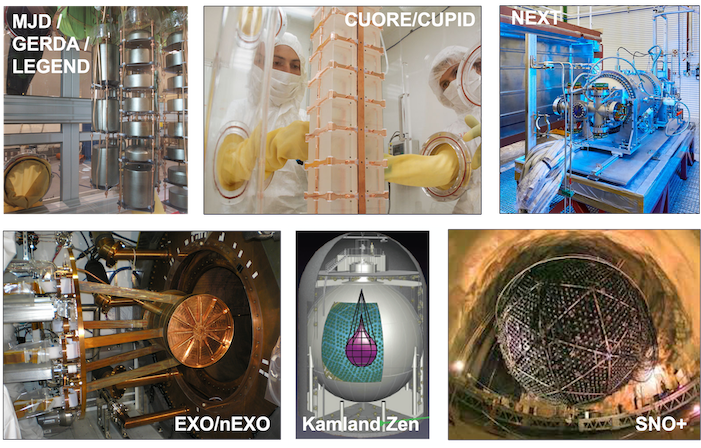}
\par\end{centering}
\caption{Photos of some of the current generation of \ndbd experiments described in this report.
\label{fig:BaIonAtom}}

\end{figure}

\subsubsection{\textsc{CUPID}}
The CUORE Upgrade with Particle Identification (CUPID)\cite{armstrong2019cupid} is a future upgrade to the Cryogenic Underground Observatory for Rare Events (CUORE), a multinational collaborative effort to detect lepton number violation through the \ndbd of $^{130}$Te. Approximately one-third of the institutions in CUPID are U.S. universities and national laboratories involving faculty, students, and research scientists across the United States, with responsibilities in management, remote monitoring and operations, detector design and R\&D, sensor testing, software development, and modeling detector performance.

The baseline design for CUPID features an array of 1596 scintillating crystal bolometers and 1710 light detectors, each instrumented with germanium neutron transmutation doped (NTD) sensors, and organized into 57 towers. While the current design is based on a full complement of Li$_2$MoO$_4$ (LMO) crystals, one of the key scientific features of the detector design is the ability to flexibly incorporate multiple isotopes. The new detector will be installed in an upgraded cryostat at Gran Sasso National Laboratories (LNGS), taking advantage of the existing infrastructure and facilities developed for use in CUORE.

CUPID builds on the success of the CUORE~\cite{arnaboldi2004cuore}, CUPID-0~\cite{azzolini2018first}, CUPID-Mo~\cite{armengaud2020cupid}, and CROSS~\cite{armatol2021cupid} experiments, including years-long, stable operation of the CUORE detector at base temperatures on the order of 10 mK. In addition to the current work on CUPID, a future, ton-scale version of the CUPID concept, CUPID-1T, is under conceptual development. CUPID and CUPID-1T\footnote{White Paper: Toward CUPID-1T~\cite{armatol2022toward}} both boast the potential to probe the entire region of the inverted hierarchy of the neutrino mass splitting, and sensitivity to a variety of beyond the Standard Model processes, including symmetry violation and dark matter searches, with CUPID-1T additionally carrying the potential to reach into the normal-ordering region of the neutrino mass splitting.

\subsubsection{\textsc{KamLAND-Zen 800}}
KamLAND-Zen is a double beta decay experiment using the existing KamLAND neutrino detector with a spherical inner balloon located at the center of the detector filled with Xe-loaded liquid scintillator as the $\beta\beta$ source.  The inner balloon is surrounded by 1 kton of liquid scintillator contained in a 13-m diameter outer balloon.  The scintillation light produced in the decay processes is detected using hundreds of 17-inch and 20-inch PMTs mounted on the inside surface of the stainless-steel containment tank providing 34\% solid-angle coverage.  Surrounding the tank is a 3.2 kton water-Cherenkov outer detector.

The detector was recently upgraded to contain twice as much Xe mass for a total of 745 kg of enriched xenon and is referred to as KamLAND-800.  The experiment has been taking data since January 2019.   The first result with the upgraded detector has been published with data taken between February 2019 and May 2021 for an exposure of 970 kg yr of $^{136}$Xe.  The result is a lower limit for the \ndbd decay
half-life of T$^{0\nu}_{1/2} > 2.3 \times 10^{26}$~yr at 90\% C.L., corresponding to upper limits on the effective Majorana neutrino mass of 36 – 156 meV \cite{KamLAND-Zen:2022tow}.  A median sensitivity of $1.3 \times 10^{26}$~yr was determined using a Monte Carlo simulation of an ensemble of experiments assuming the best-fit background spectrum.
This sensitivity is the first search for \ndbd in the inverted mass ordering region and provides a limit that reaches below 50 meV for the first time.

Future upgrades are planned to improve detection of muon-induced neutrons and rejection of xenon spallation backgrounds.  These improvements and continued data-taking will continue to provide more stringent limits on the Majorana neutrino mass in the inverted mass ordering region.

\subsubsection{\textsc{LEGEND-1000}}
LEGEND-1000, the ton-scale Large Enriched Germanium Experiment for Neutrinoless $\beta \beta$ Decay, is the proposed next phase in the LEGEND experimental program to search of the \ndbd\ of $^{76}$Ge. LEGEND-1000 builds on the success of the GERDA and \textsc{Majorana Demonstrator} experiments, as well improvements developed in the ongoing LEGEND-200 experiment. The ton-scale LEGEND-1000 is designed to probe \ndbd\ decay over the full range of the allowed region for the inverted mass ordering. 

The LEGEND-1000 experiment utilizes the demonstrated low background and excellent
energy performance of high-purity p-type, ICPC Ge semiconductor detectors,
enriched to more than 90\% in $^{76}$Ge. The inverted coaxial point-contact (ICPC) detector design leads to an excellent energy resolution of 0.12\% demonstrated full-width at half-maximum (FWHM) at $Q_{\beta\beta} = 2039$\,keV and pulse shape characteristics that allow bulk \ndbd events to be distinguished from $\gamma$ ray backgrounds with multiple interaction sites. About 400 ICPC detectors with an average mass of 2.6~kg each are distributed among
four 250-kg modules to allow independent operation and phased commissioning, with the granular nature of the Ge detector array allowing rejection of background events that span multiple detectors.

The detector strings are immersed in radiopure underground-sourced LAr (UGLAr), which is reduced in cosmogenic argon isotopes,  and contained within an electroformed copper reentrant
tube. Each of the four UGLAr modules is surrounded by LAr
produced from atmospheric Ar, contained within a vacuum-insulated cryostat.
The LAr volumes are instrumented with wavelength shifting fibers and
read out by Si photomultipliers. Therefore, background interactions external to the Ge detectors are identified by
LAr scintillation light.
The cryostat is enveloped by a water tank providing
additional shielding. The baseline design assumes installation in the SNOLAB cryopit. A similar experimental setup could also be realized at the alternative LNGS site.

\subsubsection{n\textsc{EXO}}
The nEXO \ndbd decay experiment\cite{adhikari2021nexo} is designed to use a time projection chamber and 5000 kg of isotopically enriched liquid xenon (LXe) to search for the decay in the $^{136}$Xe nucleus.  The nEXO experiment builds on the expertise and success of the 100-kg scale EXO-200 experiment that first measured the \twonudbd process in $^{136}$Xe and completed it's \ndbd search with a final sensitivity of $5 \times 10^{25}$ years \cite{EXO-200:2019rkq}.  These detectors provide a large homogeneous mass of isotope and uses detection of light and charge to determine three-dimensional vertex reconstruction and well-understood energy resolution.  nEXO is designed to improve the sensitivity to \ndbd by several orders of magnitude in order to cover the allowed region of the inverted mass ordering for neutrinos.

The nEXO detector takes advantage of the benefits of a large monolithic LXe detector which allows both accurate measurement of radioactive backgrounds in the outer region while also providing a background free region in the center of the detector.  The topology of the events from the 3D reconstruction allows for the discrimination between double-$\beta$ decay events and background events which are primarily gamma-induced signals.  The gamma events deposit energy in multiple sites with a decreasing rate toward the center of the detector while the double-$\beta$ decay events deposit energy in a single site and have a uniform rate in the detector.  The energy measurement uses anti-correlation of the scintillation and ionization signals such that the energy resolution allows the rejection of the \twonudbd decay background to negligible levels.

The nEXO detector consists of a a TPC vessel (1.271 m right cylinder, vertically oriented) filled with enriched xenon and surrounded by $\sim 33,000$ kg of HFE-7000 which serves as both a thermal bath and radiation shield.  The HFE-7000 is contained within an inner vessel of a cryostat and surrounded then by vacuum insulation within an outer vessel of the cryostat.  The entire cryostat is located in an active water shield that serves as an outer detector and muon veto.  The scintillation light produced in the TPC will be collected by SiPMs located on the barrel of the TPC and the ionization will be collected by electrode strips located at the anode.  nEXO's background and design projections are based on siting the experiment in SNOLAB's cryopit.

\subsubsection{\textsc{NEXT}}
The NEXT collaboration\footnote{LOI: \href{https://www.snowmass21.org/docs/files/summaries/IF/SNOWMASS21-IF8_IF0-NF5_NF0-RF4_RF0_RGuenette-086.pdf}{NEXT Collaboration: High-pressure xenon gas time-projection chambers for neutrinoless double-beta decay searches}}\cite{adams2021sensitivity} has completed the operation of the NEXT-White demonstrator, which has shown excellent energy resolution ($<$1\% FWHM at $Q_{\beta\beta}$), a powerful topological signature to discriminate the two electrons associated with a double beta decay and excellent operational parameters, including very long electron lifetime ($>$20 ms) and stability (live time in excess of 90\%).  NEXT-White has also measured the \twonudbd mode by direct subtraction of the data acquired with enriched xenon and the data acquired with depleted xenon.

The NEXT-100 detector, currently under construction, aims to demonstrate further improvement in the HPXe performance at a large scale, to show a very low background rate ($\sim 5 \times 10^{-4}$ counts/kg/keV/year) and to perform a sensitive search of the \ndbd mode.  NEXT-100 is scheduled to run from 2023 to 2026.  A ton-scale phase called NEXT-HD will follow, based on incremental improvement of the techniques enabling NEXT-White and NEXT-100.  NEXT-HD is designed to reach a sensitivity of at least 10$^{27}$ years.  In parallel, an intense R\&D program is being carried out realize a barium tagging module that would detect Ba$^{2+}$ ions produced in double beta decay as the ultimate means of background reduction.  A ton-scale detector implementing barium tagging (NEXT-BOLD) could replace NEXT-HD before the end of the decade with an improved sensitivity covering the inverted hierarchy and advancing towards the normal ordering regime.

\subsubsection{\textsc{SNO+}}
SNO+ will search for $0\nu\beta\beta$ by ``loading'' linear alkyl benzene (LAB) liquid scintillator (LS) with natural Te. Te-loaded LAB will be deployed in the existing Sudbury Neutrino Observatory (SNO), and with an anticipated initial $^{\rm nat}$Te loading of 0.5\% (3.9 tonnes), corresponding to a total mass of $^{130}$Te of  1.3 tonnes, SNO+ has a 90\% CL sensitivity of $T_{1/2}>2\times 10^{26}$y in three years of running.  Excellent control of radio-impurities suggests that the largest backgrounds to SNO+ in the \ndbd ROI are expected to be interactions of $^8$B solar $\nu$s. The total ROI background level per year is expected to be just 9.5 events. One of the advantages of the loaded-liquid scintillator approach is the fact that many backgrounds can be measured in advance of deployment of the isotope: over half of the backgrounds have already been constrained by existing SNO+ data.

Tellurium has multiple advantages over many other isotopes, partly because of its favorable phase space and matrix elements.  With the initial 0.5\% loading of Te, SNO+ should reach into the inverted hierarchy regime of neutrino masses. An additional advantage of Te is the high 34\% natural abundance of the $\beta\beta$ isotope, $^{130}$Te. Thus SNO+ can push much higher in sensitivity without costly enrichment.  SNO+ plans therefore to go far beyond 0.5\% loading.  New R\&D on scintillator loading has allowed SNO+ to load up to 3\% Te with light quenching of around 30\%. 
 
\subsubsection{Synergistic searches at dark matter experiments} Experiments searching for dark matter with ton-scale masses of natural xenon naturally incorporate 9\% fraction of $^{136}$Xe and are thus expected to have some sensitivity to \ndbd.  Sensitivity of LZ to both neutrinoless and two-neutrino modes of double beta decay of $^{134}$Xe has also been explored~\cite{lux2021projected}.  Future very large dark matter detectors may extend this projected sensitivity still further. Both sensitivity estimates from the 50t Darwin experiment~\cite{agostini2020sensitivity}\footnote{LOI: \href{https://www.snowmass21.org/docs/files/summaries/NF/SNOWMASS21-NF5_NF4_Baudis-085.pdf}{L. Baudis and M. Schumann, Neutrino physics with the DARWIN Observatory}} and preliminary studies of the sensitivity of G3 dark matter experiment with 75 tons of active mass of natural xenon project competitive \ndbd sensitivities\footnote{LOI: \href{https://www.snowmass21.org/docs/files/summaries/NF/SNOWMASS21-NF5_NF0_Matthew_Szydagis-156.pdf}{Alex Lindote and Ibles Olcina, A 3rd generation LXe TPC dark matter experiment sensitivity to neutrino properties: magnetic moment and 0nubb decay of 136Xe}}. 
 
\subsection{Future large detectors}
\label{sub:bblarge}

Access to the normal ordering region of \ndbd parameter space will require detectors with active masses beyond the ton-scale.  For this purpose, giant, monolithic detectors with very low backgrounds, or modular systems with sufficiently large module masses for realistic deployment must be realized.  This section summarizes some of the most promising approaches to realizing future beyond-ton-scale \ndbd searches. The prospects for such experiments were the subject of extensive discussion during this Snowmass process.

\subsubsection{Very large-scale time projection chambers}

Noble element time projection chambers (TPCs) at the kiloton scale are the subject of much attention in contemporary neutrino physics~\cite{abi2020deep}.   TPCs with kiloton-scale active masses of a double beta decay isotope could enable access to half-live sensitivities in excess of 10$^{30}$~yr.  Four main approaches have been considered. Time projection chambers based on $^{136}$Xe in liquid (LXe) or gas (GXe) phase; based on  $^{136}$Xe dissolved into liquid argon; and based on other gases such as $^{82}$SeF$_6$ or other selenium or molybdenum derivatives. 

\paragraph{Xenon TPCs: large future liquid and gas phase detectors}
The plausibility of scaling pure xenon TPCs to the ton-scale was studied in Ref.~\cite{avasthi2021kiloton}.  A key focus of this work was the issue of isotope acquisition; we will discuss this further in Sec~\ref{sub:bbisotope}.  

In TPCs at the hundred-ton or larger module scale, radiogenic backgrounds are expected to be strongly self-shielded for most of the active volume, with attenuation coefficient $\mu/\rho$ = 0.038 cm$^2$/g at 2.5 MeV. This corresponds to a linear attenuation length of 8.5 cm for LXe and varying between 2.6–0.5 m for GXe at pressure between 15–50 bar, surely small compared to the  dimensions of a kton-scale Xe TPC.  Backgrounds will thus likely be dominated by other sources, in particular \twonudbd, radon decays, cosmogenic activation, and solar neutrino interactions.  

The \twonudbd background scales like b$_2\nu\propto\sigma^6$ where $\sigma$ is the energy resolution of the experiment.  Ref.~\cite{avasthi2021kiloton} estimates that if the presently optimal demonstrated resolution in LXe (0.8\%~$\sigma/E$  from Xenon1T) can be realized at this scale, an energy ROI of (0,1.5 FWHM) may enable a \ndbd search in a kiloton-scale LXe apparatus that is not compromised by irreducible \twonudbd background.  Gas detectors enjoy a finer intrinsic energy resolution due to the absence of local recombination fluctuations.  The energy resolution in xenon gas demonstrated in existing experiments of $\sigma/E$ = 0.2\% in a large GXe TPC would be more than sufficient to avoid \twonudbd backgrounds.

Backgrounds from dissolved radon are expected to be rejectable via tagging of Bi-Po coincidences.  Cosmogenic backgrounds are dominated by activation of $^{136}$Xe to $^{137}$Xe followed by $\beta$ decay with endpoint of 4.2 MeV, introducing a continuum beta electron background across the energy ROI.  Ref.~\cite{avasthi2021kiloton} estimates that a reduction in cosmogenic background rejection by a factor of around 10$\times$ beyond estimated levels in ton-scale experiments is required in order for this not to be limiting. 


\paragraph{Xenon-Doped Argon - DUNE and DarkNoon}

The DUNE-$\beta$ concept\footnote{LOI: \href{https://www.snowmass21.org/docs/files/summaries/NF/SNOWMASS21-NF5_NF10-IF8_IF0_Zennamo-175.pdf}{J. Zennamo et al., DUNE-Beta: Searching for Neutrinoless Double Beta Decay with a Large LArTPC}},\footnote{White Paper: Low Background kTon-Scale Liquid Argon Time Projection Chambers~\cite{avasthi2022low}},\footnote{White Paper: Low-Energy Physics in Neutrino LArTPCs~\cite{caratelli2022low}} explores the modification of a DUNE-like detector module in order to make it sensitive to \ndbd signals by doping liquid argon (LAr) with with xenon at the 2\% level, which would yield 300 tons of active isotope.  DUNE-$\beta$ would require an enhancement to the baseline precision of the DUNE detector design to MeV-scale energy deposits, specifically in terms of energy resolution. This would require novel readout and DAQ solutions, and possibly also optimization beyond the intrinsic energy resolution in bulk LAr with the addition of photo-ionizing dopants. Such performance enhancements could also enable additional synergistic improvements to the broader DUNE low energy physics program.

An alternative approach embodied in the DarkNoon concept\footnote{LOI: \href{https://www.snowmass21.org/docs/files/summaries/NF/SNOWMASS21-NF5_NF4-RF4_RF0_Savarese-127.pdf}{C. Savarese, The Neutrino Physics program of the Global Argon Dark Matter Collaboration}} emerging from the Global Argon Dark Matter program uses a dedicated 50t module with a 20\% admixture of xenon into liquid argon.  The background rate in the ROI was estimated to be lower than 5$\times 10^{-4}$ events/t/yr/keV, dominated by $^8$B $\nu_e$ scattering and \twonudbd events.  A dominant enabling feature of this concept is the use of high speed, sub-nanosecond time resolution SiPMs to statistically separate Cherenkov and scintillation photons in liquid argon. With this method, single vs- double electron topologies may be distinguished, providing an important background rejection tool.  

During the coming Snowmass period, the viability of these ideas may be demonstrated with R\&D focused on energy resolution, both aspects intrinsic to the working medium and limited by plausibly deployable readout schemes, Cereknov/scintillation response, and gas mixture handling at large scale.

\paragraph{SeF$_6$ time projection chambers}
Non-xenon based TPCs have been discussed, in particular using $^{82}$SeF$_6$ (or chemically similar MoF$_6$ and TeF$_6$) ~\cite{nygren2018neutrinoless}, as well as other exotic double beta decay gases including H$_2$Se and GeH$_4$\footnote{White Paper: High-pressure TPCs in pressurized caverns: opportunities in dark matter and neutrino physics~\cite{Monreal:2022crn}}.  A key benefit of selenium as a double beta decay isotope is that its Q-value lies above the gamma lines from bismuth and thalium, thus effectively mitigating most radiogenic background sources and enabling modular detectors which do not need to be individually monolithic enough to enable radiogenic suppression via self-shielding.  Free electrons do not drift in $^{82}$SeF$_6$ since it is an attaching gas; hence any TPC employing it as a working medium must be an ion TPC, sensing either positive or negative populations.  This confers important advantages, as well as technical difficulties.  Ions drift with far less diffusion than electrons, hence attempts to preserve topological event information over the long drift distances needed for beyond-ton-scale TPC experiments will likely benefit  from implementing ion readout.  Gas purity considerations are also be mitigated for ion based TPCs, since attachment on electronegative impurities does not compromise performance. On the other hand, ion TPCs require unconventional readout methodologies.  Readout via technologies such as top-metal~\cite{li2021preliminary} or chemical based approaches~\cite{DualReadout}\footnote{White Paper: The Ion Fluorescence Chamber (IFC): A new concept for directional dark matter and topologically imaging neutrinoless double beta decay searches~\cite{jones2022ion}} are under exploration, and the N$\nu$DEx project a CJPL laboratory aims to demonstrate a SeF$_6$-based TPC for \ndbd searches~\cite{NuDeXTalk}.

\subsubsection{Large multi-purpose scintillation detectors}
Large liquid scintillator-based detectors have a long history of discovery and field-leading measurements in neutrino physics, with
KamLAND-Zen, currently leading in sensitivity among current-generation \ndbd searches. As interest has turned to approaches that could explore beyond the inverted ordering parameter space, the community has begun to focus on the development of hybrid Cherenkov/scintillation detectors that would improve rejection of cosmogenic and solar neutrino-related backgrounds. These detectors could be optimized for measurements across a wide range of energies, serving as neutrino observatories pursuing a wide range of physics goals including \ndbd. There are several such proposals at varying levels of maturity, in addition to a broad and very active R\&D program.

\paragraph{\textsc{Theia}} 
The \textsc{Theia} concept envisions a large, 25--100~kton--scale detector with a novel advanced liquid scintillator target, pursuing a broad range of physics goals \cite{Theia:2019non}. 
Among these is a sensitive search for \ndbd probing light Majorana neutrino exchange at the level of $m_{\beta\beta}\sim5$~meV. 
\textsc{Theia} expands on the capabilities of previous LS-based $0\nu\beta\beta$ searches through the use of a very large multi-ton scale isotope target mass and enhanced background rejection afforded by LS and water-based LS (WbLS) tuning, advanced photon detection methods such as spectral photon sorting, fast timing, and advanced analysis approaches.

The \ndbd decay search approach with \textsc{Theia} is to deploy an inner containment volume within the larger $\mathcal{O}(100)$~kton outer volume, filled with ultra-pure LS and doped with a percent-level quantity of the \ndbd decay isotope. 
A variety of isotopes have been considered, with promising candidates including tellurium (34\% $^{130}$Te) or enriched xenon ($\sim90$\% $^{136}$Xe), loaded at the percent level by mass.  
A configuration using an 8~m radius containment volume filled with LAB+PPO scintillator, deployed within an 100-kton scale \textsc{Theia} detector with $\sim90$\% effective photodetector coverage, would achieve an energy resolution of 3\%/$\sqrt{E}$ and a background budget dominated by the elastic scatters of $^8$B solar neutrinos, and to a lesser extent \twonudbd. 
A Monte Carlo--based analysis considering all relevant backgrounds is presented in Ref.~\cite{Theia:2019non}.

With this technique, \textsc{Theia} would reach world-class sensitivity to \ndbd decay, probing $T_{1/2}^{0\nu}>10^{28}$~y, or an
effective Majorana neutrino mass at the level of $\sim5-6$~meV
depending on the target isotope, probing the nondegenerate parameter space in the case of the normal neutrino mass ordering scenario.
Importantly, this highly scalable approach offers a cost-effective path to very large target masses, for promising discovery potential or in the case of a null result, the ability to rapidly rule out large regions of allowed parameter space, complementary to targeted high-resolution and zero background \ndbd decay searches. 

\paragraph{JUNO}
The Jiangmen Underground Neutrino Observatory (JUNO) is a multi-purpose experiment focused on precision measurements of neutrino oscillations and the exploration of neutrino astrophysics~\cite{JUNO:2015zny,JUNO:2022hxd}. The experiment is currently under construction in southern China and operations are expected to begin in 2023. 
Once the primary goals of determining the neutrino mass ordering and precisely measuring the oscillation parameters have been completed, JUNO's size and energy resolution affords an excellent opportunity to conduct a second phase searching for $0\nu\beta\beta$ decay\footnote{LOI: \href{https://www.snowmass21.org/docs/files/summaries/NF/SNOWMASS21-NF5_NF10_Pedro_Ochoa-125.pdf}{J.P. Ochoa-Ricoux et al., Searching for $0\nu\beta\beta$ decays in JUNO}}, with several tens to a hundred tons of $\beta\beta$-decaying isotopes loaded in the liquid scintillator.

Preliminary studies have been performed for 50\,ton of $^{136}$Xe ($Q_{\beta\beta}=2457.8~{\rm keV}$) \cite{Zhao:2016brs}. The expected energy resolution at the $Q_{\beta\beta}$ value is better than 2\% ($1\sigma$), permitting a 110\,keV ROI (FWHM) with corresponding background index of 1.35 events per isotope-ton$\cdot$year.
In this setting, JUNO could establish a half-life limit of $T_{0\nu\beta\beta}\geq 1.8\times 10^{28}$ years based on 5 years of data collection, corresponding to an effective neutrino mass $|m_{\beta\beta}|\leq(5-12)\ {\rm meV}$  depending on the assumed values of the nuclear matrix element \cite{Zhao:2016brs}. $^{130}$Te ($Q_{\beta\beta}=2527.5~{\rm keV}$) loading offers a more cost-effective approach with similar expected sensitivities thanks to its large natural abundance. A full evaluation of the physics potential to $0\nu\beta\beta$ with $^{130}$Te loading in JUNO is underway. Further improvements can be gained by restricting the $\beta\beta$-loaded LS to a smaller nylon vessel located at the detector center, with the surrounding scintillator acting as a veto, and by using deep learning techniques for topological reconstruction or hybrid Cherenkov/scintillation detection.


Active R\&D efforts are ongoing in the collaboration to develop high Te-loaded LS with high light yield, high transparency, low background, and good long-term stability. 
The priority is to resolve the major technical issues related to isotope loading in the next few years, and to start the $0\nu\beta\beta$ upgrade in the 2030s. Test runs with the OSIRIS predetector~\cite{JUNO:2021wzm}, as well as the imminent start of JUNO operations, will provide further insights into loaded-scintillator performance and background rejection capabilities. The best knowledge available to date suggests that JUNO could eventually reach well into the $|m_{\beta\beta}|\sim{\rm meV}$ region, which would be a major step forward in the search for this rare process.

\paragraph{R\&D of loaded scintillator detectors}
A broad set of R\&D activities within the liquid scintillator community has been captured in a dedicated white paper\footnote{White Paper: Future Advances in Photon-Based Neutrino Detectors: A SNOWMASS White Paper~\cite{klein2022future}}. Here we give a brief summary of the efforts underway that are most relevant for \ndbd searches.

The goals of this extensive program are to develop:
\begin{itemize}
    \item detector designs that can be adapted for a broad set of neutrino physics measurements, like \textsc{Theia} \cite{Theia:2019non}, SLIPS \cite{Morton-Blake:2022snr} and LiquidO \cite{Cabrera:2019kxi}.
    \item novel liquid scintillator cocktails that improve isotope loading, energy resolution, and improved event topology reconstruction capabilities that can used for background reduction. These include high-fraction Te and Xe loading, liquid scintillator cocktails containing quantum dots \cite{Graham:2019zqb, Winslow:2012ey}, water-based liquid scintillator (WbLS) \cite{Yeh:2011zz}, and slow fluors \cite{Biller:2020uoi}.
    \item novel photon sensors and photon collection techniques that improve light yields and Cherenkov/ scintillation separation capabilities, such as higher-efficiency photomultiplier tubes, Large Area Picosecond Photon Detectors (LAPPDs) \cite{Adams:2015kkx}, dichroicons \cite{Kaptanoglu:2019gtg}, ARAPUCAs \cite{DUNE:2020vmp}, and distributed imaging \cite{Dalmasson:2017pow}.
    \item simulation approaches that allow for more efficient precise simulations of kiloton-scale experiments, like the \textit{Chroma} software package \cite{Seibert2011FastOM} and generative neural network-based photon detection probability simulations \cite{Mu:2021nno}.
    \item analysis techniques based on machine learning that can improve background rejection of complex background signatures, as in the recent use of KamNet in the KamLAND-Zen experiment \cite{Li:2022frp}.
    \item data acquisition techniques that can reduce the readout complexity and per-channel cost of large detectors, like onboard FPGA-based or analog pulse shape feature extraction.
\end{itemize}

Mid-scale demonstrators of these technologies will be a key step in establishing the best techniques for future large-scale neutrino observatories. Three such demonstrator experiments are underway:
\begin{itemize}
    \item{ANNIE:} The ANNIE (Accelerator Neutrino Neutron Interaction Experiment)~\cite{annieLOI,annie-results,pershingdiss} is located in the Booster Neutrino Beam at Fermilab. Though ANNIE's physics program is focused on high-energy neutrino oscillations, it will be the first mid-scale demonstration of  several techniques that are of-interest to future \ndbd searches. ANNIE's current Phase 2 investigates the combination of a Gd-loaded water target read out by ultrafast LAPPDs, and a first insertion of a small WbLS-filled vessel (SANDI) is planned for the upcoming summer break, adding scintillation to the Cherenkov signal at the neutrino interaction vertex. In an eventual Phase 3, the full detection volume is to be filled with WbLS, including the first-ever deployment of a WbLS liquid re-circulation system.
    \item{NuDot:} NuDot is a mid-scale prototype designed to demonstrate timing-based Cherenkov / scintillation separation in a realistic experimental geometry, focusing on techniques applicable to searches for neutrinoless double-beta decay ($0\nu\beta\beta$). It builds on the successful demonstration of this approach in the FlatDot test stand~\cite{flatdot} and is currently undergoing detector commissioning, with its initial physics data-taking campaign expected to begin by Summer 2022. This system focuses on testing Cherenkov/scin\-till\-a\-tion separation capability as a function of position and direction using a pointable collimated $\beta$ source, based on a design using 2"-diameter PMTs with fast-timing capability (200\,ps transit time spread). The NuDot Collaboration is also conducting R\&D measurements of quantum-dot-loaded liquid scintillator cocktails that use perovskite wavelength shifters \cite{Graham:2019zqb}.
    \item{Eos:} The proposed Eos prototype is a few-ton scale detector, designed to hold a range of novel scintillators, coupled with an array of photon detection options and the ability to deploy a range of low-energy calibration sources.  Eos will be constructed, calibrated and tested in Berkeley.  Assuming a successful surface deployment,  Eos could later be re-deployed underground, for example at SURF or SNOLAB, or at an alternative off-site location such as a reactor or test beam for further characterization of detector response to a range of particle interactions. The primary goal of Eos is to validate performance predictions for large-scale hybrid neutrino detectors by performing a data-driven demonstration of low-energy event reconstruction leveraging both Cherenkov and scintillation light simultaneously.  By comparing data to model predictions, and allowing certain detector configuration parameters to vary -- such as the fraction of LS in a WbLS target cocktail, or by using PMTs with differing TTS, or deployment of dichroicons -- the predictive power of the model can be validated.  This validated microphysical model of hybrid neutrino detectors can then be used by the neutrino community for design optimization of next-generation hybrid detectors.

\end{itemize}
Taken together, this set of R\&D activities charts a path towards new technology development and eventual selection of the most successful techniques for use in future large-scale liquid scintillator detector searches for \ndbd.


\subsubsection{Barium tagging detectors}

Specific identification of a daughter nucleus from $0\nu\beta\beta$ alongside electron calorimetry with sufficient resolution to reject $2\nu\beta\beta$ can enable an ultra-low-background searches for $0\nu\beta\beta$~\cite{moe:1991ik}.  The implied elimination of radiogenic backgrounds would provide substantial increases in sensitivity with the potential to penetrate into the normal ordering parameter space before becoming background limited, as well as bolstering confidence in a tentative discovery claim.   

Single ion sensing methods with credible applicability to large detectors emerged during the past Snowmass period, for both LXe~\cite{Chambers:2018srx} and GXe~\cite{mcdonald2018demonstration} technologies.  During the coming Snowmass period, these nascent technologies will come to maturity and be demonstrated in increasingly realistic detector configurations.  Strong performance in demonstrators will set the stage for barium tagging modules that follow initial runs of non-barium-tagging ton-scale experiments.  Sustained support for R\&D is needed if this is to be realized.

Barium tagging R\&D is ongoing within both the nEXO~\cite{adhikari2021nexo} and NEXT~\cite{adams2021sensitivity} collaborations, with the favored approaches differentiated by the expected charge state of the barium daughter: Ba$^{2+}$ is dominant in gas, whereas Ba$^{1+}$ and Ba$^{0}$ are favored in liquid, a picture largely supported by studies of the recombination of radon daughters~\cite{albert:2015vma,Novella:2018ewv}.  

\paragraph{Barium tagging in liquid xenon}
Among a number of Ba tagging technologies explored by EXO-200 and nEXO, two front-running schemes have emerged\footnote{LOI: \href{https://www.snowmass21.org/docs/files/summaries/NF/SNOWMASS21-NF5_NF3-RF4_RF0-IF8_IF0_William_Fairbank-120.pdf}{W. Fairbank et al., Barium tagging for a nEXO upgrade and future 136Xe \ndbd detectors}}.  In one scheme~\cite{Chambers:2018srx,mong:2014iya}, the $^{136}$Ba daughter ion or atom is captured in a solid xenon (SXe) layer on a sapphire substrate mounted on a cryogenic probe~\cite{craycraft2019barium,twelker:2014zsa} and removed from LXe to GXe above the liquid. After closing a valve, the probe temperature and gas pressure is reduced toward the 10-30 K and near vacuum regime. To date, individual barium ions have been imaged with high signal-to-noise ratio in two of the four relevant SXe matrix sites, with work in progress on the other two matrix sites and in the one expected single-vacancy for Ba$^{+}$ . Studies of implanting Ba$^{+}$ ions in SXe on a cryogenic substrate, both in LXe and GXe, are also in progress. 

A second scheme~\cite{murray2018design} is based on extraction of the daughter Ba$^{+}$ ion in a capillary tube from LXe to GXe. After passing to vacuum in differential pumping regions while being guided by an RF ion carpet, Ba$^{+}$ capture and single-ion spectroscopic identification is made in a linear Paul trap, followed by $^{136}$Ba$^{+}$ mass identification in a multi-reflection time-of-flight mass spectrometer.  A demonstrator with Ba$^{+}$ ions from both $^{252}$Cf fission and laser ablation sources is expected to prove transport and detection of single Ba$^{+}$ ions from GXe.

To be prepared for a possible Ba tagging upgrade of nEXO in a 10+ year time frame,  R\&D is envisioned in which Ba tagging of daughters of individual beta decays of Cs isotopes is demonstrated. This would occur first in small LXe cells at an accelerator facility with 1-dimensional probe extraction from a localized decay region and later in 3-dimensional probe extraction from the full volume of a mid-sized TPC with $\sim$100 kg of LXe.  Early work at Argonne National Laboratory demonstrated that a 250 MeV $^{139}$Cs beam could be introduced through a thin Havar window into a LXe cell. The decays of the radioactive $^{139}$Ba daughters provide an additional signal for studying the extraction process.  Beams of several different Cs or La isotopes at similar energies can also be produced at TRIUMF, providing a range of possible decay topologies and Ba daughter half-lives. An alternative approach would be to use ~MeV protons to produce $^{136}$Cs in situ via the (p,n) reaction, well suited to Van de Graaff facilities where beamtime is readily available. A pilot irradiation has been carried out at the Triangle Universities Nuclear Laboratory (TUNL) demonstrating production in a gas cell, and a follow up measurement at the University of Kentucky Accelerator Laboratory (UKAL) is being considered. 

A 100 kg natural xenon TPC is being developed at Carleton University for studies of probe insertion and daughter extraction methods. Initial studies will be done with daughters of radon decay. R\&D on 3-D daughter extraction with a probe will be an important component of the next phase of preparation for engineering design of an upgrade TPC that incorporates Ba tagging. 

\paragraph{Barium tagging in xenon gas}
The Ba$^{2+}$ state expected in xenon gas has no low lying outer electronic transitions and so is not amenable to atomic fluorescence detection.  The methodology favored by NEXT\footnote{LOI: \href{https://www.snowmass21.org/docs/files/summaries/NF/SNOWMASS21-NF5_NF10-RF4_RF0-IF9_IF8_Ben_Jones-048.pdf}{NEXT Collaboration: Barium Tagging in Xenon Gas for Neutrinoless Double Beta Decay}} is single molecule fluorescence imaging (SMFI), adding fluorescence transitions via chelation with a fluorescent organic chemosensor at a suitably prepared self-assembled monolayer~\cite{elba,jones:2016qiq}. Fluorescent imaging of individual barium ions using SMFI has been demonstrated at a water-quartz boundary~\cite{mcdonald2018demonstration} and in a high pressure xenon gas environment\cite{byrnes2020demonstration} using commercially available calcium sensing agents.  These chemosenors are typically used for in-cell ion imaging and neither bind or fluoresce under the dry conditions of a TPC.  To overcome this limitation, the NEXT collaboration has developed custom  chemosensors suitable for solventless imaging based on crown-ether fluorophores bonded to rigid aromatic receptors~\cite{thapa2019barium}. Families based on turn-on fluorescence~\cite{thapa2021demonstration} and bi-color imaging~\cite{rivilla2020fluorescent} have both been demonstrated, and ion imaging in single-molecule-sensitive experiments has been accomplished via both SMFI~\cite{thapa2021demonstration} and scanning tunneling microscopy~\cite{herrero2022ba}.

R\&D is ongoing to quantify and optimize performance of single ion sensing organic monolayers.  Ion beams based on thermionically emitted barium, nuclear recoil of radium ions (a dication that serves as a barium ion proxy), and tagged online isotope beams using the ANL CARIBU facility and the ISOLDE accelerator at CERN are being exploited to study and optimize chelation and imaging efficiencies of molecular sensors in vacuum and xenon gas. Concentration of ions produced across the detector volume to the sensor, or actuation of the sensor to the ion, is also a critical area of ongoing R\&D.  The SABAT effort in Europe aims to realize a sensor that scans a fully coated ion detecting cathode with high powered lasers to realize a truly wide-field single ion microscope.  Efforts in the US favor delivery of the ion to the sensor from across the detector cathode, which may be accomplished via either nanofabricated radiofrequency carpet~\cite{jones2021dynamics,Woodruff:2019hte} or transverse transport using self assembled solid electrolyte surfaces, both active areas of R\&D. The NEXT-CRAB (Camera Readout and Barium Tagging)  experiment under construction at ANL will couple a VUV image intensified tracking system with a fully active barium tagging cathode for a first demonstration of the combined process of topological reconstruction and barium tagging of $2\nu\beta\beta$ events in natural xenon.  Meanwhile, the pBOLD program at the Laboratorio Subterraneo de Canfranc aims to use individually tagged radium recoils to characterize barium sensor response in xenon gas.  A well characterized and efficient barium ion sensing cathode proven in these prototypes will set the stage for a first ton-scale barium tagging GXe TPC module.

\subsubsection{Future directions for solid state detectors}
\paragraph{Cryogenic bolometers}
CUPID-1T~\cite{armstrong2019cupid} is a proposed future experiment that envisions using 1000\,kg of the isotope of interest to reach half-life sensitivities greater than $8\times 10^{27}$\,yrs at the $3\sigma$ level. This will require a background index of $\approx 5 \times 10^{-6}$\,cnts/(keV$\cdot$kg$\cdot$yr) and fast readout of over 10000 channels, including both phonon sensors and light sensors. Two options are under consideration for the realization of this concept: a single-cryostat option, where unenriched or larger outer crystals provide shielding for the inner lower-background crystals, and a multiple cryostat option, in which several cryostats, potentially using varying \ndbd\ isotopes, are hosted either at a single laboratory or at sites around the world. 

US groups are actively involved in leading or contributing to several key efforts to support the dual goals of multiplexed readout and background reduction. Areas of particular interest for CUPID-1T include the use of high-speed superconducting sensors for thermal or athermal phonon detection; the adaptation of multiplexed readout technologies (synergy with CMB) to macrobolometers; the development of an active $\gamma$ veto (in synergy with low-mass dark matter experiments); and the incorporation of CMOS and ASIC developments for quantum sensors (in synergy with CMB, DM, and QIS); and development of novel TES sensor fabrication techniques~\cite{hennings2020controlling}. All of these efforts, as well as international R\&D on the use of superconducting crystal coatings to enhance PSD capabilities (including work by CROSS at Canfranc) have the potential to profoundly impact the design and fabrication of bolometric detectors for fundamental science.

\paragraph{48Ca Experiments}
$^{48}$Ca is in principle a highly alluring \ndbd isotope due to its high Q value (4.273~MeV) offering a potentially major suppression of experimental radiogenic backgrounds. Challenges with realizing $^{48}$Ca  expeirments at scale derive from the difficulty of enrichment needed to extract the very low (0.187\%) natural abundance of the double beta decay isotope from natural calcium.  In addition to work on enrichment (to be described in a subsequent section), ongoing work by the CANDES collaboration has aimed to prove a viable detection methodology using CaF$_2$ scintillators.  The present generation of CANDLES is 86kg in mass, and has run for 131 days setting a half life limit of 6$\times 10^{22}$~yr~\cite{tetsuno2020status}.  Detection of scintillation alone is susceptible to scintillation quenching which limits energy resolution~\cite{ umehara2020search}, and  ongoing R\&D is underway to realize CaF$_2$ scintillating bolometers~\cite{ tetsuno2020status}.  After correcting for position dependencies, early iterations have demonstrated resolutions as good as 0.18\%, comfortably sufficient for a sensitive \ndbd search.

\paragraph{3D tracking in amorphous selenium} 
The Selena program proposes to search for \ndbd decay of $^{82}$Se with imaging sensors made from an ionization target layer of amorphous selenium (aSe) coupled to a silicon complementary metal-oxide-semiconductor (CMOS) active pixel array for charge readout.  An initial study predicts background rates at $Q_{\beta\beta}$ = 3 MeV of $^{82}$Se $<6\times10^{-5}$/keV/ton-year~\cite{Chavarria:2016hxk}, a background level possible through a combination of factors: i) the high $Q_{\beta\beta}$ is above most long-lived environmental backgrounds; ii) spatio-temporal correlations effectively reject radioactive decays in the bulk or the surfaces of the imaging modules; iii) external $\gamma$-ray backgrounds, which mostly produce single-electron events from Compton scattering or photoelectric absorption, are suppressed by the requirement that the $\beta\beta$ decay signal events have two clearly identified Bragg peaks (this selection retains 50\% of signal events while rejecting 99.9\% of the single-electron background). 
This level of background suppression requires a realistic pixel pitch of 15 $\mu$m and a livetime of the imager close to 100\%, which both appear to be within reach of current technology. The spectroscopic identification of the \ndbd decay signal over the $2\nu\beta\beta$ decay background is a bigger challenge. A partially validated energy response model predicts an energy resolution at $Q_{\beta\beta}$ of 1.1\% RMS~\cite{Li:2020ryk}. Although there is significant interference from the two-neutrino channel, a sensitivity $T_{1/2} > 5\cdot10^{28}$ y could be possible in a 100 ton-year exposure. In case of a discovery signal, Selena could thus play an important followup role, observe the process in the isotope $^{82}$Se, and confirm that it indeed occurs by the emissions of two electrons. 

\subsection{Improving performance of existing technologies}
\label{sub:bbimproved}



\subsubsection{Approaches to reducing radiogenic backgrounds}
Radiogenic backgrounds from the materials used to construct \ndbd\ experiments are being addressed through the development of ultra-low-background cables and electronics and structural materials, including active materials. Some of these are already in use in current-generation experiments, but their adoption is expected to increase at and beyond the ton scale. 

\paragraph{Cables and electronics}
Polyimide-based materials, like Kapton, are widely used in flexible cables and circuitry due to their unique electrical and mechanical characteristics, but their use in \ndbd\ experiments has been limited by relatively high $^{238}$U and $^{232}$Th backgrounds ($1.2 \times 10^4$ and $490~\mu$Bq/kg, respectively) in commercial off-the-shelf Kapton varieties. Recent efforts have identified the manufacturing additive responsible for the elevated background levels, demonstrating the production of Kapton with backgrounds two-to-three orders of magnitude cleaner than commercial-off-the-shelf options, and copper-clad polyimide laminates using the resulting material with 238U and $^{232}$Th activities of $110$ and $89~\mu$Bq/kg, respectively~\cite{ARNQUIST2020163573}. Cables made of these materials are used in the baseline designs of the LEGEND and nEXO experiments. 

Increasing channel counts, detector sizes, and low-background demands have prompted increasing use of Application Specific Integrated Circuits (ASICs) in \ndbd\ experiments. Key areas of development for use in these experiments are building radiopure readout, aggregating data inside the detector vessels to reduce the potential points of failure, and limiting power consumption to enable operation in cryogenic environments. Low-noise, precision timing, and high dynamic range capabilities are achievable, but are generally in tension with reducing power consumption and increasing bandwidth. Triggered acquisition is one approach that may resolve this problem. More extensive discussion can be found in Ref.\footnote{Frontier Report: Snowmass IF07}


\paragraph{Underground electroformed copper and alloys}
During the previous generation of \ndbd\ experiments, the \textsc{Majorana} Collaboration developed techniques to produce and assay ultrapure electroformed Cu, produced at a shallow underground site at PPNL and within a dedicated cleanroom facility on the 4850-ft level of SURF.  Over the course of 6 years, over 3000 kg of copper was grown on 60 bath-runs formed on custom mandrels. The produced Cu exceeded the 10 ksi tensile strength design criteria and was found to have a leading limit in the $^{238}$U and $^{232}$Th purity of $<0.1~\mu$Bq/kg Cu \cite{ABGRALL201622}. Improved cleaning and post-production handling techniques mitigated any surface contamination of the final parts used in the experiment \cite{doi:10.1063/1.5019001}. The LEGEND program is building upon the MAJORANA experience and existing facilities to meet its needs for R\&D and the production of underground electroformed Cu.  The next-generation nEXO program also plans to make use of underground electroformed copper.

The use of this material is constrained by the physical properties of pure copper.  A higher strength material which possesses many of the favorable attributes of copper yet remains radiopure is desired.  Exploration of chromium copper alloying techniques has provided improved mechanical performance and adequate radiopurity but the setup of underground large scale applications is currently limited by EH\&S concerns.

\paragraph{Additive manufacturing}
Additive manufacturing offers many advantages for preparation components for low-background experiments such as \ndbd. Their low overhead and supporting infrastructure would allow for 3D printers to be located underground near the experiment eliminating risks of radio-contamination during travel to the experiment site.  They are relatively simple to use, highly adaptable for a wide range of shapes and sizes and can operate within inert radon-purged environments.  Of the numerous 3D printed technologies which exist, fusion deposition modeling and stereolithography offer exciting opportunities for printing of scintillator materials.  These materials can provide not only structural support for the given experiment but also possess the ability to veto radioactive decays which occur within or interact with the scintillation medium.  While there are numerous advantages for 3D printed low-background materials, there is considerable R\&D effort needed to address outstanding questions to their use.  Examples of targeted questions include: Can their scintillation performance match non-3D printed components?  Can one prepare or acquire low-background material for use in 3D printers?  How well do 3D printed components hold up in extreme environments such as cryogenic liquid noble gases? Though these efforts are in their early stages, favorable answers to these open questions could make them an important tool to addressing radiogenic backgrounds in beyond-the-ton-scale experiments. 

\subsubsection{Approaches to reducing cosmogenic backgrounds}

As radiogenic backgrounds are  eliminated through improved material screening, self-shielding, and active technological approaches, backgrounds of non-radiogenic origin are expected to become more substantial.  Of these, notable examples include cosmogenic neutron capture to produce problematic long lived isotopes such as $^{68}$Ge and $^{60}$Co in germanium~\cite{mei2009cosmogenic, cebrian2006cosmogenic} and $^{137}$Xe in xenon~\cite{ albert2016cosmogenic, baudis2015cosmogenic}. In tellurium, a variety of long-lived radioisotopes produced in cosmogenic activation become relevant at kiloton scales~\cite{ lozza2015cosmogenic}.  Protection of materials from activation during a cool-down period prior to installation, as well as extensive muon veto and neutron moderation in situ~\cite{ pandola2007monte}, will be crucial for future very large experiments. Strategies to mitigate the cosmogenic background by admixture of a neutron absorbing / tagging isotope~\cite{ rogers2020mitigation} may also be a promising path to reduce neutron capture backgrounds.  

\subsubsection{New background sources: ($\alpha$,n) reactions and solar neutrinos}

At large detector scales,  new classes of background which were previously negligible are expected to become limiting.  One notable class of backgrounds that is beginning to receive attention is that of ($\alpha$,n) reactions\footnote{LOI: \href{https://www.snowmass21.org/docs/files/summaries/CF/SNOWMASS21-CF1_CF0-NF5_NF0-RF4_RF0-AF5_AF0-IF9_IF0_Shawn_Westerdale-052.pdf}{S. Westerdale, Neutron yield in ($\alpha$, n)-reactions in rare-event searches}}~\cite{kudryavtsev2020neutron}, which can introduce neutron activations even in the hypothetical scenario where shielding from cosmic ray activity is absolute.  Detailed experimental characterization and simulation of  ($\alpha$,n) backgrounds will receive increasing attention in the coming Snowmass period.

Another background source that will transition from negligible to important as experiments pass the 10~ton scale are backgrounds associated with solar neutrino interactions~\cite{ de2011solar, ejiri2016solar, ejiri2014charged}. These have been considered by several collaborations and generally found to be irrelevant at the ton-scale (see e.g.~\cite{ albert2018sensitivity}), though at larger scales they will eventually become problematic.  The background due to solar neutrinos may be rejected by fast timing in large liquid scintillator experiments~\cite{elagin2017separating}, or potentially by directional reconstruction in TPC detectors, or barium tagging, among other possible mitigation strategies.

\subsubsection{Deepening understanding of detector micro-physics}

As electronic readout elements continue to improve, noise associated with readout and data acquisition eventually becomes sub-dominant to fluctuations and uncertainties associated with the detailed micro-physics of the working detector medium.  Fundamental studies of the microphysics of light, charge, and phonon production in \ndbd active media are mandatory in order to maximize the achievable precision of running and proposed technologies.  

In time projection chambers, this project includes (but is not limited to) understanding light and charge production via both scintillation~\cite{anton2020measurement}, electroluminescence~\cite{monteiro2007secondary} and sub-leading emission mechanisms including neutral bremstrahlung~\cite{henriques2022neutral}; Understanding charge drift and diffusion~\cite{albert2017measurement,simon2018electron} in both pure xenon and mixtures~\cite{mcdonald2019electron,azevedo2016homeopathic}.  Charge emission and migration from and along surfaces in noble media is also an evolving area of understanding~\cite{akerib2020investigation,sorensen2018two,bodnia2021electric}.   Continued development of microscopic and semi-empirical models informed by new data from a wider array of precision experiments and test stands will remain critical for achieving optimal performance. Notable cross-collaboration efforts include the NEST framework~\cite{szydagis2011nest,mock2014modeling}\footnote{LOI: \href{https://www.snowmass21.org/docs/files/summaries/IF/SNOWMASS21-IF8_IF0-NF5_NF10-CF1_CF0-CompF5_CompF7_Matthew_Szydagis-104.pdf}{Velan et al., NEST, The Noble Element Simulation Technique:
A Multi-Disciplinary Monte Carlo Tool and Framework for Noble Elements}} and microscopic models of electron transport in gases in the MagBoltz / PyBoltz family of simulation codes~\cite{schindler2010calculation,al2020electron}.

For liquid scintillators detectors,  fundamental work on metal-loading and the resultant properties~\cite{biller2017new,shimizu2019double}\footnote{LOI: \href{https://www.snowmass21.org/docs/files/summaries/NF/SNOWMASS21-NF10_NF5_Steve_Biller-059.pdf}{S. Biller, A Method to Load Tellurium in Liquid Scintillator to Study Neutrinoless Double Beta Decay}} of the working medium is ongoing.  Success of future large experiments relies on achieving the required combination of loading fraction, optical transparency~\cite{hans2015purification} and management of scintillation quenching~\cite{hans2020light} to appropriate levels.  In addition to xenon- and tellurium loaded approaches pursued by Kamland-Zen and SNO+ respectively, work on loading with other double beta metals has also been explored, including Neodynium~\cite{barabanov2012nd}, Molybdenum~\cite{gehman2010solubility} and Tin~\cite{hwang2009search}. Other areas of study include increasing the use of spatiotemporal information in discriminating backgrounds; recent success has been achieved using a specially-formulated neural network for the KamLAND-Zen experiment \cite{2022arXiv220301870L}. Direct detection of Cherenkov light, which may be used for background identification, has been achieved in several test-stands \cite{chess, chesslappd, flatdot, LI2016303}. Further studies of the microphysical properties of novel scintillators and photodetectors are planned in the EOS and NuDot experiments. Simulations of these processes are being improved via increasing use of the \textit{Chroma} parallel GPU-based optical Monte Carl software package \cite{chroma}.

In the area of cryogenic bolometry, R\&D on $^{100}$Mo-based scintillating bolometers is ongoing in order to realize CUPID-1T.  Specific areas of development include coating of crystals with superconducting layers to improve pulse shape discrimination~\cite{armatol2021cupid}, implemenation of active gamma ray vetos (with strong synergy to dark matter experiments), realization of high speed superconducting sensors including TES and MKIDs, development of readout multiplexing solutions,  and cryogenic ASIC R\&D.  Development of each of these devices represents a coupled problem of understanding material micro-physics and optimizing materials and interfaces in the presence of those intrinsic behaviours.

In germanium, important topics of study include novel detector geometries and improved control of impurity gradients, allowing for the development of larger-mass detectors without sacrificing energy reconstruction capabilities. These efforts have resulted in the inverted-coaxial point contact detector style used in the baseline design of LEGEND \cite{Salathe2017}. In addition to reduced backgrounds in $^{76}$Ge-based \ndbd\ experiments, these designs will enable improved Ge-based radioassay sensitivities. Improved microphysical models of charge transport in the detectors, particularly near the detector surfaces, are another area of development aimed at reducing backgrounds in \ndbd\ experiments, with strong synergy to dedicated dark matter experiments \cite{TUBE, GALATEA_scan}. Simulations that correctly model the charge drift and resulting waveforms from these surface background events are under active development. 

\subsection{Isotope procurement}       
\label{sub:bbisotope}

Acquisition of isotopes for future beyond-ton scale neutrinoless double beta decay experiments presents substantial challenges.  Plausibility of isotopic acquisition appears likely to be a driver for technology choices in the field in the post-ton-scale phases. Enrichment of large quantities of stable isotope is typically achieved by centrifugation using a cascade of centrifuges that must be finely-tuned for an isotope of interest.  Enrichment by electromagnetic isotope separation (EMIS) gives high purity separation into all component isotopes with a rapidly re-configurable devices, but does not presently have the requisite capacity to produce ton-scale masses. Plasma isotope separation techniques are also under continuing development  with the potential to realize another technique with high throughput and quick configurability, if this R\&D is successful.  For ton-scale masses of isotopes that require enrichment for practicable use (all except potentially $^{130}$Te) centrifugation is thus the only presently available technique.  There are only a small number of suppliers with this capability in the world, and those in Russia that have supplied most \ndbd experiments in the past appear likely to be inaccessible to US experiments for some time, due to recent geopolitical events.  Beyond these suppliers, there is centriguation capacity at Oak Ridge National Laboratory stewarded by the DOE Isotope Program, as well as by the Urenco British-German-Dutch nuclear fuel consortium that operates enrichment plants in Germany, the Netherlands, United States, and United Kingdom. 

In this section we briefly outline the supply challenges and ideas for acquisition of some of the isotopes used by existing programs, ordered by the natural abundance of their double beta decay isotope.
 
\paragraph{$^{130}$Te}
Tellurium is an attractive isotope for future very large \ndbd experiments since its natural composition includes 33.8\% $^{130}$Te.  Thus use of unenriched isotope is thus most feasible for tellurium. The \textsc{Theia} concept considers loading 5\% $^{nat}$Te into an 8~m radius balloon filled with ultra-pure liquid scintillator. World annual production of tellurium is estimated to be around around 470 tons trading at \$65-80 per kg, so cost and acquisition for experiments at the tens- to hundreds-of-ton scale appear plausible.

\paragraph{$^{100}$Mo}
Molybdenum presents a potentially interesting and scalable isotope for future \ndbd experiments.  Annual world production of molybdenum is at the 300,000 ton scale, primarily for use in production of molybdenum steels.  The quantities needed for a \ndbd experiment are relatively modest on this scale. Furthermore, there is interest in enrichment of molybdenum for medical applications. In particular, $^{98}$Mo and $^{100}$Mo can be used in the production of $^{99m}$Tc, a short lived isotope which is used in medical imaging with tens of millions of medical diagnostic procedures annually.   World shortage of the short-lived isotope $^{99}$Mo that is used for most existing production has compromised the $^{99m}$Tc supply chain~\cite{ruth2020shortage}, making production via reactions such as Mo($\gamma$,n)Mo  or Mo(n,$\gamma$)Mo of special interest. If these approaches prove as economically viable as centrifugation, enrichment of the required $^{100}$Mo for a future large double beta decay experiment as a by-product is a potentially appealing prospect.
 
\paragraph{$^{82}$Se}
The isotope $^{82}$Se comprises 8.7\% of natural selenium, whose annual world production is around 2,200 tons per year. Selenium enrichment follows a similar process to uranium enrichment for power or weapons applications, proceeding by centrifugation of the hexafluoride compound in the gaseous state~\cite{beeman2015double}. Concepts for future \ndbd experiments based on selenium may be based either on direct use of the enriched hexafluoride gas, or through conversion of the hexafluoride into the metal for use in amorphous layers.  Based on availability and similarity of enrichment to existing industrial processes for uranium, scalable acquisition and enrichment of selenium both appear viable, though scalability of the relevant detector technologies has yet to be demonstrated.

\paragraph{$^{136}$Xe}
Xenon is produced primarily through liquefaction of air required to supply oxygen for the steel industry, with typical production around 100 tons per year, though this has recently been impacted by disruptions from geopolitical events.  The isotope $^{136}$Xe represents an 8.9\% isotopic fraction of natural xenon, so practical neutrinoless double beta decay searches to date have relied on enrichment to a 90\% level, which can be achieved by centrifugation~\cite{Tikhomirov:2000td}.  

Future $^{136}$Xe experiments would either load a large quantity of xenon into a liquid argon or liquid scintillator, or otherwise realize a monolithic TPC at the hundred ton to kiloton scale.  To enable such detectors, new xenon acquisition methods would be required.   Ref.~\cite{avasthi2021kiloton} outlines that the total mass of Xe in the atmosphere is approximately 200 Mtonnes, providing an in-principle ample supply.  Through xenon absorption from atmospheric air on selective metal organic frameworks followed by thermal swing absorption, it may be possible to dramatically increase the xenon supply.  Xenon is also produced with a modest 50\% enrichment of $^{136}$Xe in spent nuclear fuel~\cite{Hayes:2012sg}. There is around 6kg of $^{136}$Xe produced per ton of uranium used.  Since around 60,000 tons of uranium are used per year for nuclear power generation, an additional supply of up to 720 tons per year of 50\% enriched $^{136}$Xe may in principle be accessible by reprocessing this fuel.  Programmatic issues associated with reprocessing fuel in the context of realizing  physics experiments are, naturally, non-trivial.
 
\paragraph{$^{76}$Ge}
World production of germanium is around 120 tons per year, at a market price of \$1000/kg, of which 7.8\% is the double beta decay isotope $^{76}$Ge.  The primary industrial use of germanium is in the semiconductor industry, which drives production. Commercial enrichment of $^{72}$Ge for the semiconductor industry for pre-amorphisation implantation means that $^{76}$Ge may be produced as a by-product, easing the supply challenges somewhat for this isotope.  \ndbd experiments based on germanium diodes are planned up to the ton-scale with LEGEND, though scalability of these techniques to tens or hundreds of tons appears both difficult and expensive. The cost of germanium, its supply chain, and the difficulty of scaling a concept based on individual crystals to the multi-ton scale suggest that the need for hundred-ton scale quantities for next-generation \ndbd experiments is unlikely.
 
\paragraph{$^{48}$Ca }
Calcium is the fifth most abundant element in the Earth's crust (4.1\%) and tens of thousands of tons are extracted per year, making the supply almost inexhaustible from the viewpoint of double beta decay experiments.  The isotope is especially technologically appealing since its high Q-value is far above most radiogenic backgrounds, which is expected to dramatically simplify background rejection in large-scale experiments.  On the other hand, the small isotopic abundance of the double beta decay isotope $^{48}$Ca (0.187\%), as well as the lack of an obviously scalable enrichment method have prevented $^{48}$Ca from being a leading candidate for large-scale experiments to date.  Promising ongoing work on enrichment by laser isotope separation (LIS)~\cite{Matsuoka:2020spy} and crown ether resin~\cite{Umehara:2015cna} methods have both demonstrated some isotopic separation in R\&D studies, though the plausibility of application at industrial scales required for enrichment of many tons of isotope has yet to be proven.  If a method for enrichment of $^{48}$Ca can be proven scalable it would surely become a highly compelling isotope for future double beta decays searches.

\subsection{Facilities for 0$\nu\beta\beta$ experiments}
\label{sub:facilities}
\subsubsection{Underground laboratory facilities}

Experiments searching for \ndbd are susceptible to backgrounds from cosmogenic activation of detector materials, and as such the detectors must be situated deep underground.  In addition to requiring underground facilities for experimental operation, current and future efforts will require auxiliary underground facilities for radioassay, material production, and storage.  There is an extensive ``Underground Frontier'' effort within the Snowmass process, and so we do not re-tread the scope of this effort here.  We note though, that there is a clear need for at least two, and potentially more, underground sites for ton-scale or larger \ndbd experiments within the next decade that will explore the inverted mass ordering region of half-life sensitivities. Furthermore, preparations will be needed for mounting future experiments at even larger scales (either through single large-scale monolithic experiments or modular and potentially multi-site approaches) that will be required if sensitivities to probe to normal mass ordering region of parameter space prove to be required.   Table~\ref{tab:ugSites} lists the underground laboratories that have housed or are proposed to house \ndbd experiments to date.

\begin{table}[]
\begin{tabular}{|p{4cm}|p{2cm}|p{5cm}|p{2cm}|p{2cm}|}
\hline
\textbf{Laboratory}                          & \textbf{Country} & \textbf{Experiment(s)}                       & \textbf{Access} & \textbf{Depth (m.w.e)} \\ \hline
Laboratoire Souterrain de Modane (LSM)       & France           & CUPID-Mo, SuperNEMO                          & Horizontal      & 4,800            \\ \hline
Laboratorio Subterraneo de Canfranc (LSC)    & Spain            & NEXT-WHITE, NEXT-100, NEXT-HD module 1      & Horizontal      & 2450             \\ \hline
Yangyang Underground Laboratory              & South Korea      & AMoRE                                        & Horizontal      & 2000             \\ \hline
Kamioka Observatory                          & Japan            & KamLAND-Zen, KamLAND2-Zen, CANDLES           & Horizontal      & 2700             \\ \hline
China Jinping Underground Laboratory (CJPL)  & China            & PandaX-III                                   & Horizontal      & 6700             \\ \hline
Sudbury Neutrino Observatory (SNOLAB)        & Canada           & SNO+, nEXO, LEGEND-1000                      & Vertical        & 6010             \\ \hline
Sanford Underground Research Facility (SURF) & USA              & Majorana Demonstrator, Theia                 & Vertical        & 4300             \\ \hline
Gran Sasso National Laboratory (LNGS)        & Italy            & CUORE, CUPID, GERDA, LEGEND-200, LEGEND-1000 & Horizontal      & 3400             \\ \hline
Waste Isolation Pilot Plant (WIPP)*          & USA              & EXO-200                                      & Vertical        & 2000             \\ \hline
\end{tabular}
\caption{Underground sites that have housed past and are proposed to house future underground \ndbd experiments.\label{tab:ugSites}}
\end{table}

\subsubsection{Assay capabilities}

Precision material characterization for radio-impurities is critical to the success of all \ndbd programs.  Measurements with ICMPS, GDMS, neutron activation analysis and germanium counting are required to vet all materials used in the construction of detectors.  Extensive radiopurity databases~\cite{cooley2018radiopurity} are an invaluable tool for the community to share information and maximally profit from synergistic efforts.  

The need for assay facilities is cross-cutting across all low background / underground physics programs, and will be discussed extensively elsewhere in the Snowmass process.  Such facilities are an invaluable and non-negotiable requirement for the success of the world \ndbd program.

\subsection{Other approaches to testing for Majorana neutrinos}
Neutrinoless double beta decay is the most sensitive known way to search for Majorana neutrinos.  However, other approaches have also been considered and pursued. While less sensitive under the minimal light Majorana neutrino exchange mechanism, such alternative techniques to test for Majorana fermions may offer enhanced sensitivity in the case of non-standard mechanisms.

\label{sub:bbother}
\subsubsection{Neutrinoless double electron capture}
Neutrinoless double electron capture~\cite{blaum2020neutrinoless,kotila2014neutrinoless}  $0\nu$ EC EC  is an analog of neutrinoless double beta decay wherein the two positrons that would otherwise have been emitted in a $0\nu\beta^+\beta ^+$ process are instead replaced by two electrons captured from the inner atomic shells by the nucleus, thus turning two neutrons into two protons. As with \ndbd, this process occurs if and only if the neutrino is a Majorana fermion.  Experimental searches focus on geochemical methods, $\gamma$ ray spectroscopy, or integrated calorimetry of the decay.

While in general $0\nu$ EC EC  is expected to be slow relative to \ndbd, early speculations suggested that there may be an important resonant enhancement active in certain isotopes~\cite{krivoruchenko2011resonance,eliseev2011resonant} that could make the projected decay rate competitive with or even faster than that of \ndbd in the commonly studied isotopes.  Following accurate Penning trap measurements to establish the relevant Q-values with sufficient precision to assess possible enhancements, it now appears unlikely that sufficient  enhancement factors are available to make $0\nu$ EC EC an appealing experimental prospect in most cases.  For example, among the near resonant isotopes discussed in Ref.~\cite{blaum2020neutrinoless}, almost all have project half-lives in excess of $10^{31}$~yr, whereas the best current experimental limits are around 10$^{21}$~yr.

The standard-model counterpart of $0\nu$ EC EC is two neutrino double electron capture, or $2\nu$ EC EC, wherein two electrons are captured and two neutrinos emitted.  This process was first observed in 2019 for the isotope $^{124}$Xe by the XENON1T collaboration~\cite{aprile2019observation}.  It has also been searched for, but not yet observed, in a wide variety of other isotopes. A full review of experimental status of searches for both $0\nu$ EC EC and $2\nu$ EC EC  is given in Ref~\cite{blaum2020neutrinoless}.
\label{sub:DoubleCapture}

\subsubsection{Majorana particles at colliders}

The effects of the nature of neutrino mass drop precipitously with energy, since at high energies helicity becomes equivalent to chirality, requiring the observational effects of lepton number conservation to become equivalent to those of angular momentum conservation.  Thus, the Majorana nature of light neutrinos is typically not directly testable at collider experiments. However, the mechanisms by which light Majorana neutrinos obtain their mass - for example, the Type-I seasaw mechanism, often  imply the existence of additional heavy leptonic degrees of freedom. If the new beyond-standard-model (BSM) neutrino states are sufficiently heavy, their Majorana nature can be introduce non-zero rates of lepton number violating processes with the heavy state as an intermediate~\cite{kovalenko2009lepton}.  Given an observation of a unstable heavy neutrino, its Majorana nature can also be probed directly through decay kinematics~\cite{balantekin2019addressing}.  Searching for such particles at colliders is an active area that can be thought of as an indirect test of the nature of neutrino mass~\cite{das2017bounds,atre2009search,alva2015heavy,han2006signatures}.
\newpage
\section{Neutrino Electromagnetic Properties}
\label{sec:Electromagnetic}

\begin{figure}[b!]
\centering
\subfigure[]{\label{fig:smemvert}
\boxed{
\includegraphics*[width=0.3\linewidth]{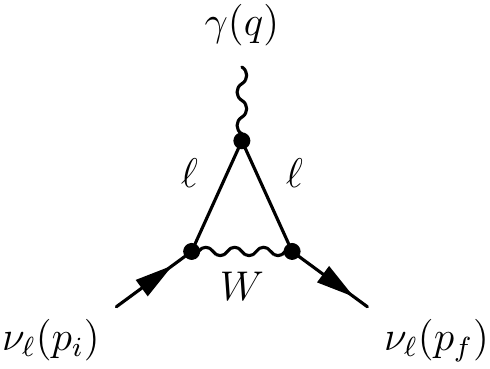}
\quad
\includegraphics*[width=0.3\linewidth]{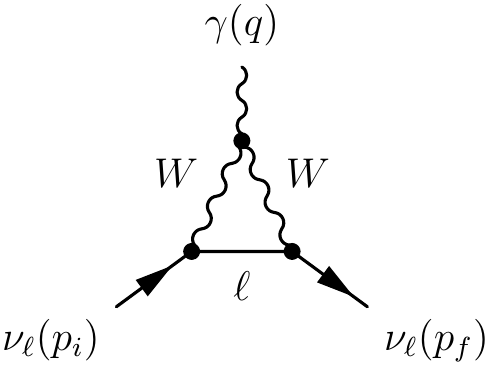}
}
}
\subfigure[]{\label{fig:genemvert}
\boxed{
\includegraphics*[width=0.3\linewidth]{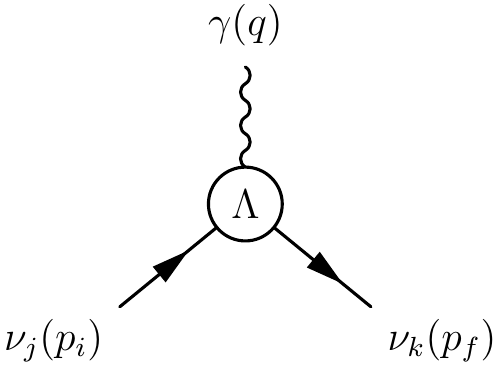}
}
}
\caption{ \label{fig:emvert}
\subref{fig:smemvert}
One-loop Feynman diagrams that generate the charge radii of flavor neutrinos in the SM, with $\ell=e,\mu,\tau$.
\subref{fig:genemvert}
General neutrino electromagnetic vertex of massive neutrinos
($j,k=1,2,3,\ldots$).
}
\end{figure}

In the Standard Model (SM) neutrinos are electrically neutral and
do not interact with photons at the tree level.
However, radiative corrections generate neutrino interactions
with photons through loops involving the charged leptons and the $W$ boson,
as shown by the one-loop diagrams in Fig.~\ref{fig:smemvert}.
The corresponding neutrino-photon interactions
are described by charge radii of the flavor neutrinos
(see the review in Ref.~\cite{Giunti:2014ixa}).
Therefore,
even in the SM, where neutrinos are neutral and massless,
there are non-zero neutrino electromagnetic properties.

Physics beyond the SM can generate
more general electromagnetic neutrino interactions
of massive neutrinos
which are described by the diagram in Fig.~\ref{fig:genemvert}
with the vertex function
$\Lambda_{kj}^{\alpha}(q)$
that depends on the four-momentum transfer
$q = p_{i} - p_{f}$.
Taking into account general principles such as
Lorentz invariance and electromagnetic gauge invariance,
for ultrarelativistic left-handed neutrinos at low-$q^2$,
the electromagnetic vertex function can be written as
\begin{equation}
\Lambda_{kj}^{\alpha}(q)
=
\gamma^{\alpha}
\left(
Q_{\nu_{kj}}
+
\dfrac{q^2}{6}\, \langle r^2 \rangle_{\nu_{kj}}
\right)
-
i \sigma^{\alpha\beta} q_{\beta} \, \mu_{\nu_{kj}}
,
\label{vertfun}
\end{equation}
where
$Q_{\nu_{kj}}$,
$\langle r^2 \rangle_{\nu_{kj}}$, and
$\mu_{\nu_{kj}}$
are, respectively,
the neutrino
electric charges,
charge radii, and
magnetic moments\footnote{
For simplicity, in Eq.~\eqref{vertfun}
we consider effective charge radii
that include possible anapole moments
and effective magnetic moments
that include possible electric moments
(see, e.g., Ref.~\cite{Giunti:2014ixa}).
We also neglect the neutrino mass differences in the interaction.
}.
Note that these electromagnetic properties are defined in the
mass basis and all of them can have diagonal components for $j=k$
and off-diagonal (transition) components for $j \neq k$
in the case of Dirac neutrinos.
The matrix of each electromagnetic property is Hermitian:
$X_{\nu_{kj}}=X_{\nu_{jk}}^{*}$
for
$X=Q,\langle r^2 \rangle,\mu$.
For Majorana neutrinos
the matrices of
electric charges and magnetic moments
are antisymmetric:
$X_{\nu_{kj}}=-X_{\nu_{jk}}$
for
$X=Q,\mu$.
Therefore,
Majorana neutrinos are distinguished from Dirac neutrinos
by the lack of diagonal components
for the electric charges and magnetic moments.
Moreover,
the off-diagonal
electric charges and magnetic moments
of Majorana neutrinos are purely imaginary.

The effective electromagnetic properties of flavor neutrinos
depend on neutrino mixing
and
on the process in which they are observed.
A characteristic of the electromagnetic vertex function
is that the term which depends on the
electric charges and charge radii
is helicity-conserving,
whereas the term containing the magnetic moments
is helicity-flipping.
This has important consequences for the ways in which the electromagnetic properties contribute to neutrino interactions:
the flavor-conserving contributions of the
electric charges and charge radii
add coherently with the SM contributions to the amplitudes of neutrino interaction processes,
whereas the contributions of the magnetic moments
and the flavor-violating contributions of the
electric charges and charge radii
generate additional cross-section terms.

We discuss the
neutrino charge radii,
magnetic moments, and
electric charges in the following Subsections~\ref{sub:Radius}--\ref{sub:Charge}

\begin{table}[t!]
\begin{minipage}{\textwidth}
\centering
\renewcommand{\arraystretch}{1.2}
\begin{tabular}{llcrcc}
Method & Experiment & Limit $[\text{cm}^2]$ & CL & Year & Ref.\\
\hline
\multirow{2}{*}{Reactor $\bar\nu_e \, e^-$}
&Krasnoyarsk	&$|\langle r^2 \rangle_{\nu_{e}}|<7.3\times10^{-32}$		&90\%	&1992&
\cite{Vidyakin:1992nf}\\
&TEXONO		&$\langle r^2 \rangle_{\nu_{e}}\in(-4.2,6.6)\times10^{-32}$	&90\%	&2009&
\cite{Deniz:2009mu}~\footnote{\label{f2}Corrected by a factor of two due to a different convention.}\\
\hline
\multirow{2}{*}{Accelerator $\nu_e \, e^-$}
&LAMPF		&$\langle r^2 \rangle_{\nu_{e}}\in(-7.12,10.88)\times10^{-32}$	&90\%	&1992&
\cite{Allen:1992qe}~\footref{f2}\\
&LSND		&$\langle r^2 \rangle_{\nu_{e}}\in(-5.94,8.28)\times10^{-32}$	&90\%	&2001&
\cite{Auerbach:2001wg}~\footref{f2}\\
\hline
\multirow{2}{*}{Accelerator $\nu_{\mu} \, e^-$}
&BNL-E734	&$\langle r^2 \rangle_{\nu_{\mu}}\in(-5.7,1.1)\times10^{-32}$	&90\%	&1990&
\cite{Ahrens:1990fp}~\footref{f2}\textsuperscript{,}\footnote{Corrected in Ref.~\cite{Hirsch:2002uv}.}\\
&CHARM-II	&$|\langle r^2 \rangle_{\nu_{\mu}}|<1.2\times10^{-32}$	&90\%	&1994&
\cite{Vilain:1994hm}~\footref{f2}\\
\hline
CEvNS
&
\begin{tabular}{c}
COHERENT
\\[-0.1cm]
\cite{Akimov:2017ade,Akimov:2020pdx}
\end{tabular}
&
$
\begin{array}{c} \displaystyle
\langle r^2 \rangle_{\nu_{e}}\in(-68,11)\times10^{-32}
\\ \displaystyle
\langle r^2 \rangle_{\nu_{\mu}}\in(-68,12)\times10^{-32}
\\ \displaystyle
|\langle r^2 \rangle_{\nu_{e\mu}}|<33\times10^{-32}
\\ \displaystyle
|\langle r^2 \rangle_{\nu_{e\tau}}|<40\times10^{-32}
\\ \displaystyle
|\langle r^2 \rangle_{\nu_{\mu\tau}}|<40\times10^{-32}
\end{array}
$
&95\%	&2020&
\cite{Cadeddu:2020lky}\\
\hline
\end{tabular}
\end{minipage}
\caption{ \label{tab:chr}
Experimental limits for the neutrino charge radii.
}
\end{table}

\subsection{Charge Radius}
\label{sub:Radius}

In the SM neutrinos are massless and
the generation lepton numbers are conserved.
Therefore, the charge radii are well-defined in the flavor basis.
The charge radii of the flavor neutrinos
generated by one-loop radiative corrections are given
by~\cite{Bernabeu:2002pd}~\footnote{
The difference by a factor of 2 with respect to
the definition of the charge radius
in Ref.~\cite{Bernabeu:2002pd}
is explained in Ref.~\cite{Cadeddu:2018dux}.
}
\begin{equation}
\langle r^2 \rangle_{\nu_{\alpha}}^{\text{SM}}
=
-
\frac{G_{\text{F}}}{2\sqrt{2}\pi^{2}}
\left[
3-2\ln\left(\frac{m_{\alpha}^{2}}{m^{2}_{W}}\right)
\right]
=
\left\{
\begin{array}{rl} \displaystyle
- 0.83 \times 10^{-32} \, \text{cm}^{2}
& \displaystyle
(\nu_{e})
\\ \displaystyle
- 0.48 \times 10^{-32} \, \text{cm}^{2}
& \displaystyle
(\nu_{\mu})
\\ \displaystyle
- 0.30 \times 10^{-32} \, \text{cm}^{2}
& \displaystyle
(\nu_{\tau})
\end{array}
\right.
\label{SMchr}
\end{equation}
where $m_{W}$ is the $W$ boson mass
and $m_{\alpha}$ are the charged lepton masses for $\alpha=e,\mu,\tau$.
The values of
$\langle r^2 \rangle_{\nu_{e}}^{\text{SM}}$
and
$\langle r^2 \rangle_{\nu_{\mu}}^{\text{SM}}$
are only about one order of magnitude
lower than the current experimental upper bounds listed in Table~\ref{tab:chr}.
Therefore,
an experimental effort is mandatory in order to try to improve
the experimental sensitivity and try to discover
the neutrino electromagnetic interactions
predicted by the SM.

Finding possible deviations from the SM values would be a discovery
of new physics beyond the SM.
In this case,
since neutrinos are massive and the generation lepton numbers are not conserved,
the effective charge radii of flavor neutrinos are related to the fundamental
charge radii of massive neutrinos by~\cite{Kouzakov:2017hbc}
\begin{equation}
\langle r^2 \rangle_{\nu_{\ell'\ell}}
=
\sum_{k,j}
U_{\ell' k}
\langle r^2 \rangle_{\nu_{kj}}
U_{\ell j}^{*}
,
\label{chrflavor}
\end{equation}
where $U$ is the neutrino mixing matrix.
Note that the unitary transformation from the mass to the flavor basis
preserves the hermiticity of the matrix of charge radii:
$\langle r^2 \rangle_{\nu_{\ell'\ell}}=\langle r^2 \rangle_{\nu_{\ell\ell'}}^{*}$.
Therefore the diagonal charge radii of flavor neutrinos
$\langle r^2 \rangle_{\nu_{e}}\equiv\langle r^2 \rangle_{\nu_{ee}}$,
$\langle r^2 \rangle_{\nu_{\mu}}\equiv\langle r^2 \rangle_{\nu_{\mu\mu}}$, and
$\langle r^2 \rangle_{\nu_{\tau}}\equiv\langle r^2 \rangle_{\nu_{\tau\tau}}$
are real.

The traditional way to obtain limits on
$\langle r^2 \rangle_{\nu_{e}}$
is the observation of neutrino-electron elastic scattering
using the intense fluxes of reactor $\bar\nu_e$'s.
The limits on
$\langle r^2 \rangle_{\nu_{\mu}}$
have been obtained with neutrino-electron elastic scattering
of accelerator $\nu_\mu$'s produced mainly by pion decay.
These methods have been recently complemented by
the measurement of coherent elastic neutrino-nucleus scattering (CEvNS)
of low-energy electron and muon neutrinos
produced by pion and muon decays at rest
or reactor $\bar\nu_e$'s.
Note that it is possible to probe not only the diagonal
neutrino charge radii, but also the off-diagonal
ones~\cite{Kouzakov:2017hbc,Cadeddu:2018dux}.

Future probes of the neutrino charge radii will take place at
reactor and accelerator experiments
that can measure neutrino-electron elastic scattering
(e.g. MINERvA, SBND, and DUNE~\cite{Mathur:2021trm})
and/or
CEvNS~\footnote{\label{fn:CEvNS}
LOIs:
\href{https://www.snowmass21.org/docs/files/summaries/NF/SNOWMASS21-NF9_NF5-CF1_CF0-IF8_IF0_JNewby-108.pdf}{``ORNL Neutrino Sources for Future Experiments''},
\href{https://www.snowmass21.org/docs/files/summaries/UF/SNOWMASS21-UF4_UF3-NF5_NF6-CF1_CF0-IF3_IF0-CompF2_CompF3-CommF5_CommF0-006.pdf}{``Advanced Germanium Detectors and Technologies for Underground Physics''},
\href{https://www.snowmass21.org/docs/files/summaries/NF/SNOWMASS21-NF5_NF3-022.pdf}{``Investigation of Neutrino Properties with Global Analysis of CEvNS Data''}.
White paper:
``Coherent elastic neutrino-nucleus scattering: Terrestrial and astrophysical applications''~\cite{Abdullah:2022zue}.}.
Decreasing the neutrino flux uncertainties~\footnote{See the NF06 discussion on neutrino flux uncertainties.}
will be essential for improving the sensitivity.

\begin{table}[t!]
\begin{minipage}{\textwidth}
\centering
\renewcommand{\arraystretch}{1.2}
\begin{tabular}{llcrcc}
Method & Experiment & Limit $[\mu_{\text{B}}]$ & CL & Year& Ref.\\
\hline
\multirow{5}{*}{Reactor $\bar\nu_e \, e^-$}
&Krasnoyarsk		&$\mu_{\nu_e} < 2.4 \times 10^{-10}$	&90\%	&1992&\cite{Vidyakin:1992nf}\\
&Rovno			&$\mu_{\nu_e} < 1.9 \times 10^{-10}$	&95\%	&1993&\cite{Derbin:1993wy}\\
&MUNU			&$\mu_{\nu_e} < 9   \times 10^{-11}$	&90\%	&2005&\cite{Daraktchieva:2005kn}\\
&TEXONO			&$\mu_{\nu_e} < 7.4 \times 10^{-11}$	&90\%	&2006&\cite{Wong:2006nx}\\
&GEMMA			&$\mu_{\nu_e} < 2.9 \times 10^{-11}$	&90\%	&2012&\cite{Beda:2012zz}\\
\hline
\multirow{1}{*}{Accelerator $\nu_e \, e^-$}
&LAMPF			&$\mu_{\nu_e} < 1.1 \times 10^{-9}$		&90\%	&1992&\cite{Allen:1992qe}\\
\hline
\multirow{3}{*}{Accelerator $(\nu_{\mu},\bar\nu_{\mu}) \, e^-$}
&BNL-E734		&$\mu_{\nu_{\mu}} < 8.5 \times 10^{-10}$	&90\%	&1990&\cite{Ahrens:1990fp}\\
&LAMPF			&$\mu_{\nu_{\mu}} < 7.4 \times 10^{-10}$	&90\%	&1992&\cite{Allen:1992qe}\\
&LSND			&$\mu_{\nu_{\mu}} < 6.8 \times 10^{-10}$	&90\%	&2001&\cite{Auerbach:2001wg}\\
\hline
Beam Dump
&BEBC~\cite{BEBCWA66:1986err}	&$\mu_{\nu_{\tau}} < 5.4 \times 10^{-7}$	&90\%	&1991&\cite{Cooper-Sarkar:1991vsl}\\
\hline
Accelerator $(\nu_{\tau},\bar\nu_{\tau}) \, e^-$
&DONUT			&$\mu_{\nu_{\tau}} < 3.9 \times 10^{-7}$	&90\%	&2001&\cite{Schwienhorst:2001sj}\\
\hline
\multirow{3}{*}{CEvNS}
&
\begin{tabular}{c}
COHERENT
\\[-0.1cm]
\cite{Akimov:2017ade,Akimov:2020pdx}
\end{tabular}
&
$
\begin{array}{c} \displaystyle
\mu_{\nu_e} < 44 \times 10^{-10}
\\ \displaystyle
\mu_{\nu_{\mu}} < 34 \times 10^{-10}
\end{array}
$
& 95\%
& 2020
& \cite{Cadeddu:2020lky}
\\[-0.3cm]
&
\multicolumn{5}{c}{\rule{0.8\linewidth}{0.5pt}}
\\[-0.2cm]
& CONUS
& $\mu_{\nu_e} < 7.5 \times 10^{-11}$
& 90\%
& 2022
& \cite{CONUS:2022qbb}
\\
& Dresden-II~\cite{Colaresi:2022obx}
& $\mu_{\nu_e} \lesssim 3 \times 10^{-10}$
& 90\%
& 2022
& \cite{Coloma:2022avw}
\\
\hline
\multirow{4}{*}{Solar $\nu_e \, e^-$}
&Super-Kamiokande	&$\mu_{\text{S}}^{(E_{\nu} \gtrsim 5 \, \text{MeV})} < 1.1 \times 10^{-10}$	&90\%	&2004&\cite{Liu:2004ny}
\\
&Borexino		&$\mu_{\text{S}}^{(E_{\nu} \lesssim 1 \, \text{MeV})} < 2.8 \times 10^{-11}$	&90\%	&2017&\cite{Borexino:2017fbd}
\\
&XMASS-I		&$\mu_{\text{S}}^{(E_{\nu} \lesssim 1 \, \text{MeV})} < 1.8 \times 10^{-10}$	&90\%	&2020&\cite{XMASS:2020zke}
\\
&PandaX-II		&$\mu_{\text{S}}^{(E_{\nu} \lesssim 1 \, \text{MeV})} < 4.9 \times 10^{-11}$	&90\%	&2020&\cite{PandaX-II:2020udv}
\\
\hline
\end{tabular}
\end{minipage}
\caption{ \label{tab:mag}
Experimental limits for the neutrino magnetic moments.
}
\end{table}

\subsection{Magnetic Moment}
\label{sub:Magnetic}

The magnetic and electric dipole moments
are the most studied neutrino electromagnetic properties,
because they can be generated in many models with massive neutrinos
by effective dimension-5 operators involving left-handed and right-handed neutrinos.
In the simplest extension of the Standard Model with three massive Dirac neutrinos,
the diagonal magnetic moments
are given by~\cite{Fujikawa:1980yx,Pal:1981rm,Shrock:1982sc}
\begin{equation}
\mu^{\text{Dirac}}_{\nu_{kk}}
\simeq
3 \times 10^{-19}
\left( \frac{m_{\nu_{k}}}{\text{eV}} \right) \mu_{\text{B}}
,
\label{Dirac_mu}
\end{equation}
where $m_{\nu_{k}}$ are the neutrino masses for $k=1,2,3$
and
$\mu_{\text{B}} \equiv e / 2 m_{e}$ is the Bohr magneton
($e$ is the elementary electric charge and $m_{e}$ is the electron mass).
Therefore,
in the simplest Dirac extension of the Standard Model
the diagonal magnetic moments are strongly suppressed
by the smallness of neutrino masses and are many orders of magnitude
smaller than the current experimental bounds listed in Table~\ref{tab:chr}.
The transition magnetic moments are further suppressed by about
four orders of magnitude
(see the review in Ref.~\cite{Giunti:2014ixa}).

Majorana neutrinos can have only transition magnetic moments
that are as suppressed as the Dirac transition magnetic moments
in the minimal extension of the SM with Majorana neutrino masses.
However,
in more elaborate models
the Majorana magnetic moments can be enhanced by several orders of magnitude,
leading to values that are not too far from
the current experimental bounds in Table~\ref{tab:chr}.
Therefore,
it is important to develop new experiments with
improved sensitivity to the neutrino magnetic moments.

\begin{figure}[t!]
\centering
\subfigure[]{\label{fig:Balantekin-1312-6858-f2}
\boxed{
\includegraphics*[width=0.45\linewidth, viewport=60 104 2249 1719]{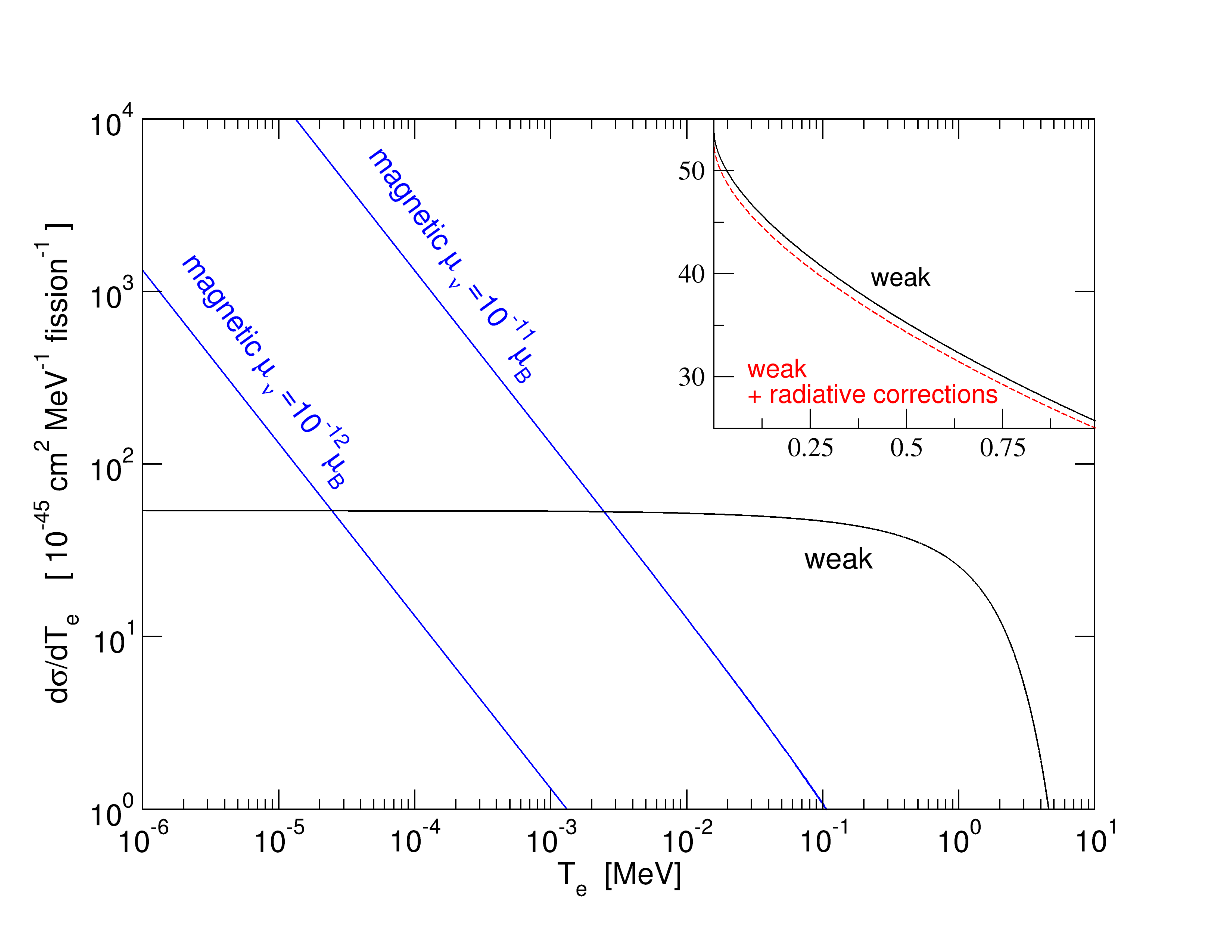}
}
}
\subfigure[]{\label{fig:Raffelt-hep-ph-9903472-f11}
\boxed{
\includegraphics*[width=0.45\linewidth]{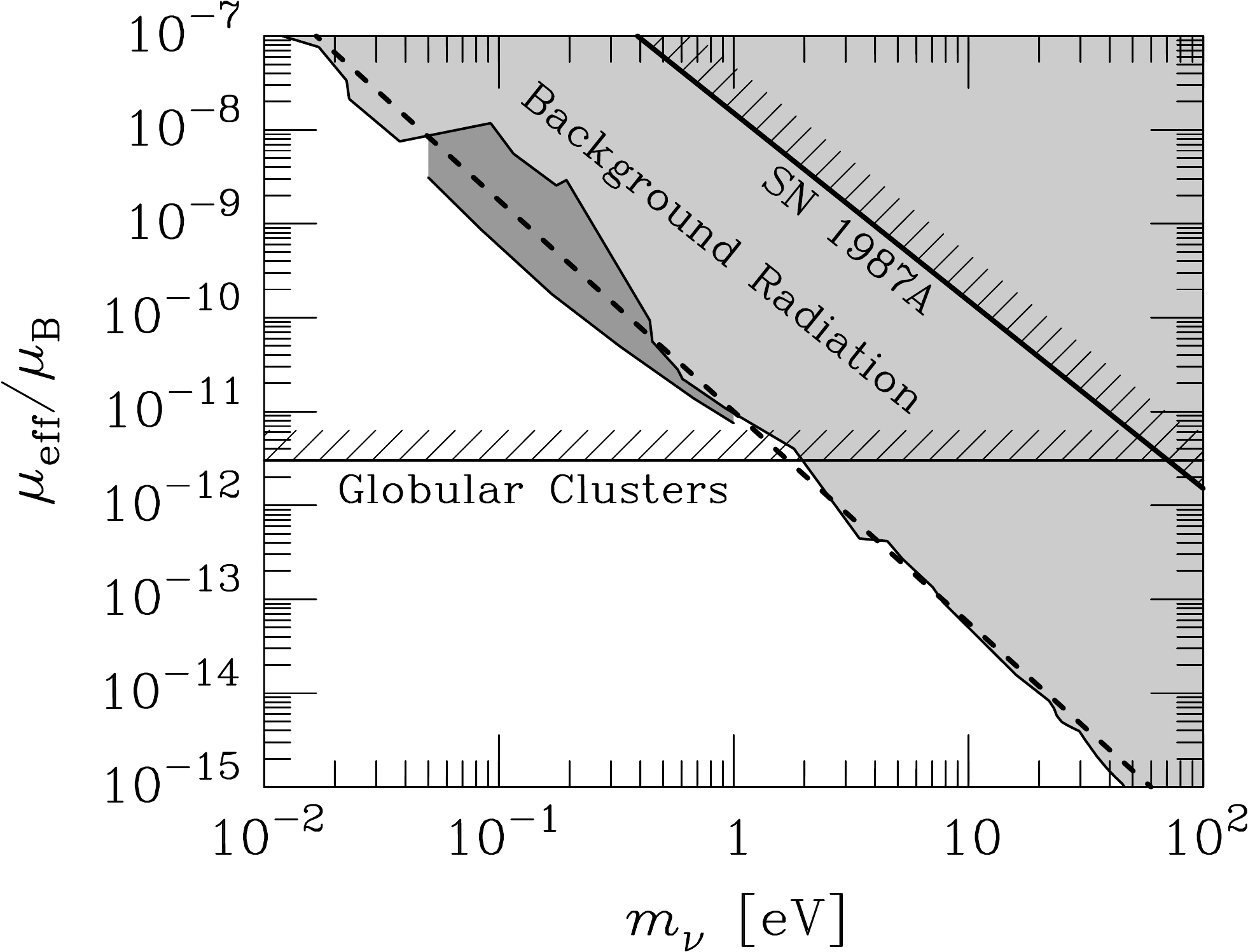}
}
}
\caption{ \label{fig:mag}
\subref{fig:Balantekin-1312-6858-f2}
Comparison of the SM weak-interaction cross section and
the magnetic moment cross section
in neutrino-electron elastic scattering~\cite{Balantekin:2013sda}.
\subref{fig:Raffelt-hep-ph-9903472-f11}
Astrophysical limits on the effective neutrino magnetic moment~\cite{Raffelt:1999tx}.
}
\end{figure}

The experimental determination of the magnetic moment
in laboratory experiments is made through the observation of
neutrino-electron elastic scattering or CEvNS~\footref{fn:CEvNS},
with the cross section
\begin{equation}
\left(\frac{d\sigma_{\nu_{\ell}X}}{dT_{X}}\right)_{\text{mag}}
=
\frac{\pi\alpha^{2}}{m_{e}^{2}}
\left(\frac{1}{T_{X}}-\frac{1}{E_{\nu}}\right)
Q_{X}
\left(\frac{\mu_{\nu_{\ell}}}{\mu_{\text{B}}}\right)^{2}
,
\label{magcs}
\end{equation}
where $T_{X}$ and $Q_{X}$ are the recoil kinetic energy and electric charge
of the target $X$
($X=e$ for neutrino-electron elastic scattering
and
$X=\mathcal{N}$ for CEvNS with the target nucleus
$\mathcal{N}$)~\footnote{The electron mass in Eq.~\eqref{magcs}
is due to the definition of the Bohr magneton,
independently of the target.}.
The effective magnetic moment of the flavor neutrino $\nu_{\ell}$
is given by
\begin{equation}
\mu_{\nu_{\ell}}^{2}
=
\sum_{k,j}
U_{\ell k} \, \mu^2_{kj} \, U_{\ell j}^{*}
.
\label{effmu}
\end{equation}
The effects of the magnetic moments can be probed
by measuring neutrino-electron elastic scattering or CEvNS
at very low values of the recoil kinetic energy $T_{X}$
of the target, where the cross section \eqref{magcs}
can increase above the SM cross section,
as illustrated in Fig.~\ref{fig:Balantekin-1312-6858-f2}
for neutrino-electron elastic scattering.
Table~\ref{tab:chr} show the current experimental bounds
obtained with this method in laboratory experiments.
Figure~\ref{fig:Raffelt-hep-ph-9903472-f11}
show the approximate bounds on the effective neutrino magnetic moment
obtained from astrophysical constraints on neutrino radiative decay
(see Section~\ref{sub:Lifetime}).

The GEMMA experiment~\cite{Beda:2012zz} obtained the most stringent limit on $\mu_{\nu_{e}}$ in Table~\ref{tab:mag}
by observing neutrino-electron elastic scattering
at the Kalinin Nuclear Power Plant in Russia,
using a 1.5 kg high-purity germanium (HPGe) detector with an energy threshold of 2.8~keV.
The GEMMA collaboration is preparing a new experiment (GEMMA~II)~\cite{Alekseev:2017yla}
with four point-contact HPGe detectors of 400~g each and an energy threshold of about 300 eV
placed below a 3.1~GW reactor at a distance from the core center of 10.7 m,
where the $\bar\nu_{e}$ flux is $5.2 \times 10^{13}~\text{cm}^{-2}~\text{s}^{-1}$.
The expected sensitivity for $\mu_{\nu_{e}}$ is about $1 \times 10^{-11}~\mu_{\text{B}}$
in three years of operation.

The TEXONO experiment~\cite{Wong:2006nx} obtained the limit on $\mu_{\nu_{e}}$ in Table~\ref{tab:mag}
by observing neutrino-electron elastic scattering
at the Kuo-Sheng nuclear power station
with a 1.06~kg HPGe detector at a
distance of 28~m from a 2.9~GW reactor core.
Future searches of $\mu_{\nu_{e}}$ are planned in the TEXONO research program~\footnote{LOI: \href{https://www.snowmass21.org/docs/files/summaries/NF/SNOWMASS21-NF5_NF6_Henry_Wong-057.pdf}{``Neutrino Physics at the Kuo-Sheng Reactor Laboratory with the TEXONO
Research Program -- Highlights of Status and Plans''}.}
at the Kuo-Sheng Reactor Neutrino Laboratory (KSNL) in Taiwan
by observing neutrino-electron elastic scattering
using HPGe detectors with an energy threshold of the order of 200~eV
and a very low background~\cite{Singh:2017bns}.

Other reactor neutrino experiments can probe $\mu_{\nu_{e}}$
with neutrino-electron elastic scattering or CEvNS~\footnote{White paper:
HEP Physics Opportunities Using Reactor
Antineutrinos~\cite{Akindele:2022sti}.}\textsuperscript{,}\footref{fn:CEvNS}.
Both $\mu_{\nu_{e}}$ and $\mu_{\nu_{\mu}}$
will be probed in CEvNS experiments
with electron and muon neutrinos
produced by pion and muon decays at rest~\footref{fn:CEvNS},
as those planned at the Spallation Neutron Source (SNS) at Oak Ridge National Laboratory~\cite{Barbeau:2021exu}
and at the European Spallation Source (ESS)~\cite{Baxter:2019mcx}.

The electron and muon neutrino magnetic moments can be probed at the Fermilab
ICARUS, SBND and DUNE~\cite{Mathur:2021trm} detectors,
which will be exposed to intense $\nu_{\mu}$ beams~\footnote{LOI: \href{https://www.snowmass21.org/docs/files/summaries/NF/SNOWMASS21-NF3_NF5-162.pdf}{``Search for Muon Neutrino Magnetic Moment at Future High Intensity Muon Neutrino Beam Experiments''}.}
and subdominant $\nu_{e}$ beams.
All the magnetic moments can be probed at the planned Forward Physics Facility at LHC~\cite{Anchordoqui:2021ghd}~\footnote{White paper:
``The Forward Physics Facility at the High-Luminosity LHC''~\cite{Feng:2022inv}.},
with special relevance for $\mu_{\nu_{\tau}}$,
that can be probed at the so far unreached level of a few $10^{-8}~\mu_{\text{B}}$.
It is also planned to search for transition magnetic moments connecting the standard active left-handed neutrinos
to sterile neutrinos beyond the SM (the so-called ``dipole portal'';
see, e.g., Ref.~\cite{Ismail:2021dyp})~\footnote{White papers:
White Paper on Light Sterile Neutrino Searches and Related Phenomenology~\cite{Acero:2022wqg},
Tau Neutrinos in the Next Decade: from GeV to EeV~\cite{Abraham:2022jse},
The Present and Future Status of Heavy Neutral Leptons~\cite{Abdullahi:2022jlv}.}.

The effects of neutrino magnetic moments are also probed with measurements of the cross section of solar neutrinos,
as done by the
Super-Kamiokande~\cite{Liu:2004ny},
Borexino~\cite{Borexino:2017fbd},
XMASS~\cite{XMASS:2020zke}, and
PandaX~\cite{PandaX-II:2020udv}
collaborations
(see Table~\ref{tab:mag}).
In this case the effective magnetic moment $\mu_{\text{S}}$ is determined by the oscillations of solar neutrinos:
\begin{equation}
(\mu_{\text{S}}^{(E_{\nu})})^2
=
\sum_{k,j=1}^{3}
P_{\nu_{e}\to\nu_{k}}^{\text{S}}(E_{\nu})
\,
|\mu_{kj}|^2
.
\label{mueffsolar}
\end{equation}
There is a recent indication of
a possible non-zero effective magnetic moment $\mu_{\text{S}}$
that can explain the excess of electronic recoil events in
XENON1T~\cite{XENON:2020rca}:
\begin{equation}
\mu_{\text{S}}^{(E_{\nu} \lesssim 1 \, \text{MeV})}
\in
\left( 1.4 , \, 2.9 \right) \times 10^{-11} \, \mu_{\text{B}}
\quad
\text{(90\% CL)}
.
\label{XENON1T}
\end{equation}
This indication can be investigated with
new experiments sensitive to solar neutrinos,
e.g,
the Jinping neutrino experiment~\cite{Yue:2021vjg},
JUNO~\cite{Ye:2021zso},
LUX-ZEPLIN~\cite{LZ:2021xov}
and
DARWIN~\footnote{LOI: \href{https://www.snowmass21.org/docs/files/summaries/NF/SNOWMASS21-NF5_NF4-084.pdf}{``Neutrino physics with the DARWIN Observatory''}.}.

\begin{table}[t!]
\begin{minipage}{\textwidth}
\centering
\renewcommand{\arraystretch}{1.2}
\begin{tabular}{llcrcc}
Method & Experiment & Limit $[e]$ & CL & Year & Ref.\\
\hline
Neutrality of matter
& Bressi et al.~\cite{Bressi:2011yfa}
& $Q_{\nu_{e}} \in (-3.8,2.6) \times 10^{-21}$
& 68\%
& 2014
& \cite{Giunti:2014ixa}
\\
\hline
\multirow{3}{*}{Reactor $\bar\nu_e \, e^-$}
& TEXONO~\cite{TEXONO:2002pra}
& $|Q_{\nu_{e}}|< 3.7 \times 10^{-12}$
& 90\%
& 2006
& \cite{Gninenko:2006fi}
\\
& GEMMA~\cite{Beda:2012zz}
& $|Q_{\nu_{e}}|< 1.5 \times 10^{-12}$
& 90\%
& 2013
& \cite{Studenikin:2013my}
\\
& TEXONO
& $|Q_{\nu_{e}}|< 1.0 \times 10^{-12}$
& 90\%
& 2014
& \cite{Chen:2014dsa}
\\
\hline
Accelerator $(\nu_{\mu},\bar\nu_{\mu}) \, e^-$
& LSND~\cite{Auerbach:2001wg}
& $|Q_{\nu_{\mu}}|< 3 \times 10^{-9}$
& 90\%
& 2020
& \cite{Das:2020egb}
\\
\hline
Beam Dump
& BEBC~\cite{BEBCWA66:1986err}
& $|Q_{\nu_{\tau}}|< 4 \times 10^{-4}$
& 90\%
& 1993
& \cite{Babu:1993yh}
\\
\hline
Accelerator $(\nu_{\tau},\bar\nu_{\tau}) \, e^-$
& DONUT~\cite{Schwienhorst:2001sj}
& $|Q_{\nu_{\tau}}|< 4 \times 10^{-6}$
& 90\%
& 2020
& \cite{Das:2020egb}
\\
\hline
\multirow{2}{*}{CEvNS}
&
\begin{tabular}{c}
COHERENT
\\[-0.1cm]
\cite{Akimov:2017ade,Akimov:2020pdx}
\end{tabular}
&
$
\begin{array}{c} \displaystyle
Q_{\nu_e} \in (-14,34) \times 10^{-8}
\\ \displaystyle
Q_{\nu_e} \in (-10,12) \times 10^{-8}
\\ \displaystyle
|Q_{\nu_{e\mu}}| < 17 \times 10^{-8}
\\ \displaystyle
|Q_{\nu_{e\tau}}| < 27 \times 10^{-8}
\\ \displaystyle
|Q_{\nu_{\mu\tau}}| < 20 \times 10^{-8}
\end{array}
$
&95\%	&2020&
\cite{Cadeddu:2020lky}
\\[-0.3cm]
&
\multicolumn{5}{c}{\rule{0.8\linewidth}{0.5pt}}
\\[-0.2cm]
& CONUS
& $|Q_{\nu_{e}}| < 3.3 \times 10^{-12}$
& 90\%
& 2022
& \cite{CONUS:2022qbb}
\\
\hline
\multirow{1}{*}{Solar $\nu_e \, e^-$}
& XMASS-I
&
$\begin{array}{c} 
\displaystyle |Q_{\nu_{e}}| < 7.3 \times 10^{-12}\\
\displaystyle |Q_{\nu_{\mu}}|,|Q_{\nu_{\tau}}| < 1.1 \times 10^{-11}\\
\end{array}$
& 90\%
& 2020
& \cite{XMASS:2020zke}
\\
\hline
\end{tabular}
\end{minipage}
\caption{ \label{tab:ech}
Experimental limits for the neutrino electric charges.
}
\end{table}
\subsection{Electric Charge}
\label{sub:Charge}

In theories beyond the SM,
neutrinos can have very small electric charges (millicharges) 
(see the reviews in Refs.~\cite{Giunti:2014ixa,Das:2020egb}).
The relation between the effective electric charges of flavor neutrinos
and the electric charges of massive neutrinos is similar to that for the charge radii in Eq.~\eqref{chrflavor}~\cite{Kouzakov:2017hbc}:
\begin{equation}
Q_{\nu_{\ell'\ell}}
=
\sum_{k,j}
U_{\ell' k}
Q_{\nu_{kj}}
U_{\ell j}^{*}
.
\label{echflavor}
\end{equation}
The hermiticity of the matrix of electric charges
is preserved by the unitary transformation.
For simplicity we denote
the real diagonal electric charges of the flavor neutrinos as
$Q_{\nu_{\ell}} \equiv Q_{\nu_{\ell\ell}}$.

As shown in Table~\ref{tab:ech},
the strongest bound on a neutrino electric charge is
$|Q_{\nu_{e}}| \lesssim 3 \times 10^{-21}~e$
obtained from the neutrality of matter~\cite{Raffelt:1999gv,Giunti:2014ixa}.
Weaker bounds on the electric charges of the flavor neutrinos
have been obtained with the scattering experiments listed in Table~\ref{tab:ech}.

The electric charges of neutrinos can be probed in future CEvNS experiments~\footref{fn:CEvNS},
in reactor experiments,
in accelerator experiments~\footnote{LOI: \href{https://www.snowmass21.org/docs/files/summaries/RF/SNOWMASS21-EF9_EF10_NF3_NF5_CF1_CF3_CF7_TF7_TF8_TF9_AF5_UF3_Yu-Dai_Tsai-114.pdf}{``Accelerator Probes of Millicharged Particles and Dark Matter''}.}
(e.g. DUNE~\cite{Mathur:2021trm}),
and in experiments sensitive to solar neutrinos
(e.g. LUX-ZEPLIN~\cite{LZ:2021xov}).
\newpage

\section{Other Neutrino Properties}
\label{sec:Other}

Neutrinos can have several properties beyond those predicted by the SM,
that are not discussed in the previous Sections and in other Topical Groups.
In the following Subsections we discuss briefly the neutrino lifetime
and the possibilities of testing with neutrinos Lorentz and CPT violation
and the violation of the equivalence principle.

\subsection{Neutrino Lifetime}
\label{sub:Lifetime}

In theories beyond the SM,
since neutrinos are massive and the neutrino masses are different,
heavier neutrinos can decay into lighter neutrinos
with emission of a massless or very light vector or scalar boson.

Neutrino magnetic moments induce the radiative decay
$\nu_{i}\to\nu_{f}+\gamma$
with the decay rate~\cite{Shrock:1974nd,Marciano:1977wx,Lee:1977tib,Petcov:1976ff,Goldman:1977jx,Zatsepin:1978iy,Pal:1981rm,Shrock:1982sc}
\begin{equation}
\Gamma_{\nu_{i}\to\nu_{f}+\gamma}
=
\frac{1}{8\pi}
\left(
\frac{m_{i}^{2}-m_{f}^{2}}{m_{i}}
\right)^3
|\mu_{fi}|^2
.
\label{E010}
\end{equation}
This expression is valid for both Dirac and Majorana neutrinos,
because both can have transition magnetic moments.
In the simplest extensions of the Standard Model with massive Dirac
or Majorana neutrinos
the neutrino lifetime is about 26 orders of magnitude
larger than the age of the Universe
(see Ref.~\cite{Giunti:2014ixa}).
However, the neutrino lifetime can be shorter in
more elaborate theories.
Figure~\ref{fig:Raffelt-hep-ph-9903472-f11}
show the approximate bounds on the effective neutrino magnetic moment
obtained from the absence of decay photons
in astrophysical observations~\cite{Raffelt:1999tx}.
Future improvements may be obtained by searching for the decay photons
of the cosmic neutrino background~\cite{Bernal:2021ylz}. Cosmological constraints are discussed in Ref.\footnote{White Paper: Synergy between cosmological and laboratory searches in neutrino physics~\cite{Abazajian:2022ofy}}.

Other decay channels can be either
visible or invisible,
depending on the possibility to observe the decay products,
and can have many phenomenological consequences~\footnote{\label{fn:BSMWP}
White paper:
Beyond the Standard Model effects on Neutrino Flavor~\cite{Arguelles:2022xxa}.}
(see, e.g. the recent Refs.~\cite{Choubey:2020dhw,Chakraborty:2020cfu,Delgado:2021vha,Akita:2021hqn,Picoreti:2021yct}).

\subsection{Lorentz and CPT violation}
\label{sub:CPT}

Small violations of the Lorentz symmetry are possible in some
theories beyond the SM.
Since the CPT theorem is based on Lorentz invariance,
Lorentz and CPT violations are connected~\cite{Greenberg:2002uu}.
A useful framework to study Lorentz and CPT violations
is the Lorentz- and CPT-violating Standard-Model Extension
(SME)~\cite{Colladay:1996iz,Colladay:1998fq,Kostelecky:2008ts,Kostelecky:2011gq}.
Neutrinos can probe Lorentz and CPT violations in many different ways
(see the reviews in Refs.~\cite{Arguelles:2021kjg,Antonelli:2020nhn,Stecker:2022tzd})
\footnote{LOI:
\href{https://www.snowmass21.org/docs/files/summaries/NF/SNOWMASS21-NF3_NF5-155.pdf}{``Neutrinos as Probes for Lorentz and CPT Symmetry''}.}\textsuperscript{,}\footref{fn:BSMWP}.
Note the possibility to test Lorentz violations in $\beta$-decay experiments as KATRIN~\cite{Lehnert:2021tbv}~\footref{WP:KATRIN}.

\subsection{Violation of the equivalence principle}
\label{sub:VEP}

Neutrino oscillations are sensitive to violations of the equivalence principle
(VEP)~\cite{Gasperini:1988zf,Halprin:1995vg}~\footref{fn:BSMWP}.
For neutrinos propagating in a gravitational potential
VEP induce energy shifts between neutrino gravitational eigenstates,
which are connected to the flavor neutrinos by an unknown unitary transformation.
The differences between the energy shifts induce
energy-dependent deviations from standard oscillations
that can be probed in current and future experiments such as 
IceCube~\cite{Fiorillo:2020gsb,Esmaili:2021omm},
KM3NeT~\cite{Chianese:2021vkf},
DUNE~\cite{Diaz:2020aax}, and
solar neutrino experiments~\cite{Valdiviesso:2008vyk}.

\section{Summary and Conclusions}
\label{sec:Summary}

Direct experiments to probe the absolute mass scale of the neutrino rely on high-precision measurements of the endpoint regions of tritium $\upbeta$-decay spectra or ${}^{163}$Ho electron-capture spectra. These kinematic searches provide complementary information to more model-dependent probes based on supernovae, cosmological data, and \ndbd (Sec.~\ref{sub:OtherMass}). The KATRIN experiment has set the best direct bound on the effective neutrino-mass scale, $m_\beta < 0.8$~eV (90\% C.L.), and is continuing to take tritium-decay data while working to reduce backgrounds (Sec.~\ref{ssb:KATRIN}). Project 8 has developed a new method for tritium spectral measurement -- cyclotron radiation emission spectroscopy -- and is investigating scaling strategies and the development of an atomic source (Sec.~\ref{ssb:Project8}). Meanwhile, both the ECHo and HOLMES collaborations have made substantial progress in optimizing ${}^{163}$Ho production and developing microcalorimeters and multiplexing schemes, and ECHo is analyzing a high-statistics spectrum. Continuing improvement of the theoretical spectrum calculation will unlock high sensitivity in ${}^{163}$Ho (Sec.~\ref{sub:Capture}).

Neutrinoless double beta decay is the only truly sensitive known way to test for the Majorana nature of the neutrino, a property that, if established, would have profound consequences for particle physics and cosmology. A generation of 10-100~kg-scale detectors have proven a suite of technologies that hold promise for achieving the low backgrounds needed for ton-scale experiments (Sec~\ref{sub:bbcurton}), of order one count per ton per year in the energy ROI.  A ton-scale program employing at least one but hopefully several isotopes is being shepherded by the DOE Office of Nuclear Physics and will be a major focus for the field during the coming Snowmass period. These experiments will press beyond the existing leading limit from the Kamland-Zen experiment (2.3$\times$10$^{26}$~yr) further into the region of parameter space allowed under the inverted neutrino mass ordering, given the standard light Majorana neutrino exchange mechanism. If the neutrino masses prove to be normal ordered, still larger and lower background experiments will be required (Sec~\ref{sub:bblarge}).  Development of technologies that can meet the implied extremely challenging technological requirements will require continued effort during the coming Snowmass period if the necessary R\&D is to converge in a timely way.  

Electromagnetic neutrino interactions are described, at low energies,
by neutrino charge radii, magnetic moments and electric charges.
Even in the Standard Model,
where neutrinos are neutral and massless,
radiative corrections induce electromagnetic neutrino interactions
described by the charge radii.
Since
the current experimental limits on the neutrino charge radii are
only about one order of magnitude larger than the Standard Model prediction,
it is worthwhile to try to
improve the experimental sensitivity to the charge radii
with the aim of discovering neutrino electromagnetic interactions
and testing the Standard Model.
Other promising searches of physics beyond the Standard Model
can be pursued with searches of
neutrino electromagnetic interactions
induced by magnetic moments and electric charges,
visible or invisible neutrino decays,
effects of Lorentz and CPT violation,
and effects of a violation of the equivalence principle.

\paragraph{Acknowledgments}
The authors relied on numerous white papers submitted to the Snowmass process, which we have highlighted in footnotes and cited in the bibliography. We gratefully acknowledge the insights and backgrounds provided by the participants and speakers at NF05-sponsored  workshops and sessions; in alphabetical order, these speakers and panelists (excluding NF05 conveners) were Zara Bagdasarian, Alexander Barabash, Giovanni Benato, Steve Biller, Vedran Brdar, Thomas Brunner, Eric Church, Christine Claessens, Zohreh Davoudi, Patrick Decowski, Andr\'e de Gouv\v{e}a, Wouter Dekens, Jason Detwiler, Eleanora Di Valentino, Angelo Dragone, Steven Elliott, Jonathan Engel, William Fairbank, Loredana Gastaldo, Christopher Grant, Roxanne Guenette, Jeter Hall, Steen Hannestad, Julia Harz, Maurits W. Haverkort, Karsten Heeger, Mike Heffner, Eric Hoppe, Huan Zhong Huang, Boris Kayser, Joshua Klein, Jenni Kotila, Massimiliano Lattanzi, Kyle Leach, Aobo Li, Jonathan Link, Valentina Lozza, Michele Maltoni, Yuan Mei, Javier Men\'endez, Emanuele Mereghetti, Susanne Mertens, Omar Miranda, Benjamin Monreal, David Moore, Angelo Nucciotti, Dimitrios Papoulias, Silvia Pascoli, Michael Ramsey-Musolf, Richard Saldanha, Danielle Speller, Alexander Studenikin, Pranava Teja Surukuchi, Robert Svoboda, Brent VanDevender, Matt Wetstein, and Henry Wong. We also gratefully acknowledge the community comments that helped to improve early drafts of this report, notably from Maury Goodman, Erin Hansen, Felix Kling, Lisa Koerner, Yury Kolomensky, Joel Ullom, and Joseph Zennamo.


\addcontentsline{toc}{section}{References}
\begin{thebibliography}{100}
\providecommand{\url}[1]{\texttt{#1}}
\providecommand{\urlprefix}{URL }

\bibitem{Super-Kamiokande:1998kpq}
Y.~Fukuda \emph{et~al.}, (Super-Kamiokande), \emph{{Evidence for oscillation of
  atmospheric neutrinos}}.
  \href{http://dx.doi.org/10.1103/PhysRevLett.81.1562}{\emph{Phys. Rev. Lett.}}
  \textbf{81}:~1562~(1998) , hep-ex/9807003

\bibitem{SNO:2002tuh}
Q.~R. Ahmad \emph{et~al.}, (SNO), \emph{{Direct evidence for neutrino flavor
  transformation from neutral current interactions in the Sudbury Neutrino
  Observatory}}.
  \href{http://dx.doi.org/10.1103/PhysRevLett.89.011301}{\emph{Phys. Rev.
  Lett.}} \textbf{89}:~011301~(2002) , nucl-ex/0204008

\bibitem{Formaggio:2021nfz}
J.~A. Formaggio, A.~L.~C. de~Gouv\^ea, and R.~G.~H. Robertson, \emph{{Direct
  Measurements of Neutrino Mass}}.
  \href{http://dx.doi.org/10.1016/j.physrep.2021.02.002}{\emph{Phys. Rept.}}
  \textbf{914}:~1~(2021) , 2102.00594

\bibitem{Galeazzi:2001ih}
M.~Galeazzi, F.~Fontanelli, F.~Gatti, \emph{et~al.}, \emph{{End-point energy
  and half-life of the Re-187 beta decay}}.
  \href{http://dx.doi.org/10.1103/PhysRevC.63.014302}{\emph{Phys. Rev. C}}
  \textbf{63}:~014302~(2001)

\bibitem{Gatti:2001ty}
F.~Gatti, \emph{{Microcalorimeter measurements}}.
  \href{http://dx.doi.org/10.1016/S0920-5632(00)00954-3}{\emph{Nucl. Phys. B
  Proc. Suppl.}} \textbf{91}:~293~(2001)

\bibitem{sisti2004new}
M.~Sisti \emph{et~al.}, \emph{New limits from the milano neutrino mass
  experiment with thermal microcalorimeters}.
  \href{http://dx.doi.org/10.1016/j.nima.2003.11.273}{\emph{Nucl. Instrum.
  Meth. A}} \textbf{520}:~125~(2004)

\bibitem{Nucciotti:2015rsl}
A.~Nucciotti, \emph{{The use of low temperature detectors for direct
  measurements of the mass of the electron neutrino}}.
  \href{http://dx.doi.org/10.1155/2016/9153024}{\emph{Adv. High Energy Phys.}}
  \textbf{2016}:~9153024~(2016) , 1511.00968

\bibitem{Andreotti:2011zza}
E.~Andreotti, M.~Hult, R.~Gonzalez~de Orduna, \emph{et~al.}, \emph{{Half-life
  of the beta decay In-115 (9/2+) --\ensuremath{>} Sn-115 (3/2+)}}.
  \href{http://dx.doi.org/10.1103/PhysRevC.84.044605}{\emph{Phys. Rev. C}}
  \textbf{84}:~044605~(2011)

\bibitem{Urban:2016duq}
W.~Urban \emph{et~al.}, \emph{{Precise measurement of energies in Sn115
  following the (n,\ensuremath{\gamma}) reaction}}.
  \href{http://dx.doi.org/10.1103/PhysRevC.94.011302}{\emph{Phys. Rev. C}}
  \textbf{94}~(1):~011302~(2016)

\bibitem{deRoubin:2020eol}
A.~de~Roubin \emph{et~al.}, \emph{{High-Precision $Q$-Value Measurement
  Confirms the Potential of $^{135}$Cs for Absolute Antineutrino Mass Scale
  Determination}}.
  \href{http://dx.doi.org/10.1103/PhysRevLett.124.222503}{\emph{Phys. Rev.
  Lett.}} \textbf{124}~(22):~222503~(2020) , 2002.08282

\bibitem{Ramalho:2022fdf}
M.~Ramalho, Z.~Ge, T.~Eronen, \emph{et~al.}, \emph{Observation of an
  ultralow-$q$-value electron-capture channel decaying to $^{75}\mathrm{As}$
  via a high-precision mass measurement}.
  \href{http://dx.doi.org/10.1103/PhysRevC.106.015501}{\emph{Phys. Rev. C}}
  \textbf{106}:~015501~(2022)

\bibitem{Kopp:2009yp}
J.~Kopp and A.~Merle, \emph{{Ultra-low Q values for neutrino mass
  measurements}}.
  \href{http://dx.doi.org/10.1103/PhysRevC.81.045501}{\emph{Phys. Rev. C}}
  \textbf{81}:~045501~(2010) , 0911.3329

\bibitem{Gamage:2019xvx}
N.~D. Gamage, R.~Bhandari, M.~Horana~Gamage, \emph{et~al.},
  \emph{{Identification and investigation of possible ultra-low $Q$ value
  $\beta$ decay candidates}}.
  \href{http://dx.doi.org/10.1007/s10751-019-1588-5}{\emph{Hyperfine
  Interact.}} \textbf{240}~(1):~43~(2019)

\bibitem{Keblbeck:2022twm}
D.~K. Keblbeck, R.~Bhandari, N.~D. Gamage, \emph{et~al.}, \emph{{Updated
  evaluation of potential ultra-low Q value $\beta$-decay candidates}}
  2201.08790

\bibitem{Sandler:2019iws}
R.~Sandler, G.~Bollen, N.~D. Gamage, \emph{et~al.}, \emph{{Investigation of
  potential ultra-low $Q$-value $\beta$-decay candidates $^{89}$Sr and
  $^{139}$Ba using Penning trap mass spectrometry}}.
  \href{http://dx.doi.org/10.1103/PhysRevC.100.024309}{\emph{Phys. Rev. C}}
  \textbf{100}~(2):~024309~(2019) , 1906.03335

\bibitem{Ge:2022}
Z.~Ge, T.~Eronen, A.~de~Roubin, \emph{et~al.}, \emph{Direct determination of
  the atomic mass difference of the pairs
  $^{76}\mathrm{As}\text{\ensuremath{-}}^{76}\mathrm{Se}$ and
  $^{155}\mathrm{Tb}\text{\ensuremath{-}}^{155}\mathrm{Gd}$ rules out
  $^{76}\mathrm{As}$ and $^{155}\mathrm{Tb}$ as possible candidates for
  electron (anti)neutrino mass measurements}.
  \href{http://dx.doi.org/10.1103/PhysRevC.106.015502}{\emph{Phys. Rev. C}}
  \textbf{106}:~015502~(2022)

\bibitem{Gamage:2022thf}
N.~D. Gamage \emph{et~al.}, \emph{{Precise Q value measurements of
  $^{112,113}$Ag and $^{115}$Cd with the Canadian Penning trap for evaluation
  of potential ultra-low Q value $\beta$-decays}} 2202.12874

\bibitem{Robertson:1988xca}
R.~G.~H. Robertson and D.~A. Knapp, \emph{{Direct Measurements of Neutrino
  Mass}}.
  \href{http://dx.doi.org/10.1146/annurev.ns.38.120188.001153}{\emph{Ann. Rev.
  Nucl. Part. Sci.}} \textbf{38}:~185~(1988)

\bibitem{AshtariEsfahani:2020bfp}
A.~Ashtari~Esfahani \emph{et~al.}, \emph{{Bayesian analysis of a future $\beta$
  decay experiment's sensitivity to neutrino mass scale and ordering}}.
  \href{http://dx.doi.org/10.1103/PhysRevC.103.065501}{\emph{Phys. Rev. C}}
  \textbf{103}~(6):~065501~(2021) , 2012.14341

\bibitem{Robertson:1991vn}
R.~G.~H. Robertson, T.~J. Bowles, G.~J. Stephenson, \emph{et~al.}, \emph{{Limit
  on anti-electron-neutrino mass from observation of the beta decay of
  molecular tritium}}.
  \href{http://dx.doi.org/10.1103/PhysRevLett.67.957}{\emph{Phys. Rev. Lett.}}
  \textbf{67}:~957~(1991)

\bibitem{Stoeffl:1995wm}
W.~Stoeffl and D.~J. Decman, \emph{{Anomalous Structure in the Beta Decay of
  Gaseous Molecular Tritium}}.
  \href{http://dx.doi.org/10.1103/PhysRevLett.75.3237}{\emph{Phys. Rev. Lett.}}
  \textbf{75}:~3237~(1995)

\bibitem{Bodine:2015sma}
L.~I. Bodine, D.~S. Parno, and R.~G.~H. Robertson, \emph{{Assessment of
  molecular effects on neutrino mass measurements from tritium
  \ensuremath{\beta} decay}}.
  \href{http://dx.doi.org/10.1103/PhysRevC.91.035505}{\emph{Phys. Rev. C}}
  \textbf{91}~(3):~035505~(2015) , 1502.03497

\bibitem{Saenz:2000dul}
A.~Saenz, S.~Jonsell, and P.~Froelich, \emph{{Improved Molecular Final-State
  Distribution of HeT+ for the \ensuremath{\beta}-Decay Process of T2}}.
  \href{http://dx.doi.org/10.1103/PhysRevLett.84.242}{\emph{Phys. Rev. Lett.}}
  \textbf{84}~(2):~242~(2000)

\bibitem{Doss:2006zv}
N.~Doss, J.~Tennyson, A.~Saenz, \emph{et~al.}, \emph{{Molecular effects in
  investigations of tritium molecule beta decay endpoint experiments}}.
  \href{http://dx.doi.org/10.1103/PhysRevC.73.025502}{\emph{Phys. Rev. C}}
  \textbf{73}:~025502~(2006)

\bibitem{KATRIN:2021fgc}
M.~Aker \emph{et~al.}, (KATRIN), \emph{{Analysis methods for the first KATRIN
  neutrino-mass measurement}}.
  \href{http://dx.doi.org/10.1103/PhysRevD.104.012005}{\emph{Phys. Rev. D}}
  \textbf{104}~(1):~012005~(2021) , 2101.05253

\bibitem{TRIMS:2020nsv}
Y.~T. Lin \emph{et~al.}, (TRIMS), \emph{{Beta decay of molecular tritium}}.
  \href{http://dx.doi.org/10.1103/PhysRevLett.124.222502}{\emph{Phys. Rev.
  Lett.}} \textbf{124}~(22):~222502~(2020) , 2001.11671

\bibitem{KATRIN:2021uub}
M.~Aker \emph{et~al.}, (KATRIN), \emph{{Direct neutrino-mass measurement with
  sub-electronvolt sensitivity}}.
  \href{http://dx.doi.org/10.1038/s41567-021-01463-1}{\emph{Nature Phys.}}
  \textbf{18}~(2):~160~(2022) , 2105.08533

\bibitem{KATRIN:2021dfa}
M.~Aker \emph{et~al.}, (KATRIN), \emph{{The design, construction, and
  commissioning of the KATRIN experiment}}.
  \href{http://dx.doi.org/10.1088/1748-0221/16/08/T08015}{\emph{JINST}}
  \textbf{16}~(08):~T08015~(2021) , 2103.04755

\bibitem{Aker2019knm1}
M.~Aker \emph{et~al.}, (KATRIN), \emph{{Improved Upper Limit on the Neutrino
  Mass from a Direct Kinematic Method by KATRIN}}.
  \href{http://dx.doi.org/10.1103/PhysRevLett.123.221802}{\emph{Phys. Rev.
  Lett.}} \textbf{123}:~221802~(2019)

\bibitem{Aker2021Knm1}
M.~Aker \emph{et~al.}, (KATRIN), \emph{Analysis methods for the first {KATRIN}
  neutrino-mass measurement}.
  \href{http://dx.doi.org/10.1103/PhysRevD.104.012005}{\emph{Phys. Rev. D}}
  \textbf{104}:~012005~(2021)

\bibitem{aker2022katrin}
M.~Aker, M.~Balzer, D.~Batzler, \emph{et~al.}, \emph{Katrin: Status and
  prospects for the neutrino mass and beyond}. \emph{arXiv preprint
  arXiv:2203.08059}

\bibitem{KATRIN:2021rqj}
M.~Aker \emph{et~al.}, (KATRIN), \emph{{Precision measurement of the electron
  energy-loss function in tritium and deuterium gas for the KATRIN
  experiment}}.
  \href{http://dx.doi.org/10.1140/epjc/s10052-021-09325-z}{\emph{Eur. Phys. J.
  C}} \textbf{81}~(7):~579~(2021) , 2105.06930

\bibitem{Monreal:2009za}
B.~Monreal and J.~A. Formaggio, \emph{{Relativistic Cyclotron Radiation
  Detection of Tritium Decay Electrons as a New Technique for Measuring the
  Neutrino Mass}}.
  \href{http://dx.doi.org/10.1103/PhysRevD.80.051301}{\emph{Phys. Rev. D}}
  \textbf{80}:~051301~(2009) , 0904.2860

\bibitem{AshtariEsfahani:2019yva}
A.~Ashtari~Esfahani \emph{et~al.}, \emph{{Electron Radiated Power in Cyclotron
  Radiation Emission Spectroscopy Experiments}}.
  \href{http://dx.doi.org/10.1103/PhysRevC.99.055501}{\emph{Phys. Rev. C}}
  \textbf{99}~(5):~055501~(2019) , 1901.02844

\bibitem{Project8:2017nal}
A.~Ashtari~Esfahani \emph{et~al.}, (Project 8), \emph{{Determining the neutrino
  mass with cyclotron radiation emission spectroscopy\textemdash{}Project 8}}.
  \href{http://dx.doi.org/10.1088/1361-6471/aa5b4f}{\emph{J. Phys. G}}
  \textbf{44}~(5):~054004~(2017) , 1703.02037

\bibitem{P8:TAUP21}
E.~Novitski, (Project 8), \emph{Project 8: the path to improved neutrino mass}.
  Talk at TAUP 2021 (2021),
  \urlprefix\url{https://indico.ific.uv.es/event/6178/contributions/15484/}

\bibitem{esfahani2022project}
A.~Ashtari~Esfahani, S.~B{\"o}ser, N.~Buzinsky, \emph{et~al.}, \emph{The
  project 8 neutrino mass experiment}. \emph{arXiv preprint arXiv:2203.07349}

\bibitem{Croce:2015kwa}
M.~P. Croce \emph{et~al.}, \emph{{Development of holmium-163 electron-capture
  spectroscopy with transition-edge sensors}}.
  \href{http://dx.doi.org/10.1007/s10909-015-1451-2}{\emph{J. Low Temp. Phys.}}
  \textbf{184}~(3-4):~958~(2016) , 1510.03874

\bibitem{ECHo:2015qgh}
S.~Eliseev \emph{et~al.}, (ECHo), \emph{{Direct Measurement of the Mass
  Difference of $^{163}$Ho and $^{163}$Dy Solves the $Q$-Value Puzzle for the
  Neutrino Mass Determination}}.
  \href{http://dx.doi.org/10.1103/PhysRevLett.115.062501}{\emph{Phys. Rev.
  Lett.}} \textbf{115}~(6):~062501~(2015) , 1604.04210

\bibitem{Robertson:2014fka}
R.~G.~H. Robertson, \emph{{Examination of the calorimetric spectrum to
  determine the neutrino mass in low-energy electron capture decay}}.
  \href{http://dx.doi.org/10.1103/PhysRevC.91.035504}{\emph{Phys. Rev. C}}
  \textbf{91}~(3):~035504~(2015) , 1411.2906

\bibitem{Faessler:2015txa}
A.~Faessler, C.~Enss, L.~Gastaldo, \emph{et~al.}, \emph{{Determination of the
  neutrino mass by electron capture in $^{163}$Ho and the role of the
  three-hole states in $^{163}$Dy}}.
  \href{http://dx.doi.org/10.1103/PhysRevC.91.064302}{\emph{Phys. Rev. C}}
  \textbf{91}~(6):~064302~(2015) , 1503.02282

\bibitem{DeRujula:2016fdu}
A.~De~R\'ujula and M.~Lusignoli, \emph{{The calorimetric spectrum of the
  electron-capture decay of $^{163}$Ho. The spectral endpoint region}}.
  \href{http://dx.doi.org/10.1007/JHEP05(2016)015}{\emph{JHEP}}
  \textbf{05}:~015~(2016) , 1601.04990

\bibitem{Brass:2017kov}
M.~Bra\ss{}, C.~Enss, L.~Gastaldo, \emph{et~al.}, \emph{{$\textit{Ab initio}$
  calculation of the calorimetric electron capture spectrum of $^{163}$Holmium:
  Intra-atomic decay into bound-states}}.
  \href{http://dx.doi.org/10.1103/PhysRevC.97.054620}{\emph{Phys. Rev. C}}
  \textbf{97}~(5):~054620~(2018) , 1711.10309

\bibitem{Brab:2020uzx}
M.~Bra\ss{} and M.~W. Haverkort, \emph{{$Ab initio$ calculation of the electron
  capture spectrum of $^{163}$Ho: Auger--Meitner decay into continuum states}}.
  \href{http://dx.doi.org/10.1088/1367-2630/abac72}{\emph{New J. Phys.}}
  \textbf{22}~(9):~093018~(2020) , 2002.05989

\bibitem{Koehler:2018hzx}
K.~E. Koehler, M.~A. Famiano, C.~J. Fontes, \emph{et~al.}, \emph{{First
  Calorimetric Measurement of Electron Capture in ${}^{193}$Pt with a
  Transition Edge Sensor}}.
  \href{http://dx.doi.org/10.1007/s10909-018-1984-2}{\emph{J. Low Temp. Phys.}}
  \textbf{193}~(5-6):~1151~(2018) , 1803.06370

\bibitem{Ullom:2022kai}
J.~Ullom \emph{et~al.}, \emph{{Measuring the electron neutrino mass using the
  electron capture decay of 163Ho}}. In \emph{{2022 Snowmass Summer Study}}
  (2022), 2203.07572

\bibitem{Gastaldo:2017edk}
L.~Gastaldo \emph{et~al.}, \emph{{The electron capture in$^{163}$Ho experiment
  \textendash{} ECHo}}.
  \href{http://dx.doi.org/10.1140/epjst/e2017-70071-y}{\emph{Eur. Phys. J. ST}}
  \textbf{226}~(8):~1623~(2017)

\bibitem{Mantegazzini:2021yed}
F.~Mantegazzini \emph{et~al.}, \emph{{Metallic magnetic calorimeter arrays for
  the first phase of the ECHo experiment}}.
  \href{http://dx.doi.org/10.1016/j.nima.2022.166406}{\emph{Nucl. Instrum.
  Meth. A}} \textbf{1030}:~166406~(2022) , 2111.09945

\bibitem{Hammann:2021nuj}
R.~Hammann, A.~Barth, A.~Fleischmann, \emph{et~al.}, \emph{{Data reduction for
  a calorimetrically measured $^{163}\mathrm {Ho}$ spectrum of the ECHo-1k
  experiment}}.
  \href{http://dx.doi.org/10.1140/epjc/s10052-021-09763-9}{\emph{Eur. Phys. J.
  C}} \textbf{81}~(11):~963~(2021) , 2107.13528

\bibitem{Goggelmann:2022nin}
A.~G\"oggelmann, J.~Jochum, L.~Gastaldo, \emph{et~al.}, \emph{{Study of
  naturally occurring radionuclides in the ECHo set-up}}.
  \href{http://dx.doi.org/10.1140/epjc/s10052-022-10112-7}{\emph{Eur. Phys. J.
  C}} \textbf{82}~(2):~139~(2022)

\bibitem{Dorrer:2018apa}
H.~Dorrer \emph{et~al.}, \emph{{Production, isolation and characterization of
  radiochemically pure $^{163}$Ho samples for the ECHo-project}}.
  \href{http://dx.doi.org/10.1515/ract-2017-2877}{\emph{Radiochim. Acta}}
  \textbf{106}~(7):~535~(2018)

\bibitem{Kieck:2018sap}
T.~Kieck, S.~Biebricher, C.~E. D\"ullmann, \emph{et~al.}, \emph{{Optimization
  of a laser ion source for $^{163}$Ho isotope separation}}.
  \href{http://dx.doi.org/10.1063/1.5081094}{\emph{Rev. Sci. Instrum.}}
  \textbf{90}~(5):~053304~(2019) , 1809.01358

\bibitem{Kieck:2019fcr}
T.~Kieck, H.~Dorrer, C.~E. D\"ullmann, \emph{et~al.}, \emph{{Highly efficient
  isotope separation and ion implantation of $^{163}$Ho for the ECHo project}}.
  \href{http://dx.doi.org/10.1016/j.nima.2019.162602}{\emph{Nucl. Instrum.
  Meth. A}} \textbf{945}:~162602~(2019) , 1904.05559

\bibitem{Velte:2019jvx}
C.~Velte \emph{et~al.}, \emph{{High-resolution and low-background $^{163}$Ho
  spectrum: interpretation of the resonance tails}}.
  \href{http://dx.doi.org/10.1140/epjc/s10052-019-7513-x}{\emph{Eur. Phys. J.
  C}} \textbf{79}~(12):~1026~(2019)

\bibitem{Alpert:2014lfa}
B.~Alpert \emph{et~al.}, \emph{{HOLMES - The Electron Capture Decay of
  $^{163}$Ho to Measure the Electron Neutrino Mass with sub-eV sensitivity}}.
  \href{http://dx.doi.org/10.1140/epjc/s10052-015-3329-5}{\emph{Eur. Phys. J.
  C}} \textbf{75}~(3):~112~(2015) , 1412.5060

\bibitem{Heinitz:2018bav}
S.~Heinitz \emph{et~al.}, \emph{{Production and separation of 163Ho for nuclear
  physics experiments}}.
  \href{http://dx.doi.org/10.1371/journal.pone.0200910}{\emph{PLoS One}}
  \textbf{13}~(8):~e0200910~(2018)

\bibitem{HOLMES:2019ykt}
D.~T. Becker \emph{et~al.}, (HOLMES), \emph{{Working principle and demonstrator
  of microwave-multiplexing for the HOLMES experiment microcalorimeters}}.
  \href{http://dx.doi.org/10.1088/1748-0221/14/10/P10035}{\emph{JINST}}
  \textbf{14}~(10):~P10035~(2019) , 1910.05217

\bibitem{Borghesi:2021fda}
M.~Borghesi, M.~De~Gerone, M.~Faverzani, \emph{et~al.}, \emph{{A novel approach
  for nearly-coincident events rejection}}.
  \href{http://dx.doi.org/10.1140/epjc/s10052-021-09157-x}{\emph{Eur. Phys. J.
  C}} \textbf{81}~(5):~385~(2021) , 2101.02705

\bibitem{KATRIN:2020dpx}
M.~Aker \emph{et~al.}, (KATRIN), \emph{{Bound on 3+1 Active-Sterile Neutrino
  Mixing from the First Four-Week Science Run of KATRIN}}.
  \href{http://dx.doi.org/10.1103/PhysRevLett.126.091803}{\emph{Phys. Rev.
  Lett.}} \textbf{126}~(9):~091803~(2021) , 2011.05087

\bibitem{KATRIN:2022ith}
M.~Aker \emph{et~al.}, (KATRIN), \emph{{Improved eV-scale Sterile-Neutrino
  Constraints from the Second KATRIN Measurement Campaign}} 2201.11593

\bibitem{KATRIN:2022kkv}
M.~Aker \emph{et~al.}, (KATRIN), \emph{{New Constraint on the Local Relic
  Neutrino Background Overdensity with the First KATRIN Data Runs}} 2202.04587

\bibitem{PTOLEMY:2018jst}
E.~Baracchini \emph{et~al.}, (PTOLEMY), \emph{{PTOLEMY: A Proposal for Thermal
  Relic Detection of Massive Neutrinos and Directional Detection of MeV Dark
  Matter}} 1808.01892

\bibitem{PTOLEMY:2019hkd}
M.~G. Betti \emph{et~al.}, (PTOLEMY), \emph{{Neutrino physics with the PTOLEMY
  project: active neutrino properties and the light sterile case}}.
  \href{http://dx.doi.org/10.1088/1475-7516/2019/07/047}{\emph{JCAP}}
  \textbf{07}:~047~(2019) , 1902.05508

\bibitem{LoredoLamb:1987a:2002}
T.~J. Loredo and D.~Q. Lamb, \emph{Bayesian analysis of neutrinos observed from
  supernova {SN 1987A}}.
  \href{http://dx.doi.org/10.1103/PhysRevD.65.063002}{\emph{Phys. Rev. D}}
  \textbf{65}:~063002~(2002)

\bibitem{Hansen:2019giq}
R.~S.~L. Hansen, M.~Lindner, and O.~Scholer, \emph{{Timing the neutrino signal
  of a Galactic supernova}}.
  \href{http://dx.doi.org/10.1103/PhysRevD.101.123018}{\emph{Phys. Rev. D}}
  \textbf{101}~(12):~123018~(2020) , 1904.11461

\bibitem{RoyChoudhury:2019hls}
S.~Roy~Choudhury and S.~Hannestad, \emph{{Updated results on neutrino mass and
  mass hierarchy from cosmology with Planck 2018 likelihoods}}.
  \href{http://dx.doi.org/10.1088/1475-7516/2020/07/037}{\emph{JCAP}}
  \textbf{07}:~037~(2020) , 1907.12598

\bibitem{GERDA:2020xhi}
M.~Agostini \emph{et~al.}, (GERDA), \emph{{Final Results of GERDA on the Search
  for Neutrinoless Double-$\beta$ Decay}}.
  \href{http://dx.doi.org/10.1103/PhysRevLett.125.252502}{\emph{Phys. Rev.
  Lett.}} \textbf{125}~(25):~252502~(2020) , 2009.06079

\bibitem{Esteban:2020cvm}
I.~Esteban, M.~C. Gonzalez-Garcia, M.~Maltoni, \emph{et~al.}, \emph{{The fate
  of hints: updated global analysis of three-flavor neutrino oscillations}}.
  \href{http://dx.doi.org/10.1007/JHEP09(2020)178}{\emph{JHEP}}
  \textbf{09}:~178~(2020) , 2007.14792

\bibitem{DES:2021wwk}
T.~M.~C. Abbott \emph{et~al.}, (DES), \emph{{Dark Energy Survey Year 3 results:
  Cosmological constraints from galaxy clustering and weak lensing}}.
  \href{http://dx.doi.org/10.1103/PhysRevD.105.023520}{\emph{Phys. Rev. D}}
  \textbf{105}~(2):~023520~(2022) , 2105.13549

\bibitem{KamLAND-Zen:2022tow}
S.~Abe \emph{et~al.}, (KamLAND-Zen), \emph{{First Search for the Majorana
  Nature of Neutrinos in the Inverted Mass Ordering Region with KamLAND-Zen}}
  2203.02139

\bibitem{Hara:2019bur}
H.~Hara and M.~Yoshimura, \emph{{Raman stimulated neutrino pair emission}}.
  \href{http://dx.doi.org/10.1140/epjc/s10052-019-7148-y}{\emph{Eur. Phys. J.
  C}} \textbf{79}~(8):~684~(2019) , 1904.03813

\bibitem{Tashiro:2019ghs}
M.~Tashiro, B.~P. Das, J.~Ekman, \emph{et~al.}, \emph{{Macro-coherent radiative
  emission of neutrino pair between parity-even atomic states}}.
  \href{http://dx.doi.org/10.1140/epjc/s10052-019-7430-z}{\emph{Eur. Phys. J.
  C}} \textbf{79}~(11):~907~(2019) , 1911.01639

\bibitem{Carney:2022pku}
D.~Carney, K.~G. Leach, and D.~C. Moore, \emph{{Searches for massive neutrinos
  with mechanical quantum sensors}} 2207.05883

\bibitem{Abazajian:2022ofy}
K.~N. Abazajian \emph{et~al.}, \emph{{Synergy between cosmological and
  laboratory searches in neutrino physics: a white paper}}. In \emph{{2022
  Snowmass Summer Study}} (2022), 2203.07377

\bibitem{Agostini:2022zub}
M.~Agostini, G.~Benato, J.~A. Detwiler, \emph{et~al.}, \emph{{Toward the
  discovery of matter creation with neutrinoless double-beta decay}} 2202.01787

\bibitem{schechter1982neutrinoless}
J.~Schechter and J.~W. Valle, \emph{Neutrinoless double-$\beta$ decay in su
  (2)$\times$ u (1) theories}. \emph{Physical Review D}
  \textbf{25}~(11):~2951~(1982)

\bibitem{haxton1984double}
W.~Haxton and G.~Stephenson~Jr, \emph{Double beta decay}. \emph{Progress in
  Particle and Nuclear Physics} \textbf{12}:~409~(1984)

\bibitem{agostini2017discovery}
M.~Agostini, G.~Benato, and J.~A. Detwiler, \emph{Discovery probability of
  next-generation neutrinoless double-$\beta$ decay experiments}.
  \emph{Physical Review D} \textbf{96}~(5):~053001~(2017)

\bibitem{Dolinski:2019nrj}
M.~J. Dolinski, A.~W.~P. Poon, and W.~Rodejohann, \emph{{Neutrinoless
  Double-Beta Decay: Status and Prospects}}.
  \href{http://dx.doi.org/10.1146/annurev-nucl-101918-023407}{\emph{Ann. Rev.
  Nucl. Part. Sci.}} \textbf{69}:~219~(2019) , 1902.04097

\bibitem{cirigliano2022neutrinoless}
V.~Cirigliano, Z.~Davoudi, W.~Dekens, \emph{et~al.}, \emph{Neutrinoless
  double-beta decay: A roadmap for matching theory to experiment}. \emph{arXiv
  preprint arXiv:2203.12169}

\bibitem{Pastore:2017ofx}
S.~Pastore, J.~Carlson, V.~Cirigliano, \emph{et~al.}, \emph{{Neutrinoless
  double-$\beta$ decay matrix elements in light nuclei}}.
  \href{http://dx.doi.org/10.1103/PhysRevC.97.014606}{\emph{Phys. Rev. C}}
  \textbf{97}~(1):~014606~(2018) , 1710.05026

\bibitem{Cirigliano:2019vdj}
V.~Cirigliano, W.~Dekens, J.~De~Vries, \emph{et~al.}, \emph{{Renormalized
  approach to neutrinoless double- $\beta$ decay}}.
  \href{http://dx.doi.org/10.1103/PhysRevC.100.055504}{\emph{Phys. Rev. C}}
  \textbf{100}~(5):~055504~(2019) , 1907.11254

\bibitem{Yao:2019rck}
J.~M. Yao, B.~Bally, J.~Engel, \emph{et~al.}, \emph{{Ab Initio Treatment of
  Collective Correlations and the Neutrinoless Double Beta Decay of
  $^{48}$Ca}}.
  \href{http://dx.doi.org/10.1103/PhysRevLett.124.232501}{\emph{Phys. Rev.
  Lett.}} \textbf{124}~(23):~232501~(2020) , 1908.05424

\bibitem{Yao:2020olm}
J.~M. Yao, A.~Belley, R.~Wirth, \emph{et~al.}, \emph{{$Ab initio$ benchmarks of
  neutrinoless double-$\beta$ decay in light nuclei with a chiral
  Hamiltonian}}.
  \href{http://dx.doi.org/10.1103/PhysRevC.103.014315}{\emph{Phys. Rev. C}}
  \textbf{103}~(1):~014315~(2021) , 2010.08609

\bibitem{Novario:2020dmr}
S.~Novario, P.~Gysbers, J.~Engel, \emph{et~al.}, \emph{{Coupled-Cluster
  Calculations of Neutrinoless Double-$\beta$ Decay in $^{48}$Ca}}.
  \href{http://dx.doi.org/10.1103/PhysRevLett.126.182502}{\emph{Phys. Rev.
  Lett.}} \textbf{126}~(18):~182502~(2021) , 2008.09696

\bibitem{Weiss:2021rig}
R.~Weiss, P.~Soriano, A.~Lovato, \emph{et~al.}, \emph{{Neutrinoless double-beta
  decay: combining quantum Monte Carlo and the nuclear shell model with the
  generalized contact formalism}} 2112.08146

\bibitem{Gysbers:2019uyb}
P.~Gysbers \emph{et~al.}, \emph{{Discrepancy between experimental and
  theoretical $\beta$-decay rates resolved from first principles}}.
  \href{http://dx.doi.org/10.1038/s41567-019-0450-7}{\emph{Nature Phys.}}
  \textbf{15}~(5):~428~(2019) , 1903.00047

\bibitem{Hammer:2019poc}
H.~W. Hammer, S.~K\"onig, and U.~van Kolck, \emph{{Nuclear effective field
  theory: status and perspectives}}.
  \href{http://dx.doi.org/10.1103/RevModPhys.92.025004}{\emph{Rev. Mod. Phys.}}
  \textbf{92}~(2):~025004~(2020) , 1906.12122

\bibitem{Cirigliano:2018hja}
V.~Cirigliano, W.~Dekens, J.~De~Vries, \emph{et~al.}, \emph{{New Leading
  Contribution to Neutrinoless Double-\ensuremath{\beta} Decay}}.
  \href{http://dx.doi.org/10.1103/PhysRevLett.120.202001}{\emph{Phys. Rev.
  Lett.}} \textbf{120}~(20):~202001~(2018) , 1802.10097

\bibitem{Cirigliano:2020dmx}
V.~Cirigliano, W.~Dekens, J.~de~Vries, \emph{et~al.}, \emph{{Toward Complete
  Leading-Order Predictions for Neutrinoless Double $\beta$ Decay}}.
  \href{http://dx.doi.org/10.1103/PhysRevLett.126.172002}{\emph{Phys. Rev.
  Lett.}} \textbf{126}~(17):~172002~(2021) , 2012.11602

\bibitem{Cirigliano:2021qko}
V.~Cirigliano, W.~Dekens, J.~de~Vries, \emph{et~al.}, \emph{{Determining the
  leading-order contact term in neutrinoless double $\beta$ decay}}.
  \href{http://dx.doi.org/10.1007/JHEP05(2021)289}{\emph{JHEP}}
  \textbf{05}:~289~(2021) , 2102.03371

\bibitem{Wirth:2021pij}
R.~Wirth, J.~M. Yao, and H.~Hergert, \emph{{Ab~Initio Calculation of the
  Contact Operator Contribution in the Standard Mechanism for Neutrinoless
  Double Beta Decay}}.
  \href{http://dx.doi.org/10.1103/PhysRevLett.127.242502}{\emph{Phys. Rev.
  Lett.}} \textbf{127}~(24):~242502~(2021) , 2105.05415

\bibitem{Jokiniemi:2021qqv}
L.~Jokiniemi, P.~Soriano, and J.~Men\'endez, \emph{{Impact of the leading-order
  short-range nuclear matrix element on the neutrinoless double-beta decay of
  medium-mass and heavy nuclei}}.
  \href{http://dx.doi.org/10.1016/j.physletb.2021.136720}{\emph{Phys. Lett. B}}
  \textbf{823}:~136720~(2021) , 2107.13354

\bibitem{Prezeau:2003xn}
G.~Prezeau, M.~Ramsey-Musolf, and P.~Vogel, \emph{{Neutrinoless double beta
  decay and effective field theory}}.
  \href{http://dx.doi.org/10.1103/PhysRevD.68.034016}{\emph{Phys. Rev. D}}
  \textbf{68}:~034016~(2003) , hep-ph/0303205

\bibitem{Cirigliano:2017djv}
V.~Cirigliano, W.~Dekens, J.~de~Vries, \emph{et~al.}, \emph{{Neutrinoless
  double beta decay in chiral effective field theory: lepton number violation
  at dimension seven}}.
  \href{http://dx.doi.org/10.1007/JHEP12(2017)082}{\emph{JHEP}}
  \textbf{12}:~082~(2017) , 1708.09390

\bibitem{Cirigliano:2018yza}
V.~Cirigliano, W.~Dekens, J.~de~Vries, \emph{et~al.}, \emph{{A neutrinoless
  double beta decay master formula from effective field theory}}.
  \href{http://dx.doi.org/10.1007/JHEP12(2018)097}{\emph{JHEP}}
  \textbf{12}:~097~(2018) , 1806.02780

\bibitem{Graf:2018ozy}
L.~Graf, F.~F. Deppisch, F.~Iachello, \emph{et~al.}, \emph{{Short-Range
  Neutrinoless Double Beta Decay Mechanisms}}.
  \href{http://dx.doi.org/10.1103/PhysRevD.98.095023}{\emph{Phys. Rev. D}}
  \textbf{98}~(9):~095023~(2018) , 1806.06058

\bibitem{Dekens:2020ttz}
W.~Dekens, J.~de~Vries, K.~Fuyuto, \emph{et~al.}, \emph{{Sterile neutrinos and
  neutrinoless double beta decay in effective field theory}}.
  \href{http://dx.doi.org/10.1007/JHEP06(2020)097}{\emph{JHEP}}
  \textbf{06}:~097~(2020) , 2002.07182

\bibitem{Deppisch:2020ztt}
F.~F. Deppisch, L.~Graf, F.~Iachello, \emph{et~al.}, \emph{{Analysis of light
  neutrino exchange and short-range mechanisms in $0\nu\beta\beta$ decay}}.
  \href{http://dx.doi.org/10.1103/PhysRevD.102.095016}{\emph{Phys. Rev. D}}
  \textbf{102}~(9):~095016~(2020) , 2009.10119

\bibitem{Nicholson:2018mwc}
A.~Nicholson \emph{et~al.}, \emph{{Heavy physics contributions to neutrinoless
  double beta decay from QCD}}.
  \href{http://dx.doi.org/10.1103/PhysRevLett.121.172501}{\emph{Phys. Rev.
  Lett.}} \textbf{121}~(17):~172501~(2018) , 1805.02634

\bibitem{Feng:2018pdq}
X.~Feng, L.-C. Jin, X.-Y. Tuo, \emph{et~al.}, \emph{{Light-Neutrino Exchange
  and Long-Distance Contributions to $0\nu2\beta$ Decays: An Exploratory Study
  on $\pi\pi\to ee$}}.
  \href{http://dx.doi.org/10.1103/PhysRevLett.122.022001}{\emph{Phys. Rev.
  Lett.}} \textbf{122}~(2):~022001~(2019) , 1809.10511

\bibitem{Tuo:2019bue}
X.-Y. Tuo, X.~Feng, and L.-C. Jin, \emph{{Long-distance contributions to
  neutrinoless double beta decay $\pi^- \to\pi^+ e e$}}.
  \href{http://dx.doi.org/10.1103/PhysRevD.100.094511}{\emph{Phys. Rev. D}}
  \textbf{100}~(9):~094511~(2019) , 1909.13525

\bibitem{Detmold:2020jqv}
W.~Detmold and D.~J. Murphy, (NPLQCD), \emph{{Neutrinoless Double Beta Decay
  from Lattice QCD: The Long-Distance $\pi^{-} \rightarrow \pi^{+} e^{-} e^{-}$
  Amplitude}} 2004.07404

\bibitem{Davoudi:2020gxs}
Z.~Davoudi and S.~V. Kadam, \emph{{Path from Lattice QCD to the Short-Distance
  Contribution to 0$\nu \beta \beta$ Decay with a Light Majorana Neutrino}}.
  \href{http://dx.doi.org/10.1103/PhysRevLett.126.152003}{\emph{Phys. Rev.
  Lett.}} \textbf{126}~(15):~152003~(2021) , 2012.02083

\bibitem{Davoudi:2021noh}
Z.~Davoudi and S.~V. Kadam, \emph{{On the Extraction of Low-energy Constants of
  Single- and Double-$\beta$ Decays from Lattice QCD: A Sensitivity Analysis}}
  2111.11599

\bibitem{armstrong2019cupid}
W.~Armstrong, C.~Chang, K.~Hafidi, \emph{et~al.}, \emph{{CUPID pre-CDR}}

\bibitem{arnaboldi2004cuore}
C.~Arnaboldi, F.~Avignone~Iii, J.~Beeman, \emph{et~al.}, \emph{Cuore: a
  cryogenic underground observatory for rare events}. \emph{Nuclear Instruments
  and Methods in Physics Research Section A: Accelerators, Spectrometers,
  Detectors and Associated Equipment} \textbf{518}~(3):~775~(2004)

\bibitem{azzolini2018first}
O.~Azzolini, M.~Barrera, J.~Beeman, \emph{et~al.}, \emph{First result on the
  neutrinoless double-$\beta$ decay of se 82 with cupid-0}. \emph{Physical
  review letters} \textbf{120}~(23):~232502~(2018)

\bibitem{armengaud2020cupid}
E.~Armengaud, C.~Augier, A.~Barabash, \emph{et~al.}, \emph{The cupid-mo
  experiment for neutrinoless double-beta decay: performance and prospects}.
  \emph{The European Physical Journal C} \textbf{80}~(1):~1~(2020)

\bibitem{armatol2021cupid}
\emph{A {CUPID Li2100MoO4} scintillating bolometer tested in the cross
  underground facility}

\bibitem{armatol2022toward}
A.~Armatol, C.~Augier, F.~Avignone~III, \emph{et~al.}, \emph{Toward
  {CUPID-1T}}. \emph{arXiv preprint arXiv:2203.08386}

\bibitem{adhikari2021nexo}
G.~Adhikari, S.~A. Kharusi, E.~Angelico, \emph{et~al.}, \emph{{nEXO}:
  Neutrinoless double beta decay search beyond $10^{28}$ year half-life
  sensitivity}. \emph{arXiv preprint arXiv:2106.16243}

\bibitem{EXO-200:2019rkq}
G.~Anton \emph{et~al.}, (EXO-200), \emph{{Search for Neutrinoless
  Double-$\beta$ Decay with the Complete EXO-200 Dataset}}.
  \href{http://dx.doi.org/10.1103/PhysRevLett.123.161802}{\emph{Phys. Rev.
  Lett.}} \textbf{123}~(16):~161802~(2019) , 1906.02723

\bibitem{adams2021sensitivity}
C.~Adams, V.~{\'A}lvarez, L.~Arazi, \emph{et~al.}, \emph{Sensitivity of a
  tonne-scale {NEXT} detector for neutrinoless double-beta decay searches}.
  \emph{Journal of High Energy Physics} \textbf{2021}~(8):~1~(2021)

\bibitem{lux2021projected}
D.~Akerib, A.~Al~Musalhi, S.~Alsum, \emph{et~al.}, \emph{Projected sensitivity
  of the {LUX-ZEPLIN (LZ)} experiment to the two-neutrino and neutrinoless
  double beta decays of $^{134}$ {Xe}}. \emph{arXiv preprint arXiv:2104.13374}

\bibitem{agostini2020sensitivity}
F.~Agostini, S.~Maouloud, L.~Althueser, \emph{et~al.}, \emph{Sensitivity of the
  {DARWIN} observatory to the neutrinoless double beta decay of $^{136 Xe}$}.
  \emph{The European Physical Journal C} \textbf{80}~(9):~1~(2020)

\bibitem{abi2020deep}
B.~Abi, R.~Acciarri, M.~A. Acero, \emph{et~al.}, \emph{{Deep underground
  neutrino experiment (DUNE), far detector technical design report, Volume II:
  DUNE physics}}. \emph{arXiv preprint arXiv:2002.03005}

\bibitem{avasthi2021kiloton}
A.~Avasthi, T.~Bowyer, C.~Bray, \emph{et~al.}, \emph{Kiloton-scale xenon
  detectors for neutrinoless double beta decay and other new physics searches}.
  \emph{Physical Review D} \textbf{104}~(11):~112007~(2021)

\bibitem{avasthi2022low}
A.~Avasthi, T.~Bezerra, A.~Borkum, \emph{et~al.}, \emph{Low background
  kton-scale liquid argon time projection chambers}. \emph{arXiv preprint
  arXiv:2203.08821}

\bibitem{caratelli2022low}
D.~Caratelli, W.~Foreman, A.~Friedland, \emph{et~al.}, \emph{Low-energy physics
  in neutrino {LArTPCs}}. \emph{arXiv preprint arXiv:2203.00740}

\bibitem{nygren2018neutrinoless}
\emph{Neutrinoless double beta decay with ${}^{82}${}sef}

\bibitem{Monreal:2022crn}
B.~Monreal, \emph{{High-pressure TPCs in pressurized caverns: opportunities in
  dark matter and neutrino physics}} 2203.06262

\bibitem{li2021preliminary}
Z.~Li, H.~Feng, X.~Huang, \emph{et~al.}, \emph{Preliminary test of topmetal-ii-
  sensor for x-ray polarization measurements}. \emph{Nuclear Instruments and
  Methods in Physics Research Section A: Accelerators, Spectrometers, Detectors
  and Associated Equipment} \textbf{1008}:~165430~(2021)

\bibitem{DualReadout}
{Jonathan Asaadi and others},
  {\href{https://www.snowmass21.org/docs/files/summaries/IF/SNOWMASS21-IF9_IF8-NF3_NF10-CF1_CF0-145.pdf}{
  Dual-Readout Time Projection Chamber: exploring sub-millimeter pitch for
  directional dark matter and tau identification in $\nu_{\tau}$CC
  interactions}, Snowmass 2021 Letter of Interest}

\bibitem{jones2022ion}
B.~Jones, F.~Foss, J.~Asaadi, \emph{et~al.}, \emph{The {Ion Fluorescence
  Chamber (IFC)}: A new concept for directional dark matter and topologically
  imaging neutrinoless double beta decay searches}. \emph{arXiv preprint
  arXiv:2203.10198}

\bibitem{NuDeXTalk}
{Huan Zhong Huang},
  {\href{https://indico.fnal.gov/event/46424/contributions/202839/attachments/138331/173122/China-0vbb-Perspective-2020.pdf}{
  Perspective on 0nubb program in China}, Talk at ACFI Snowmass Workshop on
  Beyond-Ton-Scale future of 0nubb}

\bibitem{Theia:2019non}
M.~Askins \emph{et~al.}, (Theia), \emph{{THEIA: an advanced optical neutrino
  detector}}.
  \href{http://dx.doi.org/10.1140/epjc/s10052-020-7977-8}{\emph{Eur. Phys. J.
  C}} \textbf{80}~(5):~416~(2020) , 1911.03501

\bibitem{JUNO:2015zny}
F.~An \emph{et~al.}, (JUNO), \emph{{Neutrino Physics with JUNO}}.
  \href{http://dx.doi.org/10.1088/0954-3899/43/3/030401}{\emph{J. Phys. G}}
  \textbf{43}~(3):~030401~(2016) , 1507.05613

\bibitem{JUNO:2022hxd}
(JUNO), \emph{{JUNO physics and detector}}.
  \href{http://dx.doi.org/10.1016/j.ppnp.2021.103927}{\emph{Prog. Part. Nucl.
  Phys.}} \textbf{123}:~103927~(2022)

\bibitem{Zhao:2016brs}
J.~Zhao, L.-J. Wen, Y.-F. Wang, \emph{et~al.}, \emph{{Physics potential of
  searching for $0\nu\beta\beta$ decays in JUNO}}.
  \href{http://dx.doi.org/10.1088/1674-1137/41/5/053001}{\emph{Chin. Phys. C}}
  \textbf{41}~(5):~053001~(2017) , 1610.07143

\bibitem{JUNO:2021wzm}
A.~Abusleme \emph{et~al.}, (JUNO), \emph{{The Design and Sensitivity of JUNO's
  scintillator radiopurity pre-detector OSIRIS}} 2103.16900

\bibitem{klein2022future}
J.~R. Klein, T.~Akindele, A.~Bernstein, \emph{et~al.}, \emph{Future advances in
  photon-based neutrino detectors: A snowmass white paper}. \emph{arXiv
  preprint arXiv:2203.07479}

\bibitem{Morton-Blake:2022snr}
I.~Morton-Blake and S.~D. Biller, \emph{{An Alternative Design for Large Scale
  Liquid Scintillator Detectors}} 2201.06498

\bibitem{Cabrera:2019kxi}
A.~Cabrera \emph{et~al.}, \emph{{Neutrino Physics with an Opaque Detector}}.
  \href{http://dx.doi.org/10.1038/s42005-021-00763-5}{\emph{Commun. Phys.}}
  \textbf{4}:~273~(2021) , 1908.02859

\bibitem{Graham:2019zqb}
E.~Graham, D.~Gooding, J.~Gruszko, \emph{et~al.}, \emph{{Light Yield of
  Perovskite Nanocrystal-Doped Liquid Scintillator}} 1908.03564

\bibitem{Winslow:2012ey}
L.~Winslow and R.~Simpson, \emph{{Characterizing Quantum-Dot-Doped Liquid
  Scintillator for Applications to Neutrino Detectors}}.
  \href{http://dx.doi.org/10.1088/1748-0221/7/07/P07010}{\emph{JINST}}
  \textbf{7}:~P07010~(2012) , 1202.4733

\bibitem{Yeh:2011zz}
M.~Yeh, S.~Hans, W.~Beriguete, \emph{et~al.}, \emph{{A new water-based liquid
  scintillator and potential applications}}.
  \href{http://dx.doi.org/10.1016/j.nima.2011.08.040}{\emph{Nucl. Instrum.
  Meth. A}} \textbf{660}:~51~(2011)

\bibitem{Biller:2020uoi}
S.~D. Biller, E.~J. Leming, and J.~L. Paton, \emph{{Slow fluors for effective
  separation of Cherenkov light in liquid scintillators}}.
  \href{http://dx.doi.org/10.1016/j.nima.2020.164106}{\emph{Nucl. Instrum.
  Meth. A}} \textbf{972}:~164106~(2020) , 2001.10825

\bibitem{Adams:2015kkx}
B.~W. Adams, A.~Elagin, H.~J. Frisch, \emph{et~al.}, \emph{{Timing
  characteristics of Large Area Picosecond Photodetectors}}.
  \href{http://dx.doi.org/10.1016/j.nima.2015.05.027}{\emph{Nuclear Instruments
  and Methods in Physics Research Section A: Accelerators, Spectrometers,
  Detectors and Associated Equipment}} \textbf{795}:~1~(2015)

\bibitem{Kaptanoglu:2019gtg}
T.~Kaptanoglu, M.~Luo, B.~Land, \emph{et~al.}, \emph{{Spectral Photon Sorting
  For Large-Scale Cherenkov and Scintillation Detectors}}.
  \href{http://dx.doi.org/10.1103/PhysRevD.101.072002}{\emph{Phys. Rev. D}}
  \textbf{101}~(7):~072002~(2020) , 1912.10333

\bibitem{DUNE:2020vmp}
D.~Totani \emph{et~al.}, (DUNE), \emph{{A measurement of absolute efficiency of
  the ARAPUCA photon detector in liquid argon}}.
  \href{http://dx.doi.org/10.1088/1748-0221/15/06/T06003}{\emph{JINST}}
  \textbf{15}~(06):~T06003~(2020) , 2008.05371

\bibitem{Dalmasson:2017pow}
J.~Dalmasson, G.~Gratta, A.~Jamil, \emph{et~al.}, \emph{{Distributed Imaging
  for Liquid Scintillation Detectors}}.
  \href{http://dx.doi.org/10.1103/PhysRevD.97.052006}{\emph{Phys. Rev. D}}
  \textbf{97}~(5):~052006~(2018) , 1711.09851

\bibitem{Seibert2011FastOM}
S.~R. Seibert and A.~Latorre, \emph{Fast optical monte carlo simulation with
  surface-based geometries using chroma} (2011)

\bibitem{Mu:2021nno}
W.~Mu, A.~I. Himmel, and B.~Ramson, \emph{{Photon detection probability
  prediction using one-dimensional generative neural network}} 2109.07277

\bibitem{Li:2022frp}
A.~Li, Z.~Fu, L.~Winslow, \emph{et~al.}, \emph{{KamNet: An Integrated
  Spatiotemporal Deep Neural Network for Rare Event Search in KamLAND-Zen}}
  2203.01870

\bibitem{annieLOI}
{\it et al.}.~I.~Anghel, \emph{{LETTER OF INTENT: The Accelerator Neutrino
  Neutron Interaction Experiment (ANNIE)}}. \emph{FNAL Report No. P-1063}
  1504.01480

\bibitem{annie-results}
{\it et al.}.~A.R.~Back, \emph{{Accelerator Neutrino Neutron Interaction
  Experiment (ANNIE): Preliminary Results and Physics Phase Proposal}}.
  \href{http://dx.doi.org/10.1088/1748-0221/14/11/P11024}{\emph{FNAL Report No.
  P-1063}} 1707.08222

\bibitem{pershingdiss}
T.~J. Pershing, \emph{{The Accelerator Neutrino-Neutron Interaction
  Experiment}}. Ph.D. thesis, University of California, Davis (2020)

\bibitem{flatdot}
J.~Gruszko, B.~Naranjo, B.~Daniel, \emph{et~al.}, \emph{{Detecting Cherenkov
  light from 1\textendash{}2 MeV electrons in linear alkylbenzene}}.
  \href{http://dx.doi.org/10.1088/1748-0221/14/02/P02005}{\emph{JINST}}
  \textbf{14}~(02):~P02005~(2019) , 1811.11144

\bibitem{moe:1991ik}
M.~K. Moe, \emph{{New approach to the detection of neutrinoless double beta
  decay}}. \href{http://dx.doi.org/10.1103/PhysRevC.44.931}{\emph{Physical
  Review}} \textbf{C44}:~931~(1991)

\bibitem{Chambers:2018srx}
C.~Chambers \emph{et~al.}, \emph{{Imaging individual barium atoms in solid
  xenon for barium tagging in nEXO}}.
  \href{http://dx.doi.org/10.1038/s41586-019-1169-4}{\emph{Nature}}
  \textbf{569}~(7755):~203~(2019)

\bibitem{mcdonald2018demonstration}
A.~McDonald, B.~Jones, D.~Nygren, \emph{et~al.}, \emph{Demonstration of
  single-barium-ion sensitivity for neutrinoless double-beta decay using
  single-molecule fluorescence imaging}. \emph{Physical review letters}
  \textbf{120}~(13):~132504~(2018)

\bibitem{albert:2015vma}
J.~B. Albert, D.~J. Auty, P.~S. Barbeau, \emph{et~al.}, (EXO-200
  Collaboration), \emph{Measurements of the ion fraction and mobility of
  $\ensuremath{\alpha}\text{\ensuremath{-}}$ and $\ensuremath{\beta}$-decay
  products in liquid xenon using the {EXO}-200 detector}.
  \href{http://dx.doi.org/10.1103/PhysRevC.92.045504}{\emph{Phys. Rev. C}}
  \textbf{92}:~045504~(2015)

\bibitem{Novella:2018ewv}
P.~Novella \emph{et~al.}, \emph{{Measurement of radon-induced backgrounds in
  the NEXT double beta decay experiment}}.
  \href{http://dx.doi.org/10.1007/JHEP10(2018)112}{\emph{JHEP}}
  \textbf{10}:~112~(2018)

\bibitem{mong:2014iya}
B.~Mong, S.~Cook, T.~Walton, \emph{et~al.}, \emph{{Spectroscopy of Ba and
  Ba$^+$ deposits in solid xenon for barium tagging in nEXO}}.
  \href{http://dx.doi.org/10.1103/PhysRevA.91.022505}{\emph{Physical Review}}
  \textbf{A91}~(2):~022505~(2015)

\bibitem{craycraft2019barium}
A.~B. Craycraft, \emph{Barium extraction from liquid xenon on a cryoprobe for
  the nEXO experiment and a nucleon decay search using EXO-200 data}. Ph.D.
  thesis, Colorado State University (2019)

\bibitem{twelker:2014zsa}
K.~Twelker, S.~Kravitz, M.~M. D{\'\i}ez, \emph{et~al.}, \emph{{An apparatus to
  manipulate and identify individual Ba ions from bulk liquid Xe}}.
  \href{http://dx.doi.org/10.1063/1.4895646}{\emph{Review of Scientific
  Instruments}} \textbf{85}:~095114~(2014)

\bibitem{murray2018design}
K.~Murray, \emph{The design and optimization of a multi-reflection
  time-of-flight mass-spectrometer for Barium tagging with nEXO and
  optimization of the 137 Xe veto with EXO-200}. Ph.D. thesis, McGill
  University (Canada) (2018)

\bibitem{elba}
D.~Nygren, \emph{Detection of the barium daughter in 136xe ->136ba + 2e- by in
  situ single-molecule fluorescence imaging} (2016), {\it{Frontier Detectors
  for Frontier Physics}}, Elba, Italy, May 24-30, To be published in Nuclear
  Instruments and Methods in Physics Research Section A: Accelerators,
  Spectrometers, Detectors and Associated Equipment, 2016

\bibitem{jones:2016qiq}
B.~J.~P. Jones, A.~D. McDonald, and D.~R. Nygren, \emph{{Single Molecule
  Fluorescence Imaging as a Technique for Barium Tagging in Neutrinoless Double
  Beta Decay}}

\bibitem{byrnes2020demonstration}
N.~Byrnes, \emph{Demonstration of single barium ion detection in high pressure
  environments for neutrinoless double beta decay using on-off barium
  chemosensors}. \emph{Bulletin of the American Physical Society} \textbf{65}

\bibitem{thapa2019barium}
P.~Thapa, I.~Arnquist, N.~Byrnes, \emph{et~al.}, \emph{Barium chemosensors with
  dry-phase fluorescence for neutrinoless double beta decay}. \emph{Scientific
  reports} \textbf{9}~(1):~1~(2019)

\bibitem{thapa2021demonstration}
P.~Thapa, N.~K. Byrnes, A.~A. Denisenko, \emph{et~al.}, \emph{Demonstration of
  selective single-barium ion detection with dry diazacrown ether naphthalimide
  turn-on chemosensors}. \emph{ACS sensors} \textbf{6}~(1):~192~(2021)

\bibitem{rivilla2020fluorescent}
I.~Rivilla, B.~Aparicio, J.~M. Bueno, \emph{et~al.}, \emph{Fluorescent bicolour
  sensor for low-background neutrinoless double $\beta$ decay experiments}.
  \emph{Nature} 1--7

\bibitem{herrero2022ba}
P.~Herrero-G{\'o}mez, J.~Calupitan, M.~Ilyn, \emph{et~al.}, \emph{Ba $^{2+}$
  ion trapping by organic submonolayer: towards an ultra-low background
  neutrinoless double beta decay detector}. \emph{arXiv preprint
  arXiv:2201.09099}

\bibitem{jones2021dynamics}
B.~Jones, A.~Raymond, K.~Woodruff, \emph{et~al.}, \emph{The dynamics of ions on
  phased radio-frequency carpets in high pressure gases and application for
  barium tagging in xenon gas time projection chambers}. \emph{arXiv preprint
  arXiv:2109.05902}

\bibitem{Woodruff:2019hte}
K.~Woodruff \emph{et~al.}, \emph{{Radio frequency and DC high voltage breakdown
  of high pressure helium, argon, and xenon}}.
  \href{http://dx.doi.org/10.1088/1748-0221/15/04/P04022}{\emph{JINST}}
  \textbf{15}~(04):~P04022~(2020)

\bibitem{hennings2020controlling}
R.~Hennings-Yeomans, C.~Chang, J.~Ding, \emph{et~al.}, \emph{Controlling t c of
  iridium films using the proximity effect}. \emph{Journal of Applied Physics}
  \textbf{128}~(15):~154501~(2020)

\bibitem{tetsuno2020status}
K.~Tetsuno, S.~Ajimura, K.~Akutagawa, \emph{et~al.}, \emph{Status of 48ca
  double beta decay search and its future prospect in candles}. In
  \emph{Journal of Physics: Conference Series}, volume 1468, 012132, IOP
  Publishing (2020)

\bibitem{umehara2020search}
S.~Umehara, M.~Nomachi, T.~Kishimoto, \emph{et~al.}, \emph{Search for
  neutrino-less double beta decay of 48ca-candles}. In \emph{Journal of
  Physics: Conference Series}, volume 1643, 012028, IOP Publishing (2020)

\bibitem{Chavarria:2016hxk}
A.~E. Chavarria, C.~Galbiati, X.~Li, \emph{et~al.}, \emph{{A high-resolution
  CMOS imaging detector for the search of neutrinoless double $\beta$ decay in
  $^{82}$Se}}.
  \href{http://dx.doi.org/10.1088/1748-0221/12/03/P03022}{\emph{JINST}}
  \textbf{12}:~P03022~(2017) , 1609.03887

\bibitem{Li:2020ryk}
X.~Li, A.~E. Chavarria, S.~Bogdanovich, \emph{et~al.}, \emph{{Measurement of
  the ionization response of amorphous selenium with 122 keV
  \ensuremath{\gamma} rays}}.
  \href{http://dx.doi.org/10.1088/1748-0221/16/06/P06018}{\emph{JINST}}
  \textbf{16}~(06):~P06018~(2021) , 2012.04079

\bibitem{ARNQUIST2020163573}
I.~J. Arnquist, C.~Beck, M.~L. {di Vacri}, \emph{et~al.}, \emph{Ultra-low
  radioactivity kapton and copper-kapton laminates}.
  \href{http://dx.doi.org/https://doi.org/10.1016/j.nima.2020.163573}{\emph{Nuclear
  Instruments and Methods in Physics Research Section A: Accelerators,
  Spectrometers, Detectors and Associated Equipment}}
  \textbf{959}:~163573~(2020) ,
  \urlprefix\url{https://www.sciencedirect.com/science/article/pii/S0168900220301480}

\bibitem{ABGRALL201622}
N.~Abgrall, I.~Arnquist, F.~Avignone, \emph{et~al.}, \emph{The {Majorana
  Demonstrator} radioassay program}.
  \href{http://dx.doi.org/https://doi.org/10.1016/j.nima.2016.04.070}{\emph{Nuclear
  Instruments and Methods in Physics Research Section A: Accelerators,
  Spectrometers, Detectors and Associated Equipment}} \textbf{828}:~22~(2016) ,
  \urlprefix\url{https://www.sciencedirect.com/science/article/pii/S0168900216302832}

\bibitem{doi:10.1063/1.5019001}
C.~D. Christofferson, N.~Abgrall, S.~I. Alvis, \emph{et~al.},
  \emph{Contamination control and assay results for the {Majorana Demonstrator}
  ultra clean components}. \href{http://dx.doi.org/10.1063/1.5019001}{\emph{AIP
  Conference Proceedings}} \textbf{1921}~(1):~060005~(2018) ,
  https://aip.scitation.org/doi/pdf/10.1063/1.5019001,
  \urlprefix\url{https://aip.scitation.org/doi/abs/10.1063/1.5019001}

\bibitem{mei2009cosmogenic}
D.-M. Mei, Z.-B. Yin, and S.~Elliott, \emph{Cosmogenic production as a
  background in searching for rare physics processes}. \emph{Astroparticle
  Physics} \textbf{31}~(6):~417~(2009)

\bibitem{cebrian2006cosmogenic}
S.~Cebri{\'a}n, J.~Amar{\'e}, B.~Beltr{\'a}n, \emph{et~al.}, \emph{Cosmogenic
  activation in germanium double beta decay experiments}. In \emph{Journal of
  Physics: Conference Series}, volume~39, 089, IOP Publishing (2006)

\bibitem{albert2016cosmogenic}
J.~Albert, D.~Auty, P.~Barbeau, \emph{et~al.}, \emph{Cosmogenic backgrounds to
  0$\nu\beta\beta$ in exo-200}. \emph{Journal of Cosmology and Astroparticle
  Physics} \textbf{2016}~(04):~029~(2016)

\bibitem{baudis2015cosmogenic}
L.~Baudis, A.~Kish, F.~Piastra, \emph{et~al.}, \emph{Cosmogenic activation of
  xenon and copper}. \emph{The European Physical Journal C}
  \textbf{75}~(10):~1~(2015)

\bibitem{lozza2015cosmogenic}
V.~Lozza and J.~Petzoldt, \emph{Cosmogenic activation of a natural tellurium
  target}. \emph{Astroparticle Physics} \textbf{61}:~62~(2015)

\bibitem{pandola2007monte}
L.~Pandola, M.~Bauer, K.~Kr{\"o}ninger, \emph{et~al.}, \emph{Monte carlo
  evaluation of the muon-induced background in the gerda double beta decay
  experiment}. \emph{Nuclear Instruments and Methods in Physics Research
  Section A: Accelerators, Spectrometers, Detectors and Associated Equipment}
  \textbf{570}~(1):~149~(2007)

\bibitem{rogers2020mitigation}
L.~Rogers, B.~Jones, A.~Laing, \emph{et~al.}, \emph{Mitigation of backgrounds
  from cosmogenic 137xe in xenon gas experiments using 3he neutron capture}.
  \emph{Journal of Physics G: Nuclear and Particle Physics}
  \textbf{47}~(7):~075001~(2020)

\bibitem{kudryavtsev2020neutron}
V.~Kudryavtsev, P.~Zakhary, and B.~Easeman, \emph{Neutron production in
  ($\alpha$, n) reactions}. \emph{Nuclear Instruments and Methods in Physics
  Research Section A: Accelerators, Spectrometers, Detectors and Associated
  Equipment} \textbf{972}:~164095~(2020)

\bibitem{de2011solar}
N.~De~Barros and K.~Zuber, \emph{Solar neutrino--electron scattering as
  background limitation for double-beta decay}. \emph{Journal of Physics G:
  Nuclear and Particle Physics} \textbf{38}~(10):~105201~(2011)

\bibitem{ejiri2016solar}
H.~Ejiri and K.~Zuber, \emph{Solar neutrino interactions with liquid
  scintillators used for double beta-decay experiments}. \emph{Journal of
  Physics G: Nuclear and Particle Physics} \textbf{43}~(4):~045201~(2016)

\bibitem{ejiri2014charged}
H.~Ejiri and S.~Elliott, \emph{Charged current neutrino cross section for solar
  neutrinos, and background to $\beta\beta(0\nu$) experiments}. \emph{Physical
  Review C} \textbf{89}~(5):~055501~(2014)

\bibitem{albert2018sensitivity}
J.~Albert, G.~Anton, I.~Arnquist, \emph{et~al.}, \emph{Sensitivity and
  discovery potential of the proposed {nEXO} experiment to neutrinoless
  double-$\beta$ decay}. \emph{Physical Review C}
  \textbf{97}~(6):~065503~(2018)

\bibitem{elagin2017separating}
A.~Elagin, H.~J. Frisch, B.~Naranjo, \emph{et~al.}, \emph{Separating
  double-beta decay events from solar neutrino interactions in a kiloton-scale
  liquid scintillator detector by fast timing}. \emph{Nuclear Instruments and
  Methods in Physics Research Section A: Accelerators, Spectrometers, Detectors
  and Associated Equipment} \textbf{849}:~102~(2017)

\bibitem{anton2020measurement}
G.~Anton, I.~Badhrees, P.~Barbeau, \emph{et~al.}, \emph{Measurement of the
  scintillation and ionization response of liquid xenon at {MeV energies in the
  EXO-200} experiment}. \emph{Physical Review C}
  \textbf{101}~(6):~065501~(2020)

\bibitem{monteiro2007secondary}
C.~Monteiro, L.~Fernandes, J.~Lopes, \emph{et~al.}, \emph{Secondary
  scintillation yield in pure xenon}. \emph{Journal of Instrumentation}
  \textbf{2}~(05):~P05001~(2007)

\bibitem{henriques2022neutral}
C.~Henriques, P.~Amedo, J.~Teixeira, \emph{et~al.}, \emph{Neutral
  bremsstrahlung emission in xenon unveiled}. \emph{arXiv preprint
  arXiv:2202.02614}

\bibitem{albert2017measurement}
J.~Albert, P.~Barbeau, D.~Beck, \emph{et~al.}, \emph{Measurement of the drift
  velocity and transverse diffusion of electrons in liquid xenon with the
  {EXO}-200 detector}. \emph{Physical Review C} \textbf{95}~(2):~025502~(2017)

\bibitem{simon2018electron}
A.~Sim{\'o}n, R.~Felkai, G.~Mart{\'\i}nez-Lema, \emph{et~al.}, \emph{Electron
  drift properties in high pressure gaseous xenon}. \emph{Journal of
  Instrumentation} \textbf{13}~(07):~P07013~(2018)

\bibitem{mcdonald2019electron}
A.~McDonald, K.~Woodruff, B.~Al~Atoum, \emph{et~al.}, \emph{Electron drift and
  longitudinal diffusion in high pressure xenon-helium gas mixtures}.
  \emph{Journal of Instrumentation} \textbf{14}~(08):~P08009~(2019)

\bibitem{azevedo2016homeopathic}
C.~Azevedo, L.~Fernandes, E.~Freitas, \emph{et~al.}, \emph{An homeopathic cure
  to pure xenon large diffusion}. \emph{Journal of Instrumentation}
  \textbf{11}~(02):~C02007~(2016)

\bibitem{akerib2020investigation}
D.~Akerib, S.~Alsum, H.~Ara{\'u}jo, \emph{et~al.}, \emph{Investigation of
  background electron emission in the {LUX} detector}. \emph{Physical Review D}
  \textbf{102}~(9):~092004~(2020)

\bibitem{sorensen2018two}
P.~Sorensen and K.~Kamdin, \emph{Two distinct components of the delayed single
  electron noise in liquid xenon emission detectors}. \emph{Journal of
  Instrumentation} \textbf{13}~(02):~P02032~(2018)

\bibitem{bodnia2021electric}
E.~Bodnia, E.~Bernard, A.~Biekert, \emph{et~al.}, \emph{The electric field
  dependence of single electron emission in the {PIXeY} two-phase xenon
  detector}. \emph{Journal of Instrumentation} \textbf{16}~(12):~P12015~(2021)

\bibitem{szydagis2011nest}
M.~Szydagis, N.~Barry, K.~Kazkaz, \emph{et~al.}, \emph{{NEST}: a comprehensive
  model for scintillation yield in liquid xenon}. \emph{Journal of
  Instrumentation} \textbf{6}~(10):~P10002~(2011)

\bibitem{mock2014modeling}
J.~Mock, N.~Barry, K.~Kazkaz, \emph{et~al.}, \emph{Modeling pulse
  characteristics in xenon with {NEST}}. \emph{Journal of Instrumentation}
  \textbf{9}~(04):~T04002~(2014)

\bibitem{schindler2010calculation}
H.~Schindler, S.~Biagi, and R.~Veenhof, \emph{Calculation of gas gain
  fluctuations in uniform fields}. \emph{Nuclear Instruments and Methods in
  Physics Research Section A: Accelerators, Spectrometers, Detectors and
  Associated Equipment} \textbf{624}~(1):~78~(2010)

\bibitem{al2020electron}
B.~Al~Atoum, S.~F. Biagi, D.~Gonz{\'a}lez-D{\'\i}az, \emph{et~al.},
  \emph{Electron transport in gaseous detectors with a {Python-based Monte
  Carlo} simulation code}. \emph{Computer Physics Communications}
  \textbf{254}:~107357~(2020)

\bibitem{biller2017new}
S.~Biller, S.~Manecki, S.~collaboration, \emph{et~al.}, \emph{A new technique
  to load ${}^{130}${Te} in liquid scintillator for neutrinoless double beta
  decay experiments}. In \emph{Journal of Physics: Conference Series}, volume
  888, 012084, IOP Publishing (2017)

\bibitem{shimizu2019double}
I.~Shimizu and M.~Chen, \emph{Double beta decay experiments with loaded liquid
  scintillator}. \emph{Frontiers in Physics} \textbf{7}:~33~(2019)

\bibitem{hans2015purification}
S.~Hans, R.~Rosero, L.~Hu, \emph{et~al.}, \emph{Purification of telluric acid
  for sno+ neutrinoless double-beta decay search}. \emph{Nuclear Instruments
  and Methods in Physics Research Section A: Accelerators, Spectrometers,
  Detectors and Associated Equipment} \textbf{795}:~132~(2015)

\bibitem{hans2020light}
S.~Hans, J.~Cumming, R.~Rosero, \emph{et~al.}, \emph{Light yield quenching and
  quenching remediation in liquid scintillator detectors}. \emph{Journal of
  Instrumentation} \textbf{15}~(12):~P12020~(2020)

\bibitem{barabanov2012nd}
I.~Barabanov, L.~Bezrukov, C.~Cattadori, \emph{et~al.}, \emph{A nd-loaded
  liquid organic scintillator for the experiment aimed at measuring double
  $\beta$ decay}. \emph{Instruments and Experimental Techniques}
  \textbf{55}~(5):~545~(2012)

\bibitem{gehman2010solubility}
V.~Gehman, P.~Doe, R.~Robertson, \emph{et~al.}, \emph{Solubility, light output
  and energy resolution studies of molybdenum-loaded liquid scintillators}.
  \emph{Nuclear Instruments and Methods in Physics Research Section A:
  Accelerators, Spectrometers, Detectors and Associated Equipment}
  \textbf{622}~(3):~602~(2010)

\bibitem{hwang2009search}
M.~Hwang, Y.~Kwon, H.~Kim, \emph{et~al.}, \emph{A search for 0$\nu\beta\beta$
  decay of ${}^{124}${Sn} with tin-loaded liquid scintillator}.
  \emph{Astroparticle Physics} \textbf{31}~(6):~412~(2009)

\bibitem{2022arXiv220301870L}
A.~{Li}, Z.~{Fu}, L.~A. {Winslow}, \emph{et~al.}, \emph{{KamNet: An Integrated
  Spatiotemporal Deep Neural Network for Rare Event Search in KamLAND-Zen}}.
  \emph{arXiv e-prints} arXiv:2203.01870, 2203.01870

\bibitem{chess}
J.~Caravaca, F.~B. Descamps, B.~J. Land, \emph{et~al.}, \emph{{Experiment to
  demonstrate separation of Cherenkov and scintillation signals}}.
  \href{http://dx.doi.org/10.1103/PhysRevC.95.055801}{\emph{Phys. Rev. C}}
  \textbf{95}~(5):~055801~(2017) , 1610.02029

\bibitem{chesslappd}
T.~Kaptanoglu, E.~J. Callaghan, M.~Yeh, \emph{et~al.}, \emph{{Cherenkov and
  scintillation separation in water-based liquid scintillator using an
  LAPPD$^{TM}$}}.
  \href{http://dx.doi.org/10.1140/epjc/s10052-022-10087-5}{\emph{Eur. Phys. J.
  C}} \textbf{82}~(2):~169~(2022) , 2110.13222

\bibitem{LI2016303}
M.~Li, Z.~Guo, M.~Yeh, \emph{et~al.}, \emph{Separation of scintillation and
  cherenkov lights in linear alkyl benzene}.
  \href{http://dx.doi.org/https://doi.org/10.1016/j.nima.2016.05.132}{\emph{Nuclear
  Instruments and Methods in Physics Research Section A: Accelerators,
  Spectrometers, Detectors and Associated Equipment}} \textbf{830}:~303~(2016)
  ,
  \urlprefix\url{https://www.sciencedirect.com/science/article/pii/S0168900216305411}

\bibitem{chroma}
S.~Seibert and A.~LaTorre, \emph{{Fast Optical Monte Carlo Simulation with
  Surface-based Geometries Using {\it Chroma}}}. \emph{Semantic Scholar}

\bibitem{Salathe2017}
M.~Salathe, R.~Cooper, H.~Crawford, \emph{et~al.}, \emph{Energy reconstruction
  of an n-type segmented inverted coaxial point-contact hpge detector}.
  \emph{Nuclear Instruments and Methods in Physics Research Section A:
  Accelerators, Spectrometers, Detectors and Associated Equipment} \textbf{868}

\bibitem{TUBE}
{The MAJORANA Collaboration}, I.~J. Arnquist, F.~T.~A. III, \emph{et~al.},
  \emph{$\alpha$-event characterization and rejection in point-contact hpge
  detectors} (2020), 2006.13179, in press

\bibitem{GALATEA_scan}
F.~Edzards, L.~Hauertmann, I.~Abt, \emph{et~al.}, \emph{Surface
  characterization of p-type point contact germanium detectors}.
  \href{http://dx.doi.org/10.3390/particles4040036}{\emph{Particles}}
  \textbf{4}:~489~(2021)

\bibitem{ruth2020shortage}
T.~J. Ruth, \emph{The shortage of technetium-99m and possible solutions}.
  \emph{Annual Review of Nuclear and Particle Science} \textbf{70}:~77~(2020)

\bibitem{beeman2015double}
J.~Beeman, F.~Bellini, P.~Benetti, \emph{et~al.}, \emph{Double-beta decay
  investigation with highly pure enriched 82 se for the lucifer experiment}.
  \emph{The European Physical Journal C} \textbf{75}~(12):~1~(2015)

\bibitem{Tikhomirov:2000td}
A.~V. Tikhomirov, \emph{{Centrifugal enrichment of stable isotopes and modern
  physical experiments}}.
  \href{http://dx.doi.org/10.1023/A:1022809128791}{\emph{Czech. J. Phys.}}
  \textbf{50}:~577~(2000)

\bibitem{Hayes:2012sg}
A.~C. Hayes and G.~Jungman, \emph{{Determining Reactor Flux from Xenon-136 and
  Cesium-135 in Spent Fuel}}.
  \href{http://dx.doi.org/10.1016/j.nima.2012.06.031}{\emph{Nucl. Instrum.
  Meth. A}} \textbf{690}:~68~(2012) , 1205.6524

\bibitem{Matsuoka:2020spy}
K.~Matsuoka, H.~Niki, I.~Ogawa, \emph{et~al.}, \emph{{The laser Isotope
  separation (LIS) methods for the enrichment of 48Ca.}}
  \href{http://dx.doi.org/10.1088/1742-6596/1468/1/012199}{\emph{J. Phys. Conf.
  Ser.}} \textbf{1468}~(1):~012199~(2020)

\bibitem{Umehara:2015cna}
S.~Umehara, T.~Kishimoto, H.~Kakubata, \emph{et~al.}, \emph{{A basic study on
  the production of enriched isotope $^{48}$Ca by using crown-ether resin}}.
  \href{http://dx.doi.org/10.1093/ptep/ptv063}{\emph{PTEP}}
  \textbf{2015}~(5):~053C03~(2015)

\bibitem{cooley2018radiopurity}
J.~Cooley, J.~Loach, and A.~Poon, \emph{The radiopurity. org material
  database}. In \emph{AIP Conference Proceedings}, volume 1921, 040001, AIP
  Publishing LLC (2018)

\bibitem{blaum2020neutrinoless}
K.~Blaum, S.~Eliseev, F.~Danevich, \emph{et~al.}, \emph{Neutrinoless
  double-electron capture}. \emph{Reviews of Modern Physics}
  \textbf{92}~(4):~045007~(2020)

\bibitem{kotila2014neutrinoless}
J.~Kotila, J.~Barea, and F.~Iachello, \emph{Neutrinoless double-electron
  capture}. \emph{Physical Review C} \textbf{89}~(6):~064319~(2014)

\bibitem{krivoruchenko2011resonance}
M.~Krivoruchenko, F.~{\v{S}}imkovic, D.~Frekers, \emph{et~al.}, \emph{Resonance
  enhancement of neutrinoless double electron capture}. \emph{Nuclear Physics
  A} \textbf{859}~(1):~140~(2011)

\bibitem{eliseev2011resonant}
S.~Eliseev, C.~Roux, K.~Blaum, \emph{et~al.}, \emph{Resonant enhancement of
  neutrinoless double-electron capture in gd 152}. \emph{Physical Review
  Letters} \textbf{106}~(5):~052504~(2011)

\bibitem{aprile2019observation}
E.~Aprile \emph{et~al.}, \emph{Observation of two-neutrino double electron
  capture in 124xe with xenon1t}. \emph{Nature} \textbf{568}~(7753)

\bibitem{kovalenko2009lepton}
S.~Kovalenko, Z.~Lu, and I.~Schmidt, \emph{Lepton number violating processes
  mediated by majorana neutrinos at hadron colliders}. \emph{Physical Review D}
  \textbf{80}~(7):~073014~(2009)

\bibitem{balantekin2019addressing}
A.~B. Balantekin, A.~De~Gouv{\^e}a, and B.~Kayser, \emph{Addressing the
  majorana vs. dirac question with neutrino decays}. \emph{Physics Letters B}
  \textbf{789}:~488~(2019)

\bibitem{das2017bounds}
A.~Das and N.~Okada, \emph{Bounds on heavy majorana neutrinos in type-i seesaw
  and implications for collider searches}. \emph{Physics Letters B}
  \textbf{774}:~32~(2017)

\bibitem{atre2009search}
A.~Atre, T.~Han, S.~Pascoli, \emph{et~al.}, \emph{The search for heavy majorana
  neutrinos}. \emph{Journal of High Energy Physics}
  \textbf{2009}~(05):~030~(2009)

\bibitem{alva2015heavy}
D.~Alva, T.~Han, and R.~Ruiz, \emph{Heavy majorana neutrinos from w$\gamma$
  fusion at hadron colliders}. \emph{Journal of High Energy Physics}
  \textbf{2015}~(2):~1~(2015)

\bibitem{han2006signatures}
T.~Han and B.~Zhang, \emph{Signatures for majorana neutrinos at hadron
  colliders}. \emph{Physical review letters} \textbf{97}~(17):~171804~(2006)

\bibitem{Giunti:2014ixa}
C.~Giunti and A.~Studenikin, \emph{{Neutrino electromagnetic interactions: a
  window to new physics}}.
  \href{http://dx.doi.org/10.1103/RevModPhys.87.531}{\emph{Rev. Mod. Phys.}}
  \textbf{87}:~531~(2015) , 1403.6344

\bibitem{Vidyakin:1992nf}
G.~S. Vidyakin, V.~N. Vyrodov, I.~I. Gurevich, \emph{et~al.},
  \emph{{Limitations on the magnetic moment and charge radius of the
  electron-anti-neutrino}}. \emph{JETP Lett.} \textbf{55}:~206~(1992)

\bibitem{Deniz:2009mu}
M.~Deniz \emph{et~al.}, (TEXONO), \emph{{Measurement of Nu(e)-bar -Electron
  Scattering Cross-Section with a CsI(Tl) Scintillating Crystal Array at the
  Kuo-Sheng Nuclear Power Reactor}}.
  \href{http://dx.doi.org/10.1103/PhysRevD.81.072001}{\emph{Phys. Rev. D}}
  \textbf{81}:~072001~(2010) , 0911.1597

\bibitem{Allen:1992qe}
R.~C. Allen \emph{et~al.}, \emph{{Study of electron-neutrino electron elastic
  scattering at LAMPF}}.
  \href{http://dx.doi.org/10.1103/PhysRevD.47.11}{\emph{Phys. Rev. D}}
  \textbf{47}:~11~(1993)

\bibitem{Auerbach:2001wg}
L.~B. Auerbach \emph{et~al.}, (LSND), \emph{{Measurement of electron - neutrino
  - electron elastic scattering}}.
  \href{http://dx.doi.org/10.1103/PhysRevD.63.112001}{\emph{Phys. Rev. D}}
  \textbf{63}:~112001~(2001) , hep-ex/0101039

\bibitem{Ahrens:1990fp}
L.~A. Ahrens \emph{et~al.}, \emph{{Determination of electroweak parameters from
  the elastic scattering of muon-neutrinos and anti-neutrinos on electrons}}.
  \href{http://dx.doi.org/10.1103/PhysRevD.41.3297}{\emph{Phys. Rev. D}}
  \textbf{41}:~3297~(1990)

\bibitem{Hirsch:2002uv}
M.~Hirsch, E.~Nardi, and D.~Restrepo, \emph{{Bounds on the tau and muon
  neutrino vector and axial vector charge radius}}.
  \href{http://dx.doi.org/10.1103/PhysRevD.67.033005}{\emph{Phys. Rev. D}}
  \textbf{67}:~033005~(2003) , hep-ph/0210137

\bibitem{Vilain:1994hm}
P.~Vilain \emph{et~al.}, (CHARM-II), \emph{{Experimental study of
  electromagnetic properties of the muon-neutrino in neutrino - electron
  scattering}}.
  \href{http://dx.doi.org/10.1016/0370-2693(94)01678-6}{\emph{Phys. Lett. B}}
  \textbf{345}:~115~(1995)

\bibitem{Akimov:2017ade}
D.~Akimov \emph{et~al.}, (COHERENT), \emph{{Observation of Coherent Elastic
  Neutrino-Nucleus Scattering}}.
  \href{http://dx.doi.org/10.1126/science.aao0990}{\emph{Science}}
  \textbf{357}~(6356):~1123~(2017) , 1708.01294

\bibitem{Akimov:2020pdx}
D.~Akimov \emph{et~al.}, (COHERENT), \emph{{First Measurement of Coherent
  Elastic Neutrino-Nucleus Scattering on Argon}}.
  \href{http://dx.doi.org/10.1103/PhysRevLett.126.012002}{\emph{Phys. Rev.
  Lett.}} \textbf{126}~(1):~012002~(2021) , 2003.10630

\bibitem{Cadeddu:2020lky}
M.~Cadeddu, F.~Dordei, C.~Giunti, \emph{et~al.}, \emph{{Physics results from
  the first COHERENT observation of coherent elastic neutrino-nucleus
  scattering in argon and their combination with cesium-iodide data}}.
  \href{http://dx.doi.org/10.1103/PhysRevD.102.015030}{\emph{Phys. Rev. D}}
  \textbf{102}~(1):~015030~(2020) , 2005.01645

\bibitem{Bernabeu:2002pd}
J.~Bernabeu, J.~Papavassiliou, and J.~Vidal, \emph{{The Neutrino charge radius
  is a physical observable}}.
  \href{http://dx.doi.org/10.1016/j.nuclphysb.2003.12.025}{\emph{Nucl. Phys.
  B}} \textbf{680}:~450~(2004) , hep-ph/0210055

\bibitem{Cadeddu:2018dux}
M.~Cadeddu, C.~Giunti, K.~A. Kouzakov, \emph{et~al.}, \emph{{Neutrino Charge
  Radii from COHERENT Elastic Neutrino-Nucleus Scattering}}.
  \href{http://dx.doi.org/10.1103/PhysRevD.98.113010}{\emph{Phys. Rev. D}}
  \textbf{98}~(11):~113010~(2018) , 1810.05606, [Erratum: Phys.Rev.D 101,
  059902 (2020)]

\bibitem{Kouzakov:2017hbc}
K.~A. Kouzakov and A.~I. Studenikin, \emph{{Electromagnetic properties of
  massive neutrinos in low-energy elastic neutrino-electron scattering}}.
  \href{http://dx.doi.org/10.1103/PhysRevD.95.055013}{\emph{Phys. Rev. D}}
  \textbf{95}~(5):~055013~(2017) , 1703.00401, [Erratum: Phys.Rev.D 96, 099904
  (2017)]

\bibitem{Mathur:2021trm}
V.~Mathur, I.~M. Shoemaker, and Z.~Tabrizi, \emph{{Using DUNE to Shed Light on
  the Electromagnetic Properties of Neutrinos}} 2111.14884

\bibitem{Abdullah:2022zue}
M.~Abdullah \emph{et~al.}, \emph{{Coherent elastic neutrino-nucleus scattering:
  Terrestrial and astrophysical applications}} (2022), 2203.07361

\bibitem{Derbin:1993wy}
A.~I. Derbin, A.~V. Chernyi, L.~A. Popeko, \emph{et~al.}, \emph{{Experiment on
  anti-neutrino scattering by electrons at a reactor of the Rovno nuclear power
  plant}}. \emph{JETP Lett.} \textbf{57}:~768~(1993)

\bibitem{Daraktchieva:2005kn}
Z.~Daraktchieva \emph{et~al.}, (MUNU), \emph{{Final results on the neutrino
  magnetic moment from the MUNU experiment}}.
  \href{http://dx.doi.org/10.1016/j.physletb.2005.04.030}{\emph{Phys. Lett. B}}
  \textbf{615}:~153~(2005) , hep-ex/0502037

\bibitem{Wong:2006nx}
H.~T. Wong \emph{et~al.}, (TEXONO), \emph{{A Search of Neutrino Magnetic
  Moments with a High-Purity Germanium Detector at the Kuo-Sheng Nuclear Power
  Station}}. \href{http://dx.doi.org/10.1103/PhysRevD.75.012001}{\emph{Phys.
  Rev. D}} \textbf{75}:~012001~(2007) , hep-ex/0605006

\bibitem{Beda:2012zz}
A.~G. Beda, V.~B. Brudanin, V.~G. Egorov, \emph{et~al.}, \emph{{The results of
  search for the neutrino magnetic moment in GEMMA experiment}}.
  \href{http://dx.doi.org/10.1155/2012/350150}{\emph{Adv. High Energy Phys.}}
  \textbf{2012}:~350150~(2012)

\bibitem{BEBCWA66:1986err}
H.~Grassler \emph{et~al.}, (BEBC WA66), \emph{{Prompt Neutrino Production in
  400-{GeV} Proton Copper Interactions}}.
  \href{http://dx.doi.org/10.1016/0550-3213(86)90246-4}{\emph{Nucl. Phys. B}}
  \textbf{273}:~253~(1986)

\bibitem{Cooper-Sarkar:1991vsl}
A.~M. Cooper-Sarkar, S.~Sarkar, J.~Guy, \emph{et~al.}, \emph{{Bound on the
  tau-neutrino magnetic moment from the BEBC beam dump experiment}}.
  \href{http://dx.doi.org/10.1016/0370-2693(92)90789-7}{\emph{Phys. Lett. B}}
  \textbf{280}:~153~(1992)

\bibitem{Schwienhorst:2001sj}
R.~Schwienhorst \emph{et~al.}, (DONUT), \emph{{A New upper limit for the tau -
  neutrino magnetic moment}}.
  \href{http://dx.doi.org/10.1016/S0370-2693(01)00746-8}{\emph{Phys. Lett. B}}
  \textbf{513}:~23~(2001) , hep-ex/0102026

\bibitem{CONUS:2022qbb}
H.~Bonet \emph{et~al.}, (CONUS), \emph{{First limits on neutrino
  electromagnetic properties from the CONUS experiment}} 2201.12257

\bibitem{Colaresi:2022obx}
J.~Colaresi, J.~I. Collar, T.~W. Hossbach, \emph{et~al.}, \emph{{Suggestive
  evidence for Coherent Elastic Neutrino-Nucleus Scattering from reactor
  antineutrinos}} 2202.09672

\bibitem{Coloma:2022avw}
P.~Coloma, I.~Esteban, M.~C. Gonzalez-Garcia, \emph{et~al.}, \emph{{Bounds on
  new physics with data of the Dresden-II reactor experiment and COHERENT}}
  2202.10829

\bibitem{Liu:2004ny}
D.~W. Liu \emph{et~al.}, (Super-Kamiokande), \emph{{Limits on the neutrino
  magnetic moment using 1496 days of Super-Kamiokande-I solar neutrino data}}.
  \href{http://dx.doi.org/10.1103/PhysRevLett.93.021802}{\emph{Phys. Rev.
  Lett.}} \textbf{93}:~021802~(2004) , hep-ex/0402015

\bibitem{Borexino:2017fbd}
M.~Agostini \emph{et~al.}, (Borexino), \emph{{Limiting neutrino magnetic
  moments with Borexino Phase-II solar neutrino data}}.
  \href{http://dx.doi.org/10.1103/PhysRevD.96.091103}{\emph{Phys. Rev. D}}
  \textbf{96}~(9):~091103~(2017) , 1707.09355

\bibitem{XMASS:2020zke}
K.~Abe \emph{et~al.}, (XMASS), \emph{{Search for exotic neutrino-electron
  interactions using solar neutrinos in XMASS-I}}.
  \href{http://dx.doi.org/10.1016/j.physletb.2020.135741}{\emph{Phys. Lett. B}}
  \textbf{809}:~135741~(2020) , 2005.11891

\bibitem{PandaX-II:2020udv}
X.~Zhou \emph{et~al.}, (PandaX-II), \emph{{A Search for Solar Axions and
  Anomalous Neutrino Magnetic Moment with the Complete PandaX-II Data}}.
  \href{http://dx.doi.org/10.1088/0256-307X/38/10/109902}{\emph{Chin. Phys.
  Lett.}} \textbf{38}~(1):~011301~(2021) , 2008.06485, [Erratum:
  Chin.Phys.Lett. 38, 109902 (2021)]

\bibitem{Fujikawa:1980yx}
K.~Fujikawa and R.~Shrock, \emph{{The Magnetic Moment of a Massive Neutrino and
  Neutrino Spin Rotation}}.
  \href{http://dx.doi.org/10.1103/PhysRevLett.45.963}{\emph{Phys. Rev. Lett.}}
  \textbf{45}:~963~(1980)

\bibitem{Pal:1981rm}
P.~B. Pal and L.~Wolfenstein, \emph{{Radiative Decays of Massive Neutrinos}}.
  \href{http://dx.doi.org/10.1103/PhysRevD.25.766}{\emph{Phys. Rev. D}}
  \textbf{25}:~766~(1982)

\bibitem{Shrock:1982sc}
R.~E. Shrock, \emph{{Electromagnetic Properties and Decays of Dirac and
  Majorana Neutrinos in a General Class of Gauge Theories}}.
  \href{http://dx.doi.org/10.1016/0550-3213(82)90273-5}{\emph{Nucl. Phys. B}}
  \textbf{206}:~359~(1982)

\bibitem{Balantekin:2013sda}
A.~B. Balantekin and N.~Vassh, \emph{{Magnetic moments of active and sterile
  neutrinos}}. \href{http://dx.doi.org/10.1103/PhysRevD.89.073013}{\emph{Phys.
  Rev. D}} \textbf{89}~(7):~073013~(2014) , 1312.6858

\bibitem{Raffelt:1999tx}
G.~G. Raffelt, \emph{{Particle physics from stars}}.
  \href{http://dx.doi.org/10.1146/annurev.nucl.49.1.163}{\emph{Ann. Rev. Nucl.
  Part. Sci.}} \textbf{49}:~163~(1999) , hep-ph/9903472

\bibitem{Alekseev:2017yla}
I.~Alekseev \emph{et~al.}, \emph{{Neutrino Physics at Kalinin Nuclear Power
  Plant: 2002 \textendash{} 2017}}.
  \href{http://dx.doi.org/10.1088/1742-6596/934/1/012006}{\emph{J. Phys. Conf.
  Ser.}} \textbf{934}~(1):~012006~(2017)

\bibitem{Singh:2017bns}
M.~K. Singh, V.~Sharma, L.~Singh, \emph{et~al.}, \emph{{Background rejection of
  TEXONO experiment to explore the sub-keV energy region with HPGe detector}}.
  \href{http://dx.doi.org/10.1007/s12648-017-1004-4}{\emph{Indian J. Phys.}}
  \textbf{91}~(10):~1277~(2017)

\bibitem{Akindele:2022sti}
O.~A. Akindele \emph{et~al.}, \emph{{High Energy Physics Opportunities Using
  Reactor Antineutrinos}} (2022), 2203.07214

\bibitem{Barbeau:2021exu}
P.~S. Barbeau, Y.~Efremenko, and K.~Scholberg, \emph{{COHERENT at the
  Spallation Neutron Source}} 2111.07033

\bibitem{Baxter:2019mcx}
D.~Baxter \emph{et~al.}, \emph{{Coherent Elastic Neutrino-Nucleus Scattering at
  the European Spallation Source}}.
  \href{http://dx.doi.org/10.1007/JHEP02(2020)123}{\emph{JHEP}}
  \textbf{02}:~123~(2020) , 1911.00762

\bibitem{Anchordoqui:2021ghd}
L.~A. Anchordoqui \emph{et~al.}, \emph{{The Forward Physics Facility: Sites,
  Experiments, and Physics Potential}} 2109.10905

\bibitem{Feng:2022inv}
J.~L. Feng \emph{et~al.}, \emph{{The Forward Physics Facility at the
  High-Luminosity LHC}}. In \emph{{2022 Snowmass Summer Study}} (2022),
  2203.05090

\bibitem{Ismail:2021dyp}
A.~Ismail, S.~Jana, and R.~M. Abraham, \emph{{Neutrino up-scattering via the
  dipole portal at forward LHC detectors}}.
  \href{http://dx.doi.org/10.1103/PhysRevD.105.055008}{\emph{Phys. Rev. D}}
  \textbf{105}~(5):~055008~(2022) , 2109.05032

\bibitem{Acero:2022wqg}
M.~A. Acero \emph{et~al.}, \emph{{White Paper on Light Sterile Neutrino
  Searches and Related Phenomenology}} (2022), 2203.07323

\bibitem{Abraham:2022jse}
R.~M. Abraham \emph{et~al.}, \emph{{Tau Neutrinos in the Next Decade: from GeV
  to EeV}} (2022), 2203.05591

\bibitem{Abdullahi:2022jlv}
A.~M. Abdullahi \emph{et~al.}, \emph{{The Present and Future Status of Heavy
  Neutral Leptons}} (2022), 2203.08039

\bibitem{XENON:2020rca}
E.~Aprile \emph{et~al.}, (XENON), \emph{{Excess electronic recoil events in
  XENON1T}}. \href{http://dx.doi.org/10.1103/PhysRevD.102.072004}{\emph{Phys.
  Rev. D}} \textbf{102}~(7):~072004~(2020) , 2006.09721

\bibitem{Yue:2021vjg}
B.~Yue, J.~Liao, and J.~Ling, \emph{{Probing neutrino magnetic moment at the
  Jinping neutrino experiment}}.
  \href{http://dx.doi.org/10.1007/JHEP08(2021)068}{\emph{JHEP}}
  \textbf{08}~(068):~068~(2021) , 2102.12259

\bibitem{Ye:2021zso}
Z.~Ye, F.~Zhang, D.~Xu, \emph{et~al.}, \emph{{Unambiguously Resolving the
  Potential Neutrino Magnetic Moment Signal at Large Liquid Scintillator
  Detectors}}.
  \href{http://dx.doi.org/10.1088/0256-307X/38/11/111401}{\emph{Chin. Phys.
  Lett.}} \textbf{38}~(11):~111401~(2021) , 2103.11771

\bibitem{LZ:2021xov}
D.~S. Akerib \emph{et~al.}, (LZ), \emph{{Projected sensitivities of the
  LUX-ZEPLIN experiment to new physics via low-energy electron recoils}}.
  \href{http://dx.doi.org/10.1103/PhysRevD.104.092009}{\emph{Phys. Rev. D}}
  \textbf{104}~(9):~092009~(2021) , 2102.11740

\bibitem{Bressi:2011yfa}
G.~Bressi, G.~Carugno, F.~Della~Valle, \emph{et~al.}, \emph{{Testing the
  neutrality of matter by acoustic means in a spherical resonator}}.
  \href{http://dx.doi.org/10.1103/PhysRevA.83.052101}{\emph{Phys. Rev. A}}
  \textbf{83}~(5):~052101~(2011) , 1102.2766

\bibitem{TEXONO:2002pra}
H.~B. Li \emph{et~al.}, (TEXONO), \emph{{Limit on the electron neutrino
  magnetic moment from the Kuo-Sheng reactor neutrino experiment}}.
  \href{http://dx.doi.org/10.1103/PhysRevLett.90.131802}{\emph{Phys. Rev.
  Lett.}} \textbf{90}:~131802~(2003) , hep-ex/0212003

\bibitem{Gninenko:2006fi}
S.~N. Gninenko, N.~V. Krasnikov, and A.~Rubbia, \emph{{Search for millicharged
  particles in reactor neutrino experiments: A Probe of the PVLAS anomaly}}.
  \href{http://dx.doi.org/10.1103/PhysRevD.75.075014}{\emph{Phys. Rev. D}}
  \textbf{75}:~075014~(2007) , hep-ph/0612203

\bibitem{Studenikin:2013my}
A.~Studenikin, \emph{{New bounds on neutrino electric millicharge from limits
  on neutrino magnetic moment}}.
  \href{http://dx.doi.org/10.1209/0295-5075/107/21001}{\emph{EPL}}
  \textbf{107}~(2):~21001~(2014) , 1302.1168, [Erratum: EPL 107, 39901 (2014),
  Erratum: Europhys.Lett. 107, 39901 (2014)]

\bibitem{Chen:2014dsa}
J.-W. Chen, H.-C. Chi, H.-B. Li, \emph{et~al.}, \emph{{Constraints on
  millicharged neutrinos via analysis of data from atomic ionizations with
  germanium detectors at sub-keV sensitivities}}.
  \href{http://dx.doi.org/10.1103/PhysRevD.90.011301}{\emph{Phys. Rev. D}}
  \textbf{90}~(1):~011301~(2014) , 1405.7168

\bibitem{Das:2020egb}
A.~Das, D.~Ghosh, C.~Giunti, \emph{et~al.}, \emph{{Neutrino charge constraints
  from scattering to the weak gravity conjecture to neutron stars}}.
  \href{http://dx.doi.org/10.1103/PhysRevD.102.115009}{\emph{Phys. Rev. D}}
  \textbf{102}~(11):~115009~(2020) , 2005.12304

\bibitem{Babu:1993yh}
K.~S. Babu, T.~M. Gould, and I.~Z. Rothstein, \emph{{Closing the windows on MeV
  Tau neutrinos}}.
  \href{http://dx.doi.org/10.1016/0370-2693(94)90340-9}{\emph{Phys. Lett. B}}
  \textbf{321}:~140~(1994) , hep-ph/9310349

\bibitem{Raffelt:1999gv}
G.~G. Raffelt, \emph{{Limits on neutrino electromagnetic properties: An
  update}}. \href{http://dx.doi.org/10.1016/S0370-1573(99)00074-5}{\emph{Phys.
  Rept.}} \textbf{320}:~319~(1999)

\bibitem{Shrock:1974nd}
R.~Shrock, \emph{{Decay l0 ---\ensuremath{>} nu(lepton) gamma in gauge theories
  of weak and electromagnetic interactions}}.
  \href{http://dx.doi.org/10.1103/PhysRevD.9.743}{\emph{Phys. Rev. D}}
  \textbf{9}:~743~(1974)

\bibitem{Marciano:1977wx}
W.~J. Marciano and A.~I. Sanda, \emph{{Exotic Decays of the Muon and Heavy
  Leptons in Gauge Theories}}.
  \href{http://dx.doi.org/10.1016/0370-2693(77)90377-X}{\emph{Phys. Lett. B}}
  \textbf{67}:~303~(1977)

\bibitem{Lee:1977tib}
B.~W. Lee and R.~E. Shrock, \emph{{Natural Suppression of Symmetry Violation in
  Gauge Theories: Muon - Lepton and Electron Lepton Number Nonconservation}}.
  \href{http://dx.doi.org/10.1103/PhysRevD.16.1444}{\emph{Phys. Rev. D}}
  \textbf{16}:~1444~(1977)

\bibitem{Petcov:1976ff}
S.~T. Petcov, \emph{{The Processes $\mu \rightarrow e + \gamma, \mu \rightarrow
  e + \overline{e}, \nu' \rightarrow \nu + \gamma$ in the Weinberg-Salam Model
  with Neutrino Mixing}}. \emph{Sov. J. Nucl. Phys.} \textbf{25}:~340~(1977) ,
  [Erratum: Sov.J.Nucl.Phys. 25, 698 (1977), Erratum: Yad.Fiz. 25, 1336 (1977)]

\bibitem{Goldman:1977jx}
J.~T. Goldman and G.~J. Stephenson, Jr., \emph{{Limits on the Mass of the
  Muon-neutrino in the Absence of Muon Lepton Number Conservation}}.
  \href{http://dx.doi.org/10.1103/PhysRevD.16.2256}{\emph{Phys. Rev. D}}
  \textbf{16}:~2256~(1977)

\bibitem{Zatsepin:1978iy}
G.~T. Zatsepin and A.~Y. Smirnov, \emph{{Neutrino Decay in Gauge Theories}}.
  \emph{Yad. Fiz.} \textbf{28}:~1569~(1978)

\bibitem{Bernal:2021ylz}
J.~L. Bernal, A.~Caputo, F.~Villaescusa-Navarro, \emph{et~al.},
  \emph{{Searching for the Radiative Decay of the Cosmic Neutrino Background
  with Line-Intensity Mapping}}.
  \href{http://dx.doi.org/10.1103/PhysRevLett.127.131102}{\emph{Phys. Rev.
  Lett.}} \textbf{127}~(13):~131102~(2021) , 2103.12099

\bibitem{Arguelles:2022xxa}
C.~A. Arg\"uelles \emph{et~al.}, \emph{{Snowmass White Paper: Beyond the
  Standard Model effects on Neutrino Flavor}}. In \emph{{2022 Snowmass Summer
  Study}} (2022), 2203.10811

\bibitem{Choubey:2020dhw}
S.~Choubey, M.~Ghosh, D.~Kempe, \emph{et~al.}, \emph{{Exploring invisible
  neutrino decay at ESSnuSB}}.
  \href{http://dx.doi.org/10.1007/JHEP05(2021)133}{\emph{JHEP}}
  \textbf{05}:~133~(2021) , 2010.16334

\bibitem{Chakraborty:2020cfu}
K.~Chakraborty, D.~Dutta, S.~Goswami, \emph{et~al.}, \emph{{Addendum to:
  Invisible neutrino decay: first vs second oscillation maximum}}.
  \href{http://dx.doi.org/10.1007/JHEP08(2021)136}{\emph{JHEP}}
  \textbf{08}:~136~(2021) , 2012.04958

\bibitem{Delgado:2021vha}
E.~A. Delgado, H.~Nunokawa, and A.~A. Quiroga, \emph{{Probing neutrino decay
  scenarios by using the Earth matter effects on supernova neutrinos}}.
  \href{http://dx.doi.org/10.1088/1475-7516/2022/01/003}{\emph{JCAP}}
  \textbf{01}~(01):~003~(2022) , 2109.02737

\bibitem{Akita:2021hqn}
K.~Akita, G.~Lambiase, and M.~Yamaguchi, \emph{{Unstable cosmic neutrino
  capture}}. \href{http://dx.doi.org/10.1007/JHEP02(2022)132}{\emph{JHEP}}
  \textbf{02}:~132~(2022) , 2109.02900

\bibitem{Picoreti:2021yct}
R.~Picoreti, D.~Pramanik, P.~C. de~Holanda, \emph{et~al.}, \emph{{Updating
  $\nu_{3}$ lifetime from solar antineutrino spectra}} 2109.13272

\bibitem{Greenberg:2002uu}
O.~W. Greenberg, \emph{{CPT violation implies violation of Lorentz
  invariance}}.
  \href{http://dx.doi.org/10.1103/PhysRevLett.89.231602}{\emph{Phys. Rev.
  Lett.}} \textbf{89}:~231602~(2002) , hep-ph/0201258

\bibitem{Colladay:1996iz}
D.~Colladay and V.~A. Kostelecky, \emph{{CPT violation and the standard
  model}}. \href{http://dx.doi.org/10.1103/PhysRevD.55.6760}{\emph{Phys. Rev.
  D}} \textbf{55}:~6760~(1997) , hep-ph/9703464

\bibitem{Colladay:1998fq}
D.~Colladay and V.~A. Kostelecky, \emph{{Lorentz violating extension of the
  standard model}}.
  \href{http://dx.doi.org/10.1103/PhysRevD.58.116002}{\emph{Phys. Rev. D}}
  \textbf{58}:~116002~(1998) , hep-ph/9809521

\bibitem{Kostelecky:2008ts}
V.~A. Kostelecky and N.~Russell, \emph{{Data Tables for Lorentz and CPT
  Violation}} 0801.0287

\bibitem{Kostelecky:2011gq}
A.~Kostelecky and M.~Mewes, \emph{{Neutrinos with Lorentz-violating operators
  of arbitrary dimension}}.
  \href{http://dx.doi.org/10.1103/PhysRevD.85.096005}{\emph{Phys. Rev. D}}
  \textbf{85}:~096005~(2012) , 1112.6395

\bibitem{Arguelles:2021kjg}
C.~A. Arg\"uelles and T.~Katori, \emph{{Lorentz Symmetry and High-Energy
  Neutrino Astronomy}}.
  \href{http://dx.doi.org/10.3390/universe7120490}{\emph{Universe}}
  \textbf{7}~(12):~490~(2021) , 2109.13973

\bibitem{Antonelli:2020nhn}
V.~Antonelli, L.~Miramonti, and M.~D.~C. Torri, \emph{{Phenomenological Effects
  of CPT and Lorentz Invariance Violation in Particle and Astroparticle
  Physics}}. \href{http://dx.doi.org/10.3390/sym12111821}{\emph{Symmetry}}
  \textbf{12}~(11):~1821~(2020) , 2110.09185

\bibitem{Stecker:2022tzd}
F.~W. Stecker, \emph{{Testing Lorentz Invariance with Neutrinos}} (2022)

\bibitem{Lehnert:2021tbv}
R.~Lehnert, \emph{{Beta-Decay Spectrum and Lorentz Violation}} 2112.13803

\bibitem{Gasperini:1988zf}
M.~Gasperini, \emph{{Testing the Principle of Equivalence with Neutrino
  Oscillations}}. \href{http://dx.doi.org/10.1103/PhysRevD.38.2635}{\emph{Phys.
  Rev. D}} \textbf{38}:~2635~(1988)

\bibitem{Halprin:1995vg}
A.~Halprin, C.~N. Leung, and J.~T. Pantaleone, \emph{{A possible violation of
  the equivalence principle by neutrinos}}.
  \href{http://dx.doi.org/10.1103/PhysRevD.53.5365}{\emph{Phys. Rev. D}}
  \textbf{53}:~5365~(1996) , hep-ph/9512220

\bibitem{Fiorillo:2020gsb}
D.~F.~G. Fiorillo, G.~Mangano, S.~Morisi, \emph{et~al.}, \emph{{IceCube
  constraints on Violation of Equivalence Principle}}. \emph{JCAP}
  \textbf{2104}:~079~(2021) , 2012.07867

\bibitem{Esmaili:2021omm}
A.~Esmaili, \emph{{Violation of Equivalence Principle in Neutrino Sector:
  Probing the Extended Parameter Space}}. \emph{JCAP} \textbf{2107}:~018~(2021)
  , 2105.08744

\bibitem{Chianese:2021vkf}
M.~Chianese, D.~F.~G. Fiorillo, G.~Mangano, \emph{et~al.}, \emph{{Sensitivity
  of KM3NeT to Violation of Equivalence Principle}}. \emph{Symmetry}
  \textbf{13}:~1353~(2021) , 2107.13013

\bibitem{Diaz:2020aax}
F.~N. D\'\i{}az, J.~Hoefken, and A.~M. Gago, \emph{{Effects of the Violation of
  the Equivalence Principle at DUNE}}.
  \href{http://dx.doi.org/10.1103/PhysRevD.102.055020}{\emph{Phys. Rev. D}}
  \textbf{102}~(5):~055020~(2020) , 2003.13712

\bibitem{Valdiviesso:2008vyk}
G.~A. Valdiviesso, M.~M. Guzzo, and P.~C. de~Holanda, \emph{{Probing new limits
  for the Violation of the Equivalence Principle in the
  solar\textendash{}reactor neutrino sector as a next to leading order
  effect}}. \href{http://dx.doi.org/10.1016/j.physletb.2011.05.057}{\emph{Phys.
  Lett. B}} \textbf{701}:~240~(2011) , 0811.2128

\end{thebibliography}
\end{document}